\documentclass[aoas,preprint]{imsart}

\usepackage{amssymb, amsfonts, amsthm}
\usepackage{graphicx,psfrag,epsf}
\usepackage{graphics}
\usepackage{enumerate}
\usepackage{booktabs}
\usepackage{multirow}
\usepackage{hyperref}
\usepackage{color}
\usepackage{bm}
\usepackage{bbm}
\usepackage{parskip}
\usepackage{tikz}
\usepackage[intlimits]{amsmath}
\usepackage{geometry}
\usepackage[export]{adjustbox}
\usepackage{natbib}
\usepackage{url} 
\usepackage{caption}
\usepackage{subcaption}
\usepackage{mathtools}
\usepackage{floatrow}
\usepackage[toc,page]{appendix}
\usepackage{multicol}

\startlocaldefs

\newcommand{\rtwo}{\mbox{${\mathbb{R}^{2}}$}}
\newcommand{\rone}{\mbox{${\mathbb{R}}$}}

\newcommand{\var}{\mathrm{Var}}

\newcommand{\rpm}{\sbox0{$1$}\sbox2{$\scriptstyle\pm$}
  \raise\dimexpr(\ht0-\ht2)/2\relax\box2 }
\newcommand{\bfs}{\mbox{${\mathbf{s}}$}}
\newcommand{\bft}{\mbox{${\mathbf{t}}$}}
\newcommand{\bfh}{\mbox{${\mathbf{h}}$}}
\newcommand{\bfx}{\mbox{${\mathbf{x}}$}}
\newcommand{\bfbeta}{\mbox{$\boldsymbol{\beta}$}}

\newcommand{\bfeps}{\mbox{$\boldsymbol{\epsilon}$}}
\newcommand{\bftheta}{\mbox{$\boldsymbol{\theta}$}}

\newfloatcommand{capbtabbox}{table}[][\FBwidth]

\theoremstyle{plain}

\theoremstyle{definition}
\newtheorem{eg}{Example}

\theoremstyle{remark}

\numberwithin{equation}{section}
\numberwithin{theo}{section}
\numberwithin{cor}{section}
\numberwithin{lem}{section}
\numberwithin{prop}{section}
\numberwithin{cor}{section}
\numberwithin{eg}{section}
\numberwithin{rem}{section}

\numberwithin{equation}{section} 

\def\Lp{\left(}
\def\Rp{\right)}
\def\LP{\left\{ } 
\def\RP{\right\}}
\endlocaldefs

\begin{document}

\begin{frontmatter}
\title{A Statistical Analysis of Noisy Crowdsourced Weather Data}
\runtitle{A Statistical Analysis of Noisy Crowdsourced Weather Data}

\begin{aug}
\author{\fnms{Arnab} \snm{Chakraborty}\ead[label=e1]{achakra6@ncsu.edu}},
\author{\fnms{Soumendra Nath} \snm{Lahiri}}
\and
\author{\fnms{Alyson} \snm{Wilson}
\ead[label=e3]{agwilso2@ncsu.edu}
\ead[label=u1,url]{http://www.foo.com}}

\runauthor{A. Chakraborty et al.}

\affiliation{Department of Statistics, North Carolina State University}

\address{5109 SAS Hall\\
2311 Stinson Dr.\\
Raleigh, NC 27695-8203\\
\printead{e1}}

\end{aug}

\begin{abstract}
Spatial prediction of weather-elements like temperature, precipitation, and barometric pressure are generally based on satellite imagery or data collected at ground-stations. None of these data provide information at a more granular or ``hyper-local'' resolution. On the other hand, crowdsourced weather data, which are captured by sensors installed on mobile devices and gathered by weather-related mobile apps like \texttt{WeatherSignal} and \texttt{AccuWeather}, can serve as potential data sources for analyzing environmental processes at a hyper-local resolution. However, due to the low quality of the sensors and the non-laboratory environment, the quality of the observations in crowdsourced data is compromised. This paper describes methods to improve hyper-local spatial prediction using this varying-quality noisy crowdsourced information. We introduce a reliability metric, namely Veracity Score (VS), to assess the quality of the crowdsourced observations using a coarser, but high-quality, reference data. A VS-based methodology to analyze noisy spatial data is proposed and evaluated through extensive simulations. The merits of the proposed approach are illustrated through case studies analyzing crowdsourced daily average ambient temperature readings for one day in the contiguous United States.
\end{abstract}


\end{frontmatter}

\section{Introduction}
\label{sec:intro}
In recent years there has been a proliferation of weather-related applications for mobile devices such as cellphones, iPods, and laptops. These applications not only provide service to the user but also collect and share spatial data on location, ambient temperature, barometric pressure, humidity, etc., captured by the small-scale sensors installed in the devices. Analyzing and understanding these crowdsourced data sets is becoming an area of increasing interest.

One use of the mobile sensor-generated data is to analyze and understand atmospheric processes at very fine spatial resolution. Most of the methodologies in literature for spatial prediction of weather elements are based on global images coming from satellites or measurements taken at meteorological stations on the ground (for example, see \citealt{thor97}; \citealt{florio04} etc.). But none of these sources are dense enough so that the variability of the process can be analyzed in `hyper-local' regions, e.g. rectangular regions inside the population centers with each sides varying approximately in between 25 to 30 miles ($0.3^\circ - 0.6^\circ$ in latitude and longitude). For instance, the ground-stations are generally situated away from localities e.g. at airports or national parks etc. Hence, weather-related analysis solely based on ground-station data does not often provide correct assessment of the variation of the underlying process in the localities. However, in disaster detection, traffic management, and many defense-related activities, prediction of the process in a very localized region (hyper-local) is often more important than the global imputation of the process over a bigger region. Crowdsourced data captured by mobile sensors can serve as a potential source in these scenarios, especially in regions where the ground weather stations are sparse but the population density and hence the density of the mobile-devices like cellphones, iPads etc. is relatively high. Recently, a handful of organizations are becoming interested in providing cost-effective hyper-local predictions of weather using sensor-generated geographical information through weather-related mobile apps. For example, the global leader in weather information, AccuWeather, launched AccUcast in 2015 (\citealt{accucast}), a feature that allows each user to share their local weather information as captured by the built-in mobile sensors. Other applications include Sunshine (\citealt{wired}) and Dark Sky (\citealt{engadget}), which turn each app-user into a ``meteorological station'' for gathering and sharing  hyper-local weather information. Mobile sensor generated weather data are already being used in traffic management, fire detection etc. In a recent article, \cite{sosko17} have used a crowdsourced mobile-sensor data in forest fire detection to densify the static geo-sensor network (SGN), which is primarily comprised of meteorological stations with high-performance sensors. Though spatial prediction of daily weather is generally based on satellite imagery or data from weather stations (\citealt{thor97}, \citealt{vancu10}, \citealt{frei14}), recent advancement of weather-related mobile apps and the concurrent business interests call for a new methodology that considers these crowdsourced weather data to generate more accurate weather prediction in hyper-local regions. In this article, we consider the daily average ambient temperature process, and show that more efficient and reasonable prediction surfaces can be created in hyper-local regions with denser but noisy crowdsourced data as compared to a global prediction surface obtained from high-quality but coarser ground-station data.

\subsection{WeatherSignal and NOAA ground-station data}
\label{subsec:OpenSignal}
\begin{figure}
  \centering
 \begin{subfigure}{0.48\textwidth}
     \centering
    \includegraphics[trim={5.5cm 4cm 3.5cm 2.5cm}, width=.58\textwidth, height=0.18\textheight]{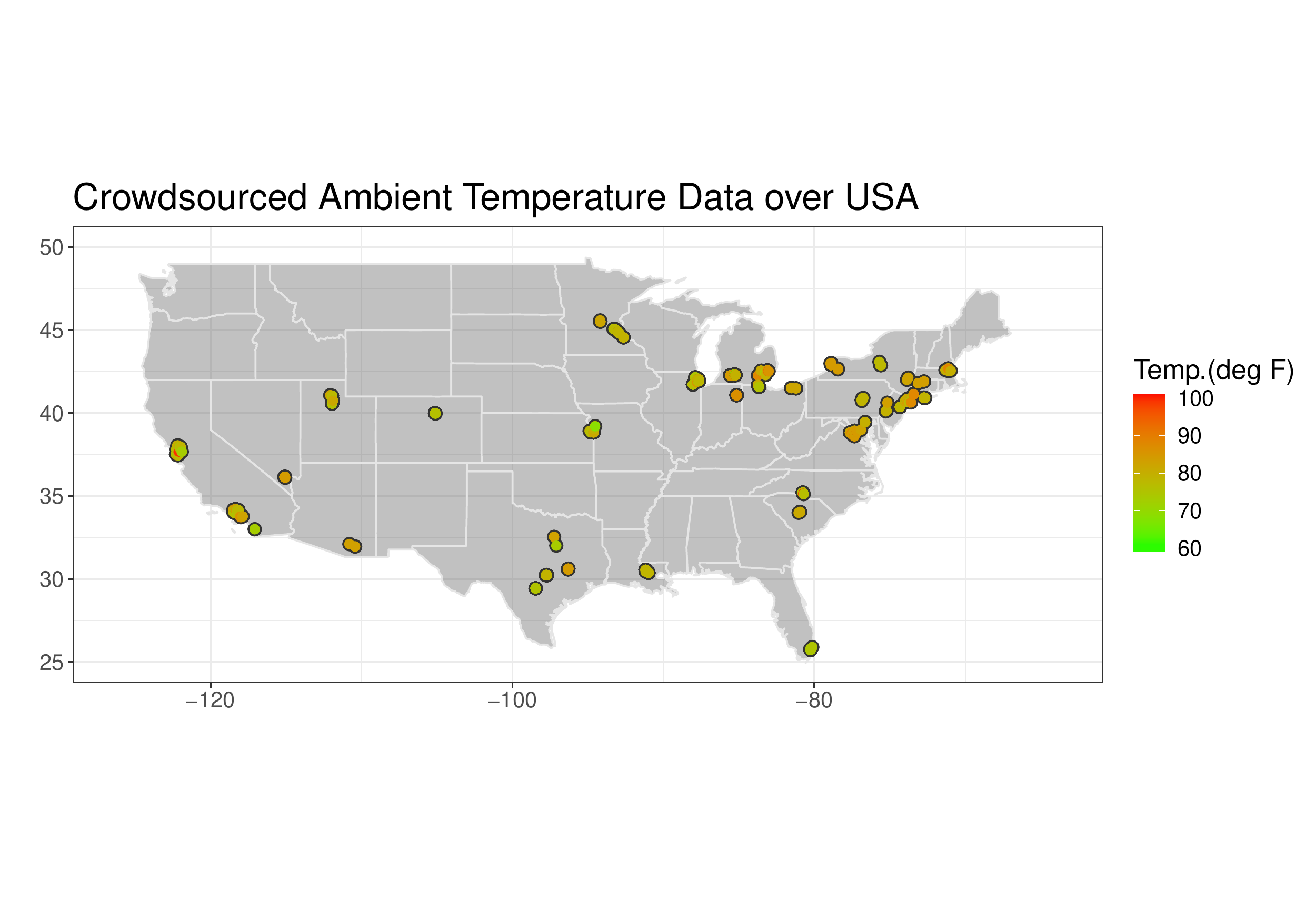}
    \subcaption{WeatherSignal (WS) data}\label{fig:CSDatAllUsa}
 \end{subfigure}\hspace{2mm}%
 \begin{subfigure}{0.48\textwidth}
     \centering
    \includegraphics[trim={3.5cm 4cm 5.5cm 2.5cm}, width=.58\textwidth, height=0.18\textheight]{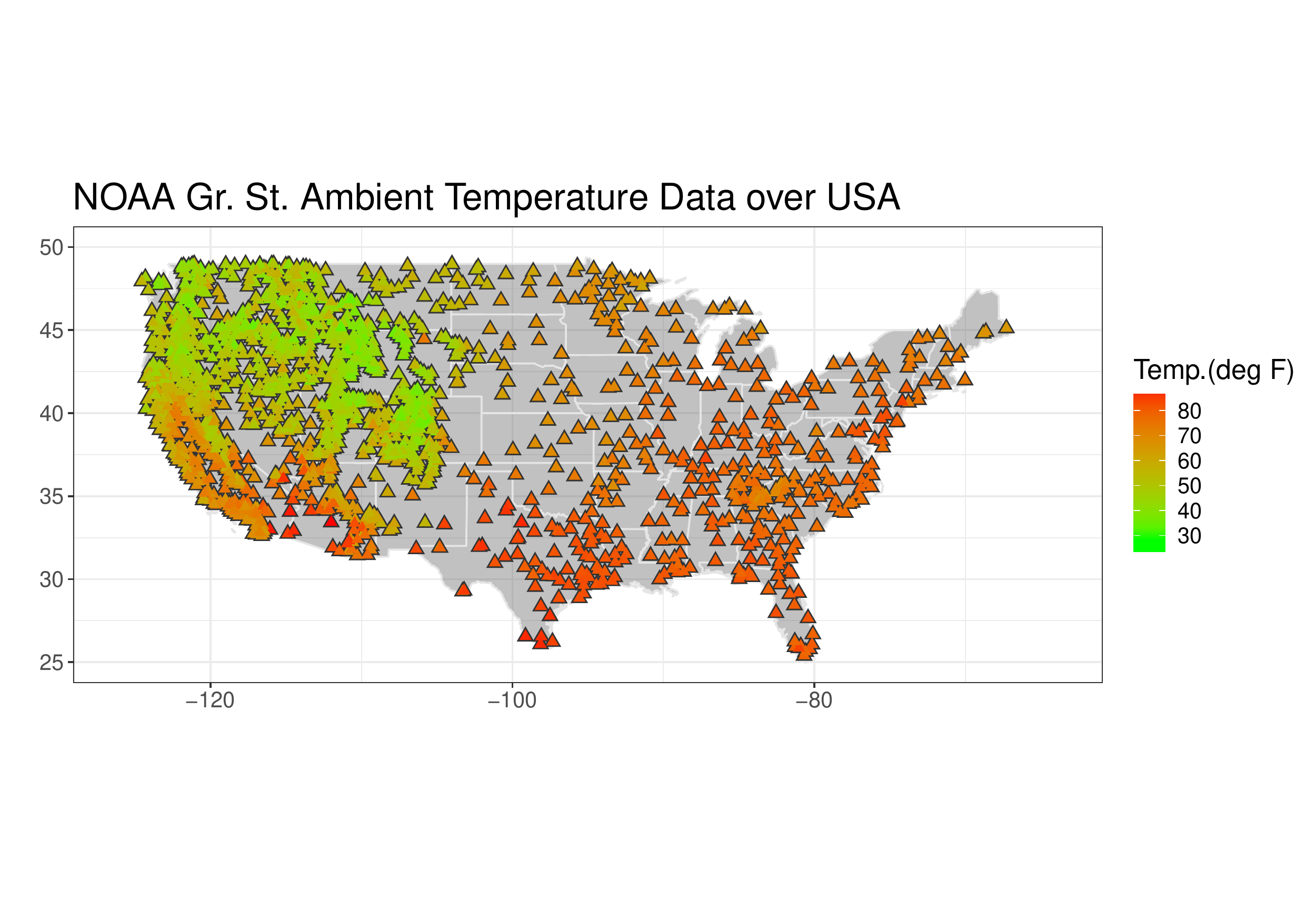}
    \subcaption{NOAA station data}\label{fig:GSDatAllUsa}
 \end{subfigure}
   \begin{subfigure}{0.25\textwidth}
     \centering
    \includegraphics[trim={3.5cm 1.2cm 3cm 1cm}, width=.8\textwidth, height=0.14\textheight]{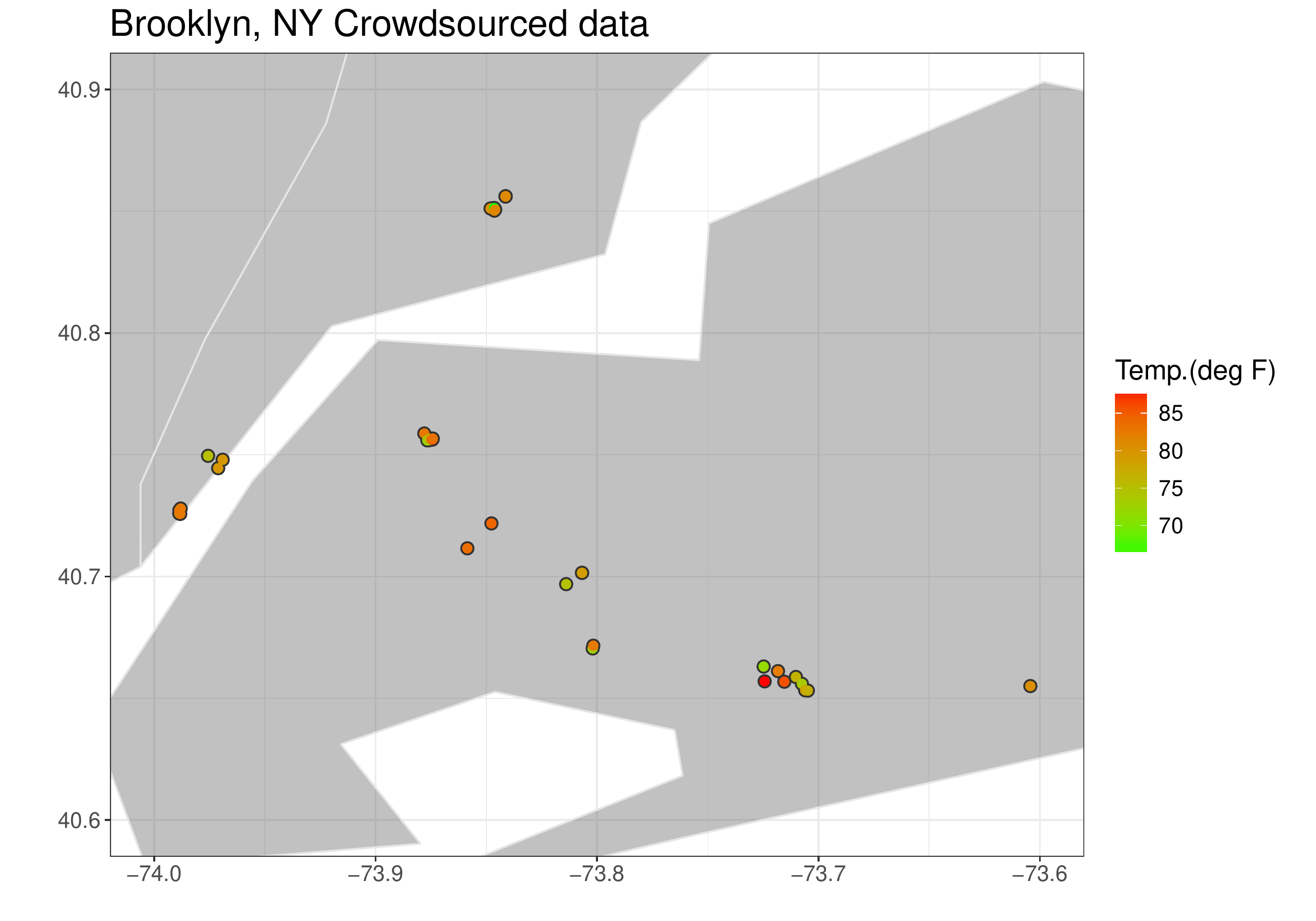}
    \subcaption{WS, Brooklyn}\label{fig:CS_NYC}
 \end{subfigure}%
 \begin{subfigure}{0.25\textwidth}
     \centering
    \includegraphics[trim={2.8cm 1.4cm 2.2cm 1cm}, width=.8\textwidth, height=0.14\textheight]{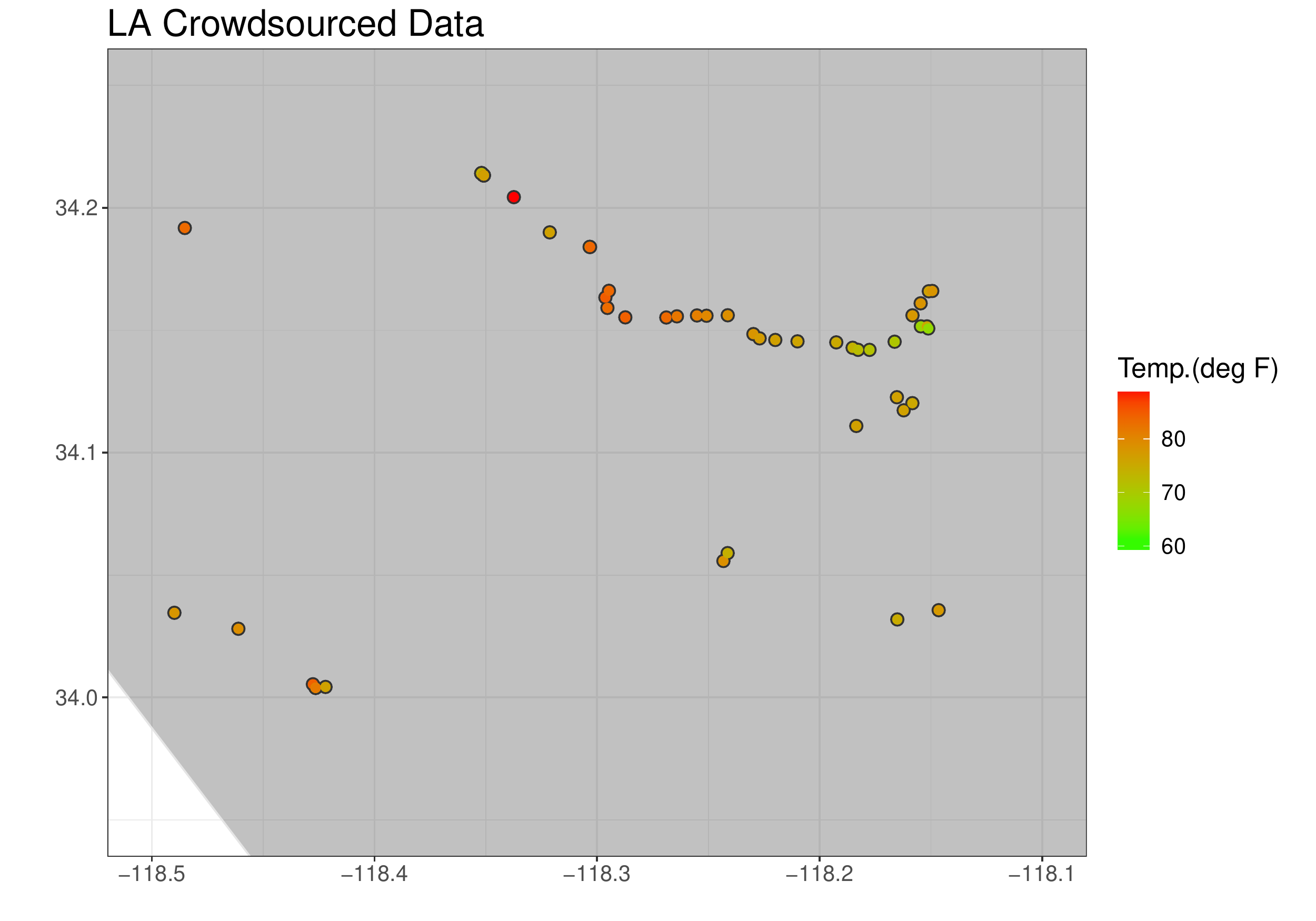}
    \subcaption{WS, LA}\label{fig:CS_LA}
 \end{subfigure}%
    \begin{subfigure}{0.25\textwidth}
     \centering
    \includegraphics[trim={3cm 1.2cm 3.5cm 1cm}, width=.8\textwidth, height=0.14\textheight]{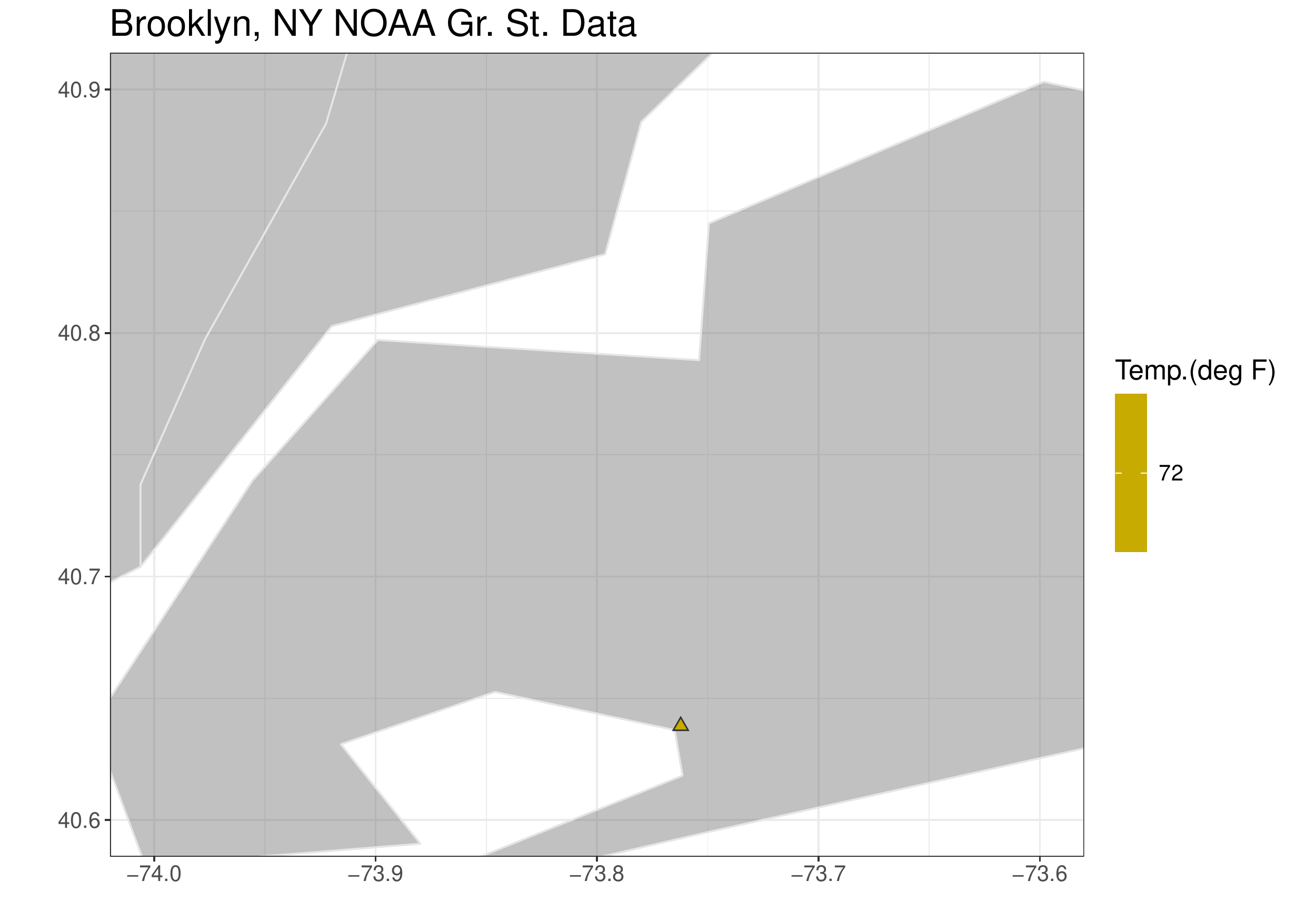}
    \subcaption{NOAA, Brooklyn}\label{fig:GS_NYC}
 \end{subfigure}%
 \begin{subfigure}{0.25\textwidth}
     \centering
    \includegraphics[trim={2.8cm 1.4cm 2.2cm 1cm}, width=.8\textwidth, height=0.14\textheight]{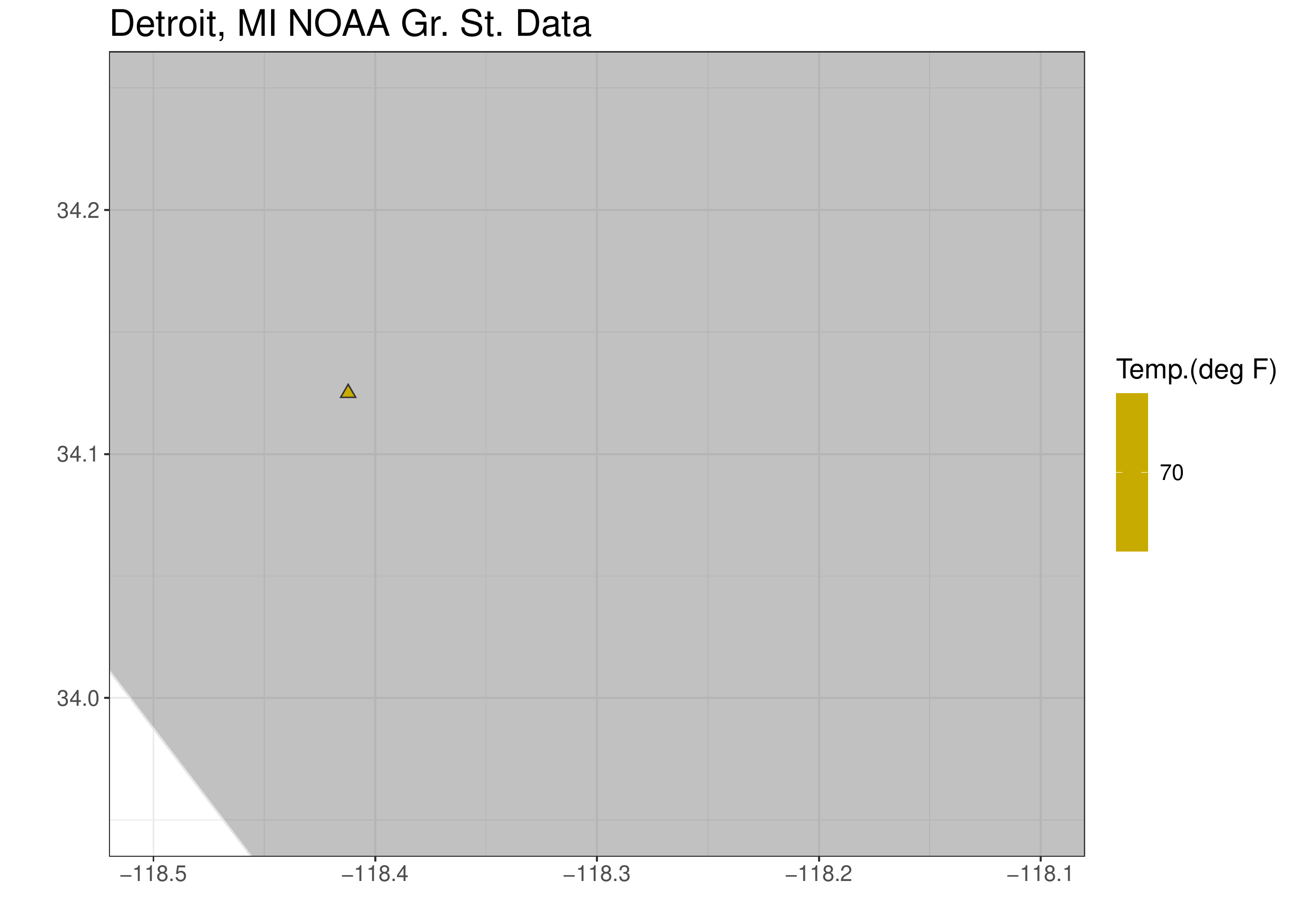}
    \subcaption{NOAA, LA}\label{fig:GS_LA}
 \end{subfigure}
 \caption{Spatial plots of the crowdsourced and NOAA ground-station data. (c) - (f) show zoomed `hyper-local' versions (each side of these regions vary from 25 to 30 miles approximately) of the crowdsourced WeatherSignal (c - d) and NOAA station data (e - f).}\label{fig:allDat}
 \end{figure}

We analyze a static crowdsourced data set consisting of geo-coded daily average ambient temperature readings over the continental United States on April 30, 2013. These data were gathered by a cellphone application named \texttt{WeatherSignal}, available both for iOS and Android. In addition to providing information on current weather and forecasts, the app also gathers geographic and weather information using cellphone sensors, leading to a huge amount of crowdsourced spatial weather data from all over the globe. The \texttt{WeatherSignal} application is operated by an organization named \href{https://opensignal.com/}{OpenSignal}. Through the research partnership program of OpenSignal, we were provided real-time (in milliseconds) ambient temperature readings captured by various mobile phones for the above-mentioned day. For each spatial location, we have temporally aggregated the temperature readings to the daily average by taking mean of the regionally estimated hourly temperatures throughout the day. The details of the aggregation are explained elaborately in Section A.1 in the supplementary material. After the aggregation, we have the crowdsourced daily average temperature readings at 1879 spatial locations in the United States, as shown in Figure~\ref{fig:CSDatAllUsa}. From the figure, it can be seen that the crowdsourced observations are clumped together in high-population density regions like Detroit, Chicago, New York, and Los Angeles etc. In Figure~\ref{fig:CS_NYC} and \ref{fig:CS_LA} we show hyper-local versions of the WeatherSignal data for two nearly square hyper-local regions at Brooklyn, NY and Los Angeles, CA.

Along with the crowdsourced data from the WeatherSignal app, we also have ground-station data on the daily average ambient temperature from the National Oceanic and Atmospheric Administration (NOAA). We used the Global Historical Climate Network Daily (\href{https://www.ncdc.noaa.gov/ghcn-daily-description}{GHCND}) data access tool to retrieve the daily ambient temperature summaries for April 30, 2013 from 2094 stations in the continental United States. We have plotted the ground-station observations in Figure~\ref{fig:GSDatAllUsa}.

Comparing Figure~\ref{fig:CSDatAllUsa} and Figure~\ref{fig:GSDatAllUsa}, we can see that the NOAA ground-station data provides much more spatial coverage than the crowdsourced data in the entire United States or large parts of United States like east-coast, mid-west etc. are considered and hence for global modeling or building a global prediction surface of the ambient temperature, the ground-station data is clearly a better choice. However, for hyper-local prediction of the spatial process, we believe that crowdsourced data has the potential to capture the local behavior of the spatial process more accurately. For example, in Figure~\ref{fig:GS_NYC} and \ref{fig:GS_LA} we have plotted the available ground-station observations in the same square neighborhoods as the crowdsourced data in Figure~\ref{fig:CS_NYC} and \ref{fig:CS_LA}. In the area around Brooklyn, NY, there are approximately 90 crowdsourced observations available, where as the number of ground-station observations is only one. Motivated by this observation, in this paper, we propose a method to improve the accuracy of the hyper-local predictions using the available crowdsourced information in addition to the ground-station data over a bigger surrounding.

\subsection{The challenge in analyzing crowdsourced mobile-sensor data}
\label{subsec:challengeInOpensignal}
The challenge in analyzing mobile sensor-generated crowdsourced data lies in the low quality and hence poor reliability of an unknown proportion of the data. When data are collected from mobile applications, the readings are prone to contamination for various reasons. The inaccurate observations can occur due external factors, low-resolution sensors, or a combination of these factors. For instance, the temperature readings can be affected by battery temperature, whether the user is indoor or outdoor, the proximity of the device to a hot or cold object, the heterogeneity of the sensors used by different devices, and many other unknown processes.

\begin{figure}
  \centering
 \begin{subfigure}{0.5\textwidth}
     \centering
    \includegraphics[trim={1.5cm .5cm 1.5cm 0.4cm}, width=.55\textwidth, height=0.19\textheight]{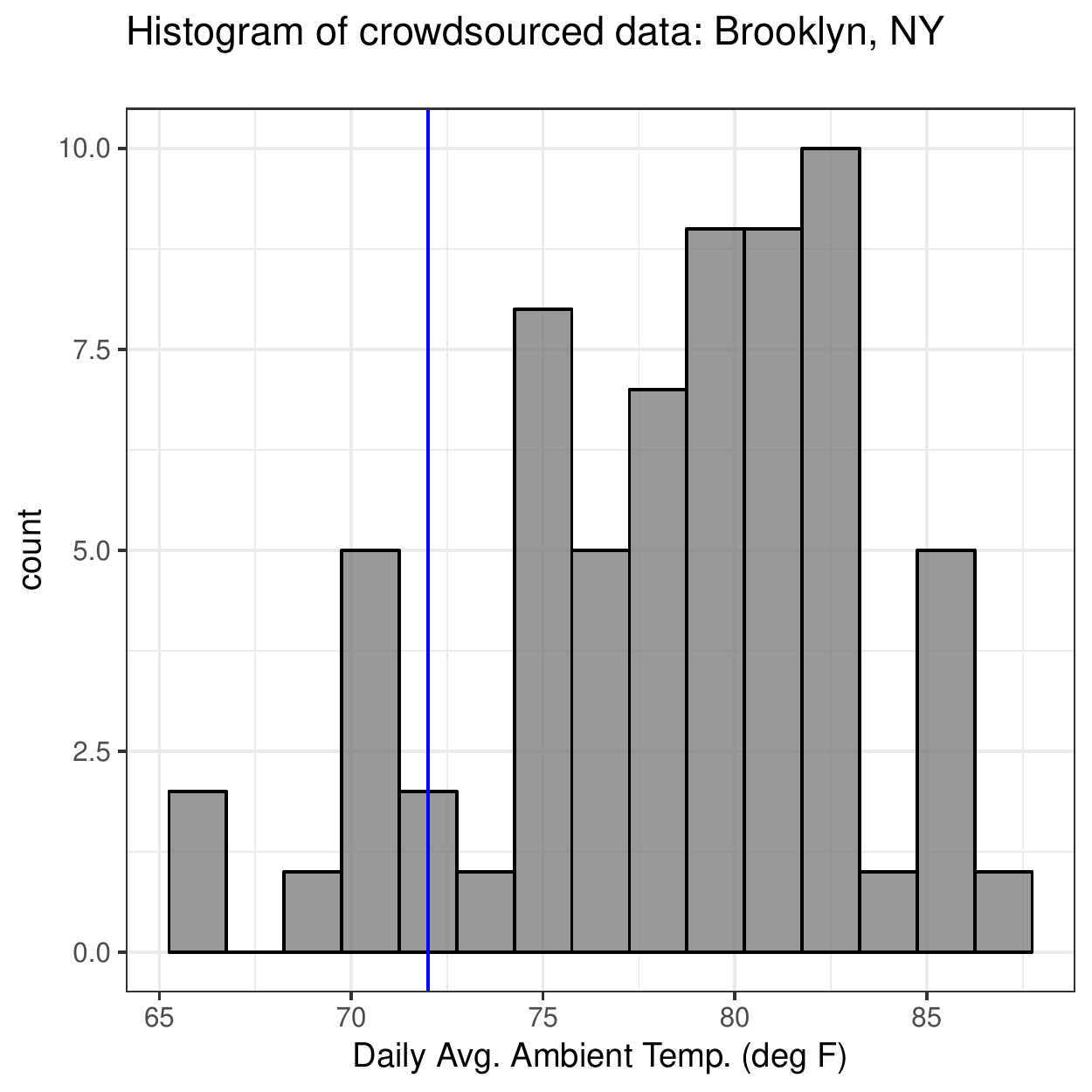}
 \end{subfigure}%
 \begin{subfigure}{0.5\textwidth}
     \centering
    \includegraphics[trim={1.5cm .5cm 1.5cm 0.4cm}, width=.55\textwidth, height=0.19\textheight]{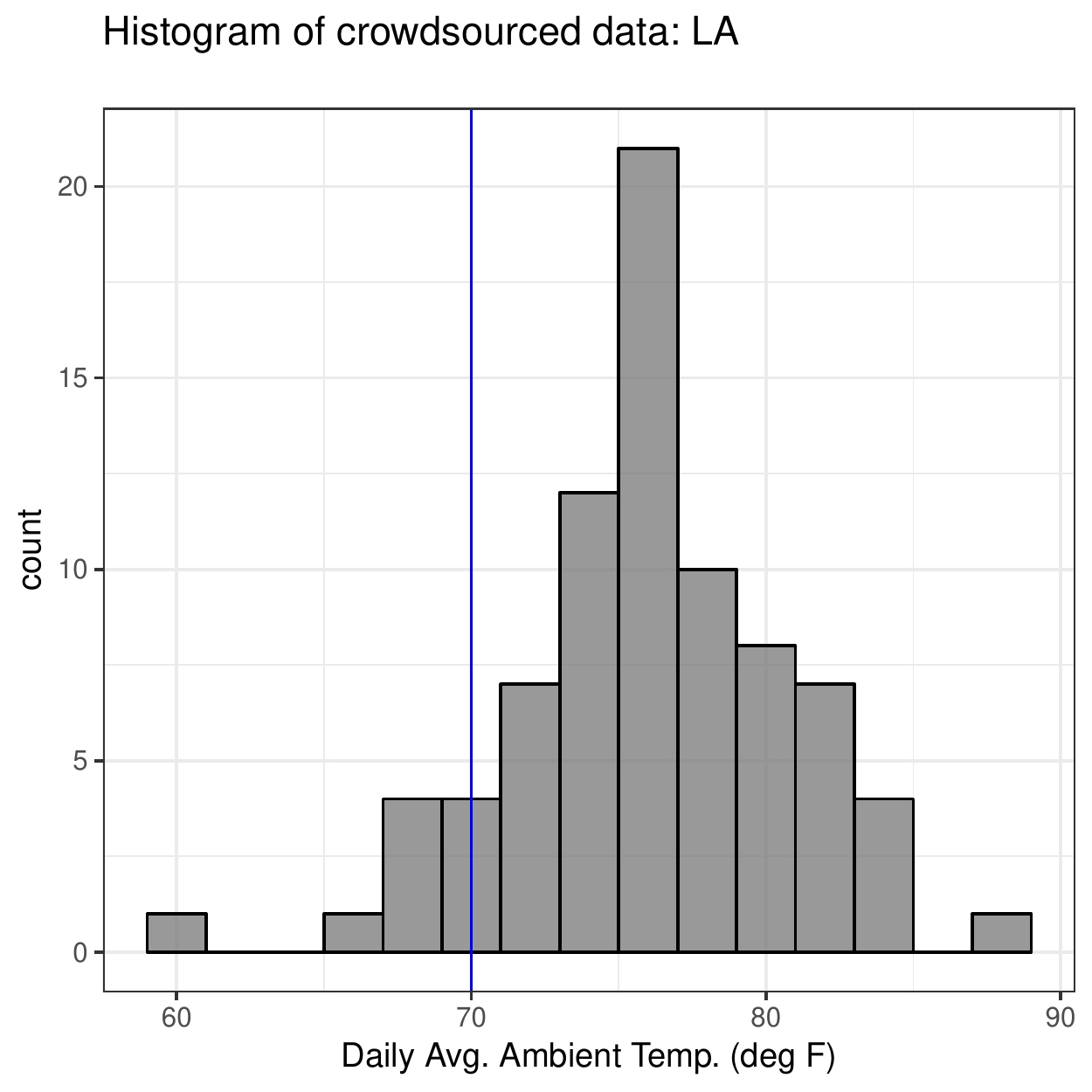}
 \end{subfigure}
 \caption{Empirical distribution of the crowdsourced average temperatures in the regions from Figure~\ref{fig:allDat} for Brooklyn, NY (left) and Los Angeles, CA (right). Blue vertical lines represent the average ground-station values in the considered regions.}\label{fig:badQuality}
 \end{figure}

To illustrate the varying quality of the observations in the WeatherSignal data, Figure~\ref{fig:badQuality} shows the temperature distribution for the two hyper-local regions shown in Figure~\ref{fig:CS_NYC} and \ref{fig:CS_LA}. The daily average temperature values in the crowdsourced data set vary from nearly $60^\circ$F to $90^\circ$F in both of the hyper-local regions for the same day.

These temperature distributions show the nature of the noise involved in the crowdsourced data. Due to the factors associated with the data collection process, a portion of the observations in the crowdsourced data are either contaminated or not representative of the ambient temperature, which is the outdoor air temperature close to the earth's surface. Such representativeness errors for weather data coming from meteorological stations have been considered previously by \cite{lorenc86}, \cite{gandin88} and \cite{lussana10}. Comparing the histograms with the single ground-station observation in both the regions, we can see that although there are large deviations, a good proportion of the crowdsourced observations are `close' to the corresponding ground-station observations ($72^\circ$F in the Brooklyn and $70^\circ$F in LA), which are collected in laboratory environment with high-quality sensors maintaining World Meteorological Organization (WMO) standards.

Building models based on the noisy crowdsourced data that ignore the reliability of the sensor-generated observations can lead to erroneous prediction. For instance, we used leave-one-out prediction of the observations in the regional block around Brooklyn (Figure~\ref{fig:CS_NYC}) using standard techniques of spatial analysis, with a reasonable mean and covariance model (discussed in Section~\ref{subsec:stanGeo}), and the errors in the predictions ranged from -30$^\circ$F to 40$^\circ$F. Similar cross-validation approach has been previously used by \cite{cressie93} and \cite{lussana10} to identify the `bad' observations. These first-stage analyses motivated us to take the quality of the observations in the WeatherSignal data into consideration. \cite{lussana10} proposed to remove observations for which the cross-validated prediction errors exceed some threshold. But, due to the inclusion of the corrupted observations at every iteration of the leave-one-out cross-validation, the predictions are not guaranteed to be a good representation of the true value at that location. Moreover, the leave-one-out cross-validation approach being computationally expensive, the method proposed by \cite{lussana10} is not readily applicable for large crowdsourced weather data coming from mobile sensors. The `absurd' observations, i.e. observations with high gross error \citep{lussana10}, can be identified using some other more scalable spatial outlier detection techniques (for example, see Chapter 1 of \citealt{cressie93}; \citealt{harris14} etc.) and thus, can be omitted from the analysis. But, in that case, it is not straightforward how to address observations with small to moderate measurement errors. For instance, using a too strict threshold on the measurement error may lead to deletion of significant number of observations, resulting in a complete loss of information for specific locations.

Hence, the new methodology should address the three following challenges. First, in addition to just identifying high-noise observations, a continuous assessment of the veracity of all the observations in a geostatistical setting is needed. Second, the definition of veracity should take into account the behavior of the process in the study region so that the ``misleading'' observations can be detected. Third, the veracity assessment of the observations should be incorporated into the subsequent analysis to allow for robust inference and efficient prediction. Though there are studies (for example, \citealt{allah13}) in the literature on quality assessment of crowdsourced data coming from volunteers or paid participants, assessment of sensor-generated data quality is not common. \cite{sosko17} mention an elementary root mean squared error approach for accuracy measurement using a reference data set from Israeli Meteorological Stations. However, neither of the above-mentioned papers provide full geostatistical inference and prediction using noisy crowdsourced data.

In this article, we make several contributions. First, we introduce a Veracity Score (VS) to measure the reliability of the crowdsourced observations on a continuous scale using a reference data set. Second, we propose a VS-based methodology to incorporate the veracity assessment into standard spatial analysis so that the effect of noisy and misleading observations is reduced, hence making the estimation and prediction more robust and efficient. Third, we show that using the VS-based technique in hyper-local regions with relatively higher number of crowdsourced observations can produce a more accurate and efficient prediction surface as compared to the global prediction surface obtained through the analysis of ground-station data alone. This paper is organized as follows. In Section \ref{sec:vs}, we introduce the veracity score and describe its elementary properties in a relevant geostatistical setting. Section~\ref{sec:vsBased} includes a brief description of the standard approach for analyzing geostatistical data, followed by a detailed description of the VS-based methodology for estimation and prediction. In Section~\ref{sec:simStudy}, we describe simulation studies to justify the superiority of VS-based methodology over the standard approach in the analysis of noisy crowdsourced data. In Section~\ref{sec:VS-OpneSignal}, we provide details of the analysis, estimation and hyper-local prediction in a case study. Finally, Section~\ref{sec:conc} summarizes our effort and discusses limitations and possible future works.

\section{Defining and Measuring Veracity}
\label{sec:vs}
In this section, we provide the intuition and motivation for veracity scoring. We denote the sample size as $n$. We denote the volume of a set $A \subset \rtwo$ as $\lvert A \rvert$, i.e., the Lebesgue measure of $A$ if it has nonzero volume and the cardinality of $A$ if $A$ is finite.

\subsection{Motivation for Veracity Scoring}
\label{subsec:PropVS}
To provide motivation for veracity scoring, consider a very simple yet practical example.
\begin{eg}\label{ex:vs-moti-one}
Let $Z_1, \dots , Z_n$ be independent noisy observations with $E\Lp Z_i \Rp = \mu$ and $\var\Lp Z_i \Rp = \sigma_i^2$ for $i \in \LP 1, \dots , n \RP$. The usual sample mean, which is also the o.l.s.\ estimator for $\mu$, is given by $\hat{\mu}_{\text{ols}} = \bar{Z}_n = n^{-1}\sum_{i=1}^n Z_i$, with $E(\hat{\mu}_{\text{ols}}) = \mu$ and $\var\Lp \hat{\mu}_{\text{ols}} \Rp = \frac{1}{n^2}\sum_{i=1}^n \sigma_i^2$. If we assume $\sigma_i^2 = C\cdot i^b$, for some constants (w.r.t. $n$) $C, b > 0$, we have
$$
\var\Lp \hat{\mu}_{\text{ols}} \Rp \approx C(b)\cdot n^b,
$$ 
for some constant (w.r.t. $n$) $C(b)$. Instead of the generic sample mean, consider a weighted average of the observations given by $\hat{\mu} = (\sum_{i=1}^n v_i Z_i)/(\sum_{i=1}^n v_i)$, where the weights $v_i = i^{-a}$ for some constant $a > 0$, i.e. the weights are inversely proportional to the variance of the noisy observations. Then,
$$
\var\Lp \hat{\mu} \Rp \approx C(a, b) \cdot n^{b-1},
$$ 
for some constant $C(a,b)$. Clearly, if $C, a$ and $b$ are constants w.r.t. to the sample size $n$, then a significant gain in efficiency ($O(n^{b-1})$ as compared to $O(n^b)$) can be achieved for large $n$ by assigning lower weights to high variance observations.
\end{eg}

If we can find a formulation of the veracity score that is inversely related to the observation noise variance, we can use it to reduce the effect of the noise in the inference and achieve a more accurate and efficient estimator.

\subsection{Preliminaries} 
\label{subsec:PrelNot} 
Let $\LP Z(\bfs_1), \dots Z(\bfs_n) \RP$ be the varying-quality observations --  for example, the crowdsourced data from cellphone sensors -- which are observed at irregularly spaced locations $\mathcal{S}_n \coloneqq \LP \bfs_1, \dots , \bfs_n \RP \subset \rtwo$. In addition, at spatial locations $\mathcal{T}_m \coloneqq \LP \bft_1, \dots , \bft_m \RP$ $\subset \rtwo$, assume that we have $\{ Y(\bft_1), \dots , Y(\bft_m) \}$, which are high-quality, reliable observations of the spatial process -- for example, measurements from the ground-stations. It is common to assume (\citealt{cressie93}, \citealt{gelfand10}) that the spatial random field of interest $\LP Y(\bfs): \bfs \in \rtwo \RP$ can be represented as
\begin{equation} \label{eq:genMod}
\begin{split}
Y(\bfs) = \mu(\bfs) + \epsilon(\bfs),
\end{split}
\end{equation}
where $\mu(\bfs)$ is a deterministic smooth mean function capturing the large scale variation of the process, i.e., $E(Y(\bfs)) = \mu(\bfs)$. Here, $\epsilon(\bfs)$ is a mean zero spatially correlated residual process which addresses the small-scale variations over the space. For the varying-quality $Z$-process, we write the decomposition in Equation~\ref{eq:genMod} as
\begin{equation} \label{eq:genMod_Z}
\begin{split}
Z(\bfs) = \mu(\bfs) + w(\bfs),
\end{split}
\end{equation} where $w(\bfs)$ is the aggregated noise associated with the observation $Z(\bfs)$. For example, if we assume that the varying-quality observations arise from an additive-multiplicative noise model as
\begin{equation} \label{eq:AddMultNoiseMod}
\begin{split}
Z(\bfs_i) = \epsilon_{M_i} Y(\bfs_i) + \epsilon_{A_i},
\end{split}
\end{equation} where $\epsilon_{M_i}$ and $\epsilon_{A_i}$ for $i \in \LP 1, \dots , n \RP$ are random variables associated with the multiplicative and additive noise in the observation $Z(\bfs_i)$. Then, the associated $w$-process will have the form $w(\bfs_i) = \mu(\bfs_i)(\epsilon_{M_i} - 1) + \epsilon_{M_i}\epsilon(\bfs_i) + \epsilon_{A_i}$. If there is no multiplicative component $\epsilon_{M_i}$ in the contamination, then $w(\bfs_i) = \epsilon(\bfs_i) + \epsilon_{A_i}$. In the next subsection, we define a score to assess the quality or reliability of the observation $Z(\bfs_i)$, namely veracity score.

\subsection{Veracity Score: Formulation and Properties}
\label{subsec:VS}
A good measure of veracity should not only identify ``absurd'' observations, but also provide a score for each observation on a continuous scale, so that the effect of the ``bad'' observations can be reduced automatically, making inference robust against the low-quality observations. Our goal is to formulate a continuous scoring procedure to measure the veracity of the observations in two different scenarios. The first scenario assumes a reference data set containing observations with high-quality but low-density in the concerned regions is available. The second scenario assumes that we do not have any high-quality reference information available.

\subsubsection{Veracity Score with Reference Data}
\label{subsubsec:vsWref}
Consider a hyper-local regional block like those in Figure~\ref{fig:CS_NYC} or \ref{fig:CS_LA}, and denote it by $\mathcal{R} \subset \rtwo$. The observation vector with locations inside $\mathcal{R}$ is given as $\mathbf{Z} \coloneqq \Lp Z(\bfs_1), \dots , Z(\bfs_n) \Rp^\prime$. Consider $\mathcal{R}$ to be the region of interest for analyzing the varying-quality observations $\mathbf{Z}$. Consider another regional block $\mathcal{D}$ such that $\mathcal{R} \subset \mathcal{D} \subset \rtwo$ and $\lvert \mathcal{R} \rvert << \lvert \mathcal{D} \rvert$. Let the reference data vector with locations inside $\mathcal{D}$ be denoted as $\mathbf{Y} \coloneqq \Lp Y(\bft_1), \dots , Y(\bft_m) \Rp^\prime$. The reference data $\mathbf{Y}$ is high-quality and hence reliable representation of the spatial process of interest, but it has low data-coverage in the hyper-local region of interest $\mathcal{R}$. So, to get a reasonable sample size for the reference data, we need to consider the larger region $\mathcal{D}$. We denote a $\delta$-neighborhood around a spatial point $\bfs \in  \rtwo$ as $\mathcal{B}_\delta(\bfs)$, with $\mathcal{B}_\delta(\bfs) \coloneqq (\bfs - \delta, \bfs + \delta]$ for some $\delta \in \rone^+$, where the subtraction and addition is component-wise.

Define the VS of the observation $Z(\bfs_i)$ as
\begin{equation}\label{eq:VerSc_wRef}
\begin{split}
V(\bfs_i) = \phi\Lp \frac{\lvert Z(\bfs_i) - \xi(\bfs_i) \rvert}{\alpha + D\Lp \boldsymbol{\xi}_i \Rp} \Rp,
\end{split}
\end{equation} 
where $\phi : \rone^+\cup \LP 0 \RP \to \rone^+ \cup \LP 0 \RP$ is some non-increasing function such that $\text{sup}_x \;\phi(x)  < \infty$. We call $\phi(\cdot)$ the veracity function with $\alpha \in \rone^+$ as a regularity parameter. By $\xi(\bfs_i)$, we denote a reasonable benchmark for the target process at $\bfs_i$, and $\boldsymbol{\xi}_i \coloneqq \Lp \xi(\bfs_{i_1}), \dots , \xi(\bfs_{i_{n(i)}}) \Rp^\prime$ where $\{ \bfs_{i_1}, \dots , \bfs_{i_{n(i)}} \}$ is the set of observation locations in the small $\delta$-neighborhood $\mathcal{B}_\delta(\bfs_i)$. Finally, $D(\bfx)$ denotes a robust measure of dispersion of the observations in the vector $\bfx$. Clearly, the VS is computed by evaluating the $\phi$-function at the \textit{scaled deviation} $\frac{\lvert Z(\bfs_i) - \xi(\bfs_i) \rvert}{\alpha + D\Lp \boldsymbol{\xi}_i \Rp}$ and due the non-increasing property of $\phi\Lp \cdot \Rp$, if the deviation is high, we have low VS and if the deviation is low, we have high VS.

Now consider the benchmark value, $\xi(\bfs)$, for the target at location $\bfs$. If we have high-quality observations of the $Y$-process from the reference data at the varying-quality data sites $\LP\bfs_1, \dots, \bfs_n  \RP$, then the obvious choice is to take $\xi(\bfs_i) = Y(\bfs_i)$. In practice, as we see in Figure~\ref{fig:CS_NYC} to \ref{fig:GS_LA}, the locations of the ground-station measurements (reference data) and the crowdsourced data (varying-quality observations) almost always differ significantly. Hence to define the benchmark at location $\bfs_i$, we propose to compute a kriging surface, $\LP \bfs, \hat{Y}(\bfs) : \bfs \in \mathcal{D}\RP$, of the $Y$-process using the observation vector $\mathbf{Y}$. Then, we define $\xi(\bfs_i)$ as
\begin{equation}\label{eq:blockTruth}
\begin{split}
\xi(\bfs_i) = \hat{Y}(\bfs_i) + (1 - \nu) \; \mathcal{C}\Lp \mathbf{Z}_i - \hat{\mathbf{Y}}_i \Rp,
\end{split}
\end{equation} where $\mathbf{Z}_i \coloneqq \Lp Z(\bfs_{i_1}), \dots , Z(\bfs_{i_{n(i)}})\Rp^\prime$ and $\hat{\mathbf{Y}}_i \coloneqq \Lp \hat{Y}(\bfs_{i_1}), \dots , \hat{Y}(\bfs_{i_{n(i)}})\Rp^\prime$. Here $\mathcal{C}(\mathbf{x})$ is a robust measure of central tendency of the values in the vector $\mathbf{x}$ and $\nu \in [0,1]$ is a mixing parameter that we discuss in detail later.

If we have a reasonable benchmark, $\xi(\bfs_i)$, for the spatial process of interest at the location $\bfs_i$, the definition of the VS in Equation~\ref{eq:VerSc_wRef} is a transformed measure of the scaled deviation of the observation $Z(\bfs_i)$ from the benchmark value. In the definition of VS, the measure of dispersion, $D\Lp \boldsymbol{\xi}_i \Rp$, in the denominator takes the variability in the $\delta$-neighborhood into account. For example, in the analysis of ambient temperature, the variation in a small neighborhood in the mountains is likely to be higher than an area close to the sea-level. Hence, the statistic $\frac{\lvert Z(\bfs_i) - \xi(\bfs_i) \rvert}{\alpha + D\Lp \boldsymbol{\xi}_i \Rp}$ measures the deviation of the observation from its benchmark relative to the local variability. In the following sections, we use interquartile range (i.e. $D(\bfx) =$ IQR$(\bfx)$) as the robust measure of dispersion in equation~\ref{eq:VerSc_wRef} and the sample median (i.e. $\mathcal{C}(\bfx) = Q_2(\bfx)$, where $Q_j$ is the $j$-th sample quartile) as the robust measure of central tendency in equation~\ref{eq:blockTruth}. There are other robust choices as well, but we use the sample quantile based statistic because it is familiar to the practitioners and easy to interpret. Also, these choices are theoretically justified as the sample quantiles are asymptotically consistent under dependence (\citealt{jkg71}, \citealt{sun06}). The parameter $\alpha$ determines the baseline of the deviation. For lower values of $\alpha$ we penalize more, and for higher values we allow for a larger deviation from the benchmark. We call $\alpha$ the \textit{baseline deviation} of the VS, and its unit is same as the process of interest, which makes the VS unit free.

We require the veracity function $\phi$ to have the following properties:
\begin{enumerate}
\item $\phi(\cdot)$ is a non-increasing function with bounded range, $\phi(x) \leq \phi(0)  < \infty$.
\item $\phi(x) \downarrow 0$ as $x \to \infty$.
\end{enumerate} 
With this formulation, lower values of the VS correspond to the low-quality or less reliable observations and high values of the VS correspond to the better quality of the observations. We use $\phi(x) = \exp\Lp -x \Rp$ for our analysis in the subsequent sections. The advantage of this function is that the VS lies naturally in $[0,1]$, and it penalizes exponentially as the scaled deviation from the benchmark value increases. We discuss other possible choices in Section B.1 in the supplementary material.

Now we try to interpret the mixing parameter $\nu$ in the definition of VS. Under the assumption that the estimated mean process $\hat{\mu}(\bfs)$ is smooth and the kriged-residual process $\hat{\epsilon}(\bfs)$ is a spatially correlated second-order stationary mean-zero process, for a small enough $\delta > 0$, we can write $Q_2\Lp \hat{\mathbf{Y}}_i \Rp \approx \hat{Y}(\bfs_i)$, as the variation of the kriged process ${\hat{Y}(\bfs)}$ inside the $\delta$-neighborhood is negligible. Hence, we can approximately rewrite the benchmark as
$$
\xi(\bfs_i) \approx \nu \; \hat{Y}(\bfs_i) + (1 - \nu) \; Q_2\Lp \mathbf{Z}_i\Rp.
$$ 
Here, to get a possible approximation the spatial process at location $\bfs_i$, instead of just using the estimated value $\hat{Y}(\bfs_i)$ from the high-quality reference data over a bigger surrounding, we want to leverage the available varying-quality observations in the hyper-local region. We propose to use a mixture of an approximation of the spatial process coming from the reference data over a bigger region $\mathcal{D}$, i.e. $\hat{Y}(\bfs_i)$ and a robust local estimate coming from the varying-quality observations in the small $\delta$-neighborhood $\mathcal{B}_\delta(\bfs_i)$ around the location of interest $\bfs_i$, i.e. $Q_2\Lp \mathbf{Z}_i\Rp$. Due to the smooth mean and spatially correlated residual process, the spatial observations in a ``small'' neighborhood are likely to behave ``similarly.'' Therefore, it is sensible to use a robust estimate of the central tendency of the varying-quality observations in that small neighborhood as the locally estimated approximation of the spatial process at $\bfs_i$. The mixing parameter $\nu$ decides the weight of mixing between the estimated process from the reference data and the local approximation from the varying-quality observations. The optimal $\nu$ balances the error in estimation from the reference data and the error in the approximation of the spatial process using the sample median in the $\delta$-neighborhood.

\subsubsection{Veracity Score without Reference Data}
\label{subsubsec:vsWOref}
We propose a similar definition of the VS when we do not have any high-quality reference observations available. In this scenario our definition of VS is
\begin{equation}\label{eq:VerSc_woRef}
\begin{split}
V(\bfs_i) = \phi\Lp \frac{\lvert Z(\bfs_i) - \mathcal{C}(\mathbf{Z}_i) \rvert}{\alpha + D(\mathbf{Z}_i)}\Rp.
\end{split}
\end{equation} The idea behind the definition given in Equation~\ref{eq:VerSc_woRef} is similar to that in Section~\ref{subsubsec:vsWref}. As we do not have information available from a high-quality reference data set, we use only the locally estimated central tendency as the proxy of the target and the local variation in the denominator to take the regional variability into account. Note that the definition of the VS in Equation~\ref{eq:VerSc_wRef} approximately equals the VS as given in Equation~\ref{eq:VerSc_woRef} if we take $\nu = 0$.

The formulations of the VS, both with and without reference data, depend on $\delta$, which is a positive scalar equal to half of the length of the neighborhood $\mathcal{B}_{\boldsymbol{\delta}}(\bfs_i)$ used to estimate the center and dispersion locally. The choice of $\delta$ should be such that the $\delta$-neighborhood $\mathcal{B}_{\boldsymbol{\delta}}(\bfs_i)$ is small as compared to the region of interest $\mathcal{R}$, but at the same time large enough to have sufficient sample size to provide a good assessment of the quality of the observations. To make the formulation of VS as well-defined, we need the number of points in the $\delta$-neighborhood, $n(i)$, larger than 2 for each $i \in \LP 1, 2, \dots , n \RP$. If we do not have enough data points to compute the measure of dispersion for an observation, we say that the VS is undefined for those observations.

Similar approach of comparing the observations with a \textit{benchmark} value has been used to detect outliers in literature (e.g., see Chapter 1 of \citealt{cressie93}; \citealt{georobMan}). \cite{lussana10} proposed a benchmark obtained through leave-one-out cross-validated prediction using the noisy observations. But, as mentioned in Section~\ref{subsec:challengeInOpensignal}, due the presence of some absurd noise in the training data of the cross-validation, the benchmarks obtained in this technique might themselves be corrupted and hence, are not necessarily robust. We prefer quantile based local summaries as benchmarks due to its scalability and computational ease, appeal to the practitioners as well as robustness and asymptotic  efficiency (see \citealt{sen68}) as compared to some other choices discussed previously.

\section{Veracity Score Methods}
\label{sec:vsBased}
Before going to the VS-based version of the spatial analysis, we briefly describe the standard approach of geostatistical analyses.

\subsection{Review of Standard Analysis of Spatial Data}
\label{subsec:stanGeo}
For this section, we use the model specified in Equations~\ref{eq:genMod} and \ref{eq:genMod_Z} as well as the notations stated in Section~\ref{subsec:PrelNot}. In geostatistics, often the smooth deterministic mean process $\{\mu(\cdot)\}$ is modeled under a spatial regression framework where the mean function is assumed to have a \textit{linear} form, $\mu(\bfs) = \mathbf{x}(\bfs)^\prime \bfbeta$, where $\mathbf{x}(\cdot) = \Lp x_1(\cdot),...,x_p(\cdot) \Rp^\prime$ is a $p$-dimensional deterministic vector process of known covariates and $\bfbeta$ denotes the unknown regression parameter vector. To make the inference feasible from only one replication of the process over the space, some stationarity assumption on the second-order structure of the residual process $\LP \epsilon(\bfs) \RP$ is required. One of the most commonly used assumptions is that $\LP \epsilon(\bfs) \RP$ is an intrinsically stationary process with an admissible parametric variogram function 
$2\gamma(\bfh; \bftheta) = \text{Var}\LP \epsilon(\bfs) - \epsilon(\bfs + \bfh) \RP$, where $\bftheta$ is the covariance parameter of interest.

For now, the description of the analysis is given without taking the noisy nature of the observations into account, so $\{w(\bfs)\}$ is assumed to be identically equal to $\{\epsilon(\bfs)\}$. Since the covariance parameter is unknown, the standard analysis starts with the estimation of the regression parameters in the linear mean model using ordinary least squares (o.l.s.).
$
\hat{\bfbeta}_{\text{ols}} = \Lp \mathbf{X}^\prime \mathbf{X} \Rp^{-1} \mathbf{X}^\prime\mathbf{Z},$ where, $\mathbf{X} \coloneqq \Lp \bfx(\bfs_1), \dots , \bfx(\bfs_n) \Rp^\prime$. Next, the de-trended observations, i.e. $\hat{\bfeps}=\mathbf{Z} - X\hat{\bfbeta}_{\text{ols}}$, are used to estimate the covariance parameter $\bftheta$ using least squares-based variogram model fitting \citep{cressie93} based on some generic nonparametric semivariogram estimator (denoted by $\hat{\gamma}\Lp \bfh \Rp$) -- e.g., the \textit{classical} or \textit{method-of-moments} semivariogram estimator proposed by \cite{matheron62}. For example, the weighted least squares (w.l.s.) estimator of $\bftheta$ is given as,
\begin{equation}\label{eq:VariogWls}
\begin{split}
\hat{\boldsymbol{\theta}}_{\text{wls}} = \underset{\boldsymbol{\theta}}{\text{argmin}} \sum_{j = 1}^k w_j \LP \hat{\gamma}(\bfh_j) - \gamma(\bfh_j; \boldsymbol{\theta}) \RP^2,
\end{split}
\end{equation} where, $w_j$ is the weight corresponding to lag $\bfh_j$ and, $\LP \bfh_1, \dots , \bfh_k \RP$ are the set of discrete lags for which the nonparametric semivariogram $\hat{\gamma}\Lp \cdot \Rp$ has been computed. For details of variogram model fitting see \cite{cressie93}, \cite{gelfand10}. Mat\'ern is a popular choice for the parametric class of admissible variograms as it provides a rich class to choose from \citep{haskard07}. A comprehensive list of parametric variogram models can be found in \cite{cressie93} and \cite{gneiting13}.

Once the covariance structure is estimated, one can try to improve the mean parameter estimates using \textit{estimated generalized least squares} (e.g.l.s.) estimator, given by $\hat{\bfbeta}_{\text{egls}} = \Lp X^\prime \hat{\Sigma}^{-1} X \Rp^{-1} X^\prime \hat{\Sigma}^{-1} \mathbf{Z}$, where $\hat{\Sigma}$ is the estimated variance of $\bfeps = \Lp \epsilon(\bfs_1), \dots , \epsilon(\bfs_n) \Rp^\prime$. However,  this introduces additional variability due to using the estimated covariance parameters in the mean estimator and is not necessarily more efficient than the o.l.s. estimator. 

The most commonly used method to predict the process at new locations is to predict the $\epsilon$-process at the given locations by the \textit{best linear unbiased predictor} (BLUP) given the observed residual vector $\hat{\bfeps}$, also known as \textit{ordinary kriging} estimator (\citealt{cressie93}, p.\ 122). The standard predictor of $Y(\bfs_0)$ is
\begin{equation}\label{eq:KrigEq_Z}
\begin{split}
\hat{Y}_\text{std}(\bfs_0) = \mathbf{x}(\bfs_0)^\prime\hat{\bfbeta}_{\text{ols}} + \hat{\epsilon}_{\text{ok}}(\bfs_0),
\end{split}
\end{equation} where $\hat{\epsilon}_{\text{ok}}(\bfs_0)$ is the ordinary kriging predictor for $\epsilon(\bfs_0)$.

The standard approach for estimation and prediction explained is not reliable for analyzing noisy spatial observations, as both the least squares-based mean parameter estimators (\citealt{huber81}) and the method-of-moments empirical semivariogram estimator are highly sensitive to the noise (\citealt{cressie80}) in the data. In the following sections, we propose a way to incorporate the VS into the analysis to make the inference and prediction robust against the noise in the data.

\subsection{Veracity score-based estimation of the mean function}
\label{subsec:RobMeanEst}
In the standard approach, as described in Section~\ref{subsec:stanGeo}, the regression parameter vector $\bfbeta$ is estimated using the o.l.s. method. For our approach, instead of simple squared error loss, motivated by Ex.~\ref{ex:vs-moti-one}, we propose to minimize a weighted version of the loss function with the veracity scores as the corresponding weights. The VS-based estimator of the mean parameter $\bfbeta$ is given as
\begin{equation}\label{eq:RobReg}
\begin{split}
\hat{\bfbeta}_{\text{vs}} = \underset{\bfbeta}{\text{argmin}} \; \sum_{i = 1}^n V(\bfs_i) \mathcal{L} \Lp  Z(\bfs_i),  \mathbf{x}(\bfs_i)^\prime \bfbeta \Rp.
\end{split}
\end{equation} For least squares-based estimators, we have $\mathcal{L}(y, u) = (y -u)^2$, the squared-error loss function. The locally estimated veracity scores lessen the effects of ``absurd'' observations in the objective function and thus make the estimation of the mean function less sensitive to the noise. The VS-based approach is adaptive to the quality of the observations and thus lessens the impact of outliers in the data. To make the estimation more robust to contamination, one can use any robust loss function instead of squared-error loss in Equation~\ref{eq:RobReg}. 
We have used an MM-type estimator with a \textit{linear quadratic quadratic} $\psi$-function for the robust regression as discussed in \cite{koller11}. The advantage of using this estimator is that in addition to penalizing less for high residuals, the parameters associated with the $\psi$-function can be tuned to improve the asymptotic efficiency for the estimators. The corresponding optimization to solve Eequation~\ref{eq:RobReg} can be executed using Iterative Re-weighted Least Squares (IRLS) as discussed in \cite{robbase09}.

The assessment of goodness of fit for the estimated linear model is essential. The usual Multiple $\text{R}^2$ is not reasonable to use, as the loss function is different from ordinary least squares. Inspired by the pseudo-$\text{R}^2_{\text{WLS}}$ coined by \cite{willet88}, we propose another variant of the coefficient of determination for VS-based regression as
$$
\begin{aligned}
\text{R}^2_{\text{vs}} = 1 - \frac{\sum_{ i = 1}^n V(\bfs_i) \mathcal{L} \Lp Z(\bfs_i), \mathbf{x}(\bfs_i)^\prime\hat{\bfbeta}_{\text{vs}} \Rp}{ \sum_{ i = 1}^n V(\bfs_i) \mathcal{L} \Lp Z(\bfs_i), \bar{Z} \Rp},
\end{aligned}
$$where $\bar{Z} = n^{-1}\sum_i Z(\bfs_i)$. The idea behind this measure is that instead of using the squared error loss to compute the total sum of squares and the residual sum of squares, the proposed $R^2_{\text{vs}}$ uses the robust loss function to measure the total variability in the data (i.e. $\sum_{ i = 1}^n V(\bfs_i) \mathcal{L} \Lp Z(\bfs_i), \bar{Z} \Rp$) and the variability that is not explained by the model (i.e., $\sum_{ i = 1}^n V(\bfs_i) \mathcal{L} \Lp Z(\bfs_i), \mathbf{x}(\bfs_i)^\prime\hat{\bfbeta}_{\text{vs}} \Rp$). Although we do not provide any theoretical justification, it appears from  explanatory analysis with synthetic data and simulations that $\text{R}^2_{\text{vs}}$ may provide an overly optimistic assessment of the goodness of the fit for the model when the Huber's loss function or MM-type estimation is used.

\subsection{Veracity score-based estimation of the covariance structure}
\label{subsec:CovEst}
To explore the second-order structure of the spatial process, we analyze the residuals obtained by de-trending the observations, $\hat{\epsilon}_{\text{vs}}(\bfs_i) = Z(\bfs_i) - \mathbf{x}(\bfs_i)^\prime \hat{\bfbeta}_{\text{vs}}$ for $i \in \LP1,2,...,n\RP$. When conducting analysis with varying-quality geostatistical data, after the robust estimation of the regression parameters, a portion of the residuals are affected by the presence of measurement error in the data, and direct analysis of these residuals can result in misleading and inefficient estimation of the covariance structure. To reduce the noise of the observed residuals, we propose a VS-based modification of residuals using a local smoothing prior estimation of the covariance parameters. When we have a high-quality reference data, we define the VS-based smoothed version of the residuals as
\begin{equation}\label{eq:NewRes_wRef}
\begin{split}
\tilde{\epsilon}(\bfs_i) = V(\bfs_i)^q \hat{\epsilon}_{\text{vs}}(\bfs_i) + (1 - V(\bfs_i)^q)Q_2(\boldsymbol{\xi}_i - \mathbf{X}_i \hat{\bfbeta}_{\text{vs}}),
\end{split}
\end{equation} where $\mathbf{X}_i \coloneqq \Lp \bfx(\bfs_{i_1}),...,\bfx(\bfs_{i_{n(i)}}) \Rp^\prime$ is the $n(i)\times p$ matrix of the covariates corresponding to the observations in $\mathcal{B}_{\boldsymbol{\delta}}(\bfs_i)$. Here, $q$ is the parameter regulating the degree of the smoothing needed. For instance, $q = 0$ implies no smoothing, and $q = 1$ implies the convex combination of the locally-corrected residual and the observed residual. As shown in Figure S4-(a) in the supplementary material, the parameter $q$ here plays the role of thresholding -- for higher $q$, only observed residuals with high VS get significant weights for the VS-based smoothing. Whereas, for smaller $q$ the formulation of the smoothed residuals in Equation~\ref{eq:NewRes_wRef} puts significant weights to even the observed residuals with low VS and thus, reducing the degree of smoothing.

If we do not have reference data available, then the analogous smoothed version of the residuals is given by
\begin{equation}\label{eq:NewRes_woRef}
\begin{split}
\tilde{\epsilon}(\bfs_i) = V(\bfs_i)^q \hat{\epsilon}_{\text{vs}}(\bfs_i) + (1 - V(\bfs_i)^q)Q_2(\hat{\boldsymbol{\epsilon}}_i),
\end{split}
\end{equation} where $\hat{\bfeps}_i = \Lp \hat{\epsilon}_{\text{vs}}(\bfs_{i_1}), \dots, \hat{\epsilon}_{\text{vs}}(\bfs_{i_{n(i)}}) \Rp^\prime$. Again note that the definition in Equation~\ref{eq:NewRes_wRef} approximately simplifies to the one in Equation~\ref{eq:NewRes_woRef} if $\nu = 0$.

For poor quality observations, when $V(\bfs_i)$ is small, the effect of the observed value of the residual $\hat{\bfeps}_{\text{vs}}(\bfs_i)$ is scaled down by $V(\bfs_i)^q$ (as $V(\bfs_i) \in (0,1]$), and the locally estimated `benchmark' value of the residual process in the small neighborhood is enforced by $\Lp 1-V(\bfs_i)^q \Rp$ in Equations~\ref{eq:NewRes_wRef} and \ref{eq:NewRes_woRef}. The effect of VS-based smoothing is illustrated on a synthetic data set in Section B.2 and Figure S.3 in the supplementary material.

We propose to use variogram model fitting with the VS-based smoothed version of the residuals, $\LP \tilde{\epsilon}(\bfs_i) \RP_{i=1}^n$, to estimate the covariance parameter $\bftheta$ robustly. First a generic nonparametric semivariogram is evaluated at discrete lags using the \textit{robust} semivariogram estimator proposed by \cite{cressie80}:
\begin{equation}
\label{eq:CrEmpSvgm}
\begin{split}
\hat{\gamma}_{\text{vs}} (\bfh_u) = \frac{\LP \frac{1}{2 \lvert N(H_u) \rvert} \sum_{\Lp \bfs_i, \bfs_j \Rp \in N(H_u)} \lvert \tilde{\epsilon}(\bfs_i) - \tilde{\epsilon}(\bfs_j) \rvert^{\frac{1}{2}} \RP^4}{0.457 + \frac{0.494}{\lvert N(H_u) \rvert}} \; , \qquad\text{for} \;\; u \in \LP 1, \dots, K \RP,
\end{split}
\end{equation} where $N(H_u) = \LP \bfh \in \mathcal{H}: \bfh \in H_u  \RP$. $H_u$ are small lag classes or \textit{bins} (see p.\ 34, \citealt{gelfand10}), which are often called \textit{tolerance regions} (see p.\ 70, \citealt{cressie93}), and these construct a partition of size $K$ of the lag-space $\mathcal{H} = \LP \bfs - \bfs^\prime : \bfs, \bfs^\prime \in \mathcal{R} \RP$. The candidate lag for the tolerance region $H_u$ is denoted by $\bfh_u$, which is often taken to be the mean of the observed lags in the bin or the centroid of the the class $H_u$. 

The parameters are estimated using method of weighted least squares as
\begin{equation}
\label{eq:VarioFitAWls}
\vspace{-2mm}
\begin{split}
\hat{\bftheta}_{\text{vs}} &= \underset{\bftheta}{\text{argmin}} \;\; Q_{\text{wls}}(\bftheta)\\ 
&=  \underset{\bftheta}{\text{argmin}} \;\; \sum_{u = 1}^K \frac{\lvert N(\bfh_u) \rvert}{\LP \gamma(\bfh_u; \bftheta) \RP^2} \LP \hat{\gamma}_{\text{vs}}(\bfh_u) - \gamma(\bfh_u; \bftheta) \RP^2,
\end{split}
\end{equation} where $\gamma(\cdot; \bftheta)$ is some pre-specified parametric admissible semivariogram model, as discussed in Section~\ref{subsec:stanGeo}. Other robust empirical variogram estimators (for example \citealt{genton98}, \citealt{lark00}) can also be used instead of the one proposed by \cite{cressie80}, as given in Equation~\ref{eq:CrEmpSvgm}. \cite{genton98} showed that the robustness properties of the empirical semivariogram proposed by \cite{cressie80} are not enough in the presence of ``absurd'' outliers in the data. But, due to the VS-based smoothing in the first stage of the covariance estimation, the very large measurement errors have already been addressed and hence, using \cite{cressie80}'s version of robust variogram estimator is reasonable here.

\subsection{Veracity score-based spatial prediction}
\label{subsec:VSKrig}
Often the aim for spatial analysis of geostatistical data is to predict the process at locations of interest or to create a prediction surface over a region of interest. To predict the $\epsilon$-process at a new location $\bfs_0$, we can use ordinary kriging with the VS-based smoothed residuals $\tilde{\bfeps} = \Lp \tilde{\epsilon}(\bfs_1), \dots , \tilde{\epsilon}(\bfs_n) \Rp^\prime$ as
\begin{equation}\label{eq:KrigEq}
\begin{split}
\tilde{\epsilon}(\bfs_0) = \LP \boldsymbol{\gamma} + \mathbf{1} \frac{\Lp 1 - \mathbf{1}^\prime \Gamma^{-1} \gamma \Rp}{\mathbf{1}^\prime \Gamma^{-1} \mathbf{1}} \RP^\prime \Gamma^{-1} \tilde{\bfeps},
\end{split}
\end{equation} where $\boldsymbol{\gamma} = \Lp \gamma(\bfs_0 - \bfs_1 ; \hat{\boldsymbol{\theta}}_{\text{vs}}),...,\gamma(\bfs_0 - \bfs_n;  \hat{\boldsymbol{\theta}}_{\text{vs}})\Rp^\prime$ and $\Lp\Gamma\Rp_{ij} = \gamma(\bfs_i - \bfs_j; \hat{\boldsymbol{\theta}}_{\text{vs}})$ (see chapter 3, \citealt{cressie93}). The residual kriging variance, which quantifies the prediction uncertainty, can be estimated as 
\begin{displaymath}
\hat{\var}\Lp \tilde{\epsilon}(\bfs_0) \Rp = \hat{\sigma}^2_{\text{ok}}(\bfs_0) =\boldsymbol{\gamma}^\prime \Gamma^{-1} \boldsymbol{\gamma} -  \frac{\Lp\mathbf{1}^\prime \Gamma^{-1} \gamma \Rp^2}{\mathbf{1}^\prime \Gamma^{-1} \mathbf{1}}. 
\end{displaymath}
Finally, we predict the process at $\bfs_0$ using the modified version of Equation~\ref{eq:KrigEq_Z} as,
\begin{equation}\label{eq:KrigEq_VS}
\begin{split}
\hat{Y}_{\text{vs}}(\bfs_0) = \mathbf{x}(\bfs_0)^\prime\hat{\bfbeta}_{\text{vs}} + \tilde{\epsilon}(\bfs_0).
\end{split}
\end{equation} In Equation~\ref{eq:KrigEq_VS}, both the mean and covariance parameters have been robustly estimated using the VS-based procedures. The smoothing parameter $q$ for the VS-based smoothing of the residuals can be chosen using cross-validation. 

There are other robust kriging approaches available in literature, for example, \cite{kunsch11} and \cite{georob18}. Both of these techniques require distributional assumption on the $\epsilon$-process. Moreover, it is not straightforward to determine how to reduce the effects of observations that are not noisy but represent some other spatial process. For example, if in a local region most of the crowdsourced ambient temperatures are captured in indoor settings, applying the robust procedures directly may lead to misleading estimation of the model parameters and hence bad prediction of the outdoor ambient temperature. On the other hand, the VS-based technique can use a benchmark value, possibly obtained from a high-quality but low-density reference data, to reduce the effects of the `misleading' observations and thus estimate and predict the process of interest efficiently. Theoretical or numerical comparison of other robust kriging methodologies with the VS-based technique in case of no available reference data is beyond the scope of this article.

\section{Simulation Study}
\label{sec:simStudy}
Our simulation study aims to justify the superiority of the VS-based estimation and prediction methods as compared to the standard approach for analyzing noisy geostatistical data. We have considered two scenarios here: the first one is when no reference data is available and the second one is when a coarser but better quality reference data is present.

\subsection{Without Reference Data}
\label{subsec:SimWoRefDat}
We take the sampling region for the varying-quality observations to be $\mathcal{R} \equiv \mathcal{R}_n \coloneqq [0, \lambda_n]^2$, where $\LP \lambda_n \RP_n$ is a sequence of positive real numbers determining the size of the sampling region. We have assumed that the varying-quality observations $\LP Z(\bfs_1), \dots , Z(\bfs_n) \RP$ are coming from an additive-multiplicative noise model as given in Equation~\ref{eq:AddMultNoiseMod}. To generate the ``true'' process for simulation purposes, we use the following spatial linear model:
\begin{equation}
\label{eq:SimMod_Y}
    \begin{split}
        Y(\bfs_i) = \beta_0 + \Lp \beta_x, \beta_y \Rp^\prime \bfs_i + \beta_h \; h(\bfs_i) + \epsilon(\bfs_i),
    \end{split}
\end{equation} where $\bfbeta \coloneqq \Lp \beta_0, \beta_x, \beta_y, \beta_h \Rp^\prime$ is the vector of regression parameters; $h(\bfs)$ is the altitude of the location $\bfs$; and $\LP \epsilon(\bfs) \RP$ is a second-order stationary spatially correlated process. 

To define the altitude function over the sampling region, we use the deterministic function $h(\bfs) = H_1 \cdot \sum_{j=1}^{H_2} w_h(j) f(\bfs; \; \boldsymbol{\mu}_j , \Sigma_j) + H_3$, where $f(\cdot; \boldsymbol{\mu}, \Sigma)$ denotes the bivariate normal density with mean $\boldsymbol{\mu}$ and covariance matrix $\Sigma$ and $\LP \Lp \boldsymbol{\mu}_j, \Sigma_j \Rp : \; j \in  \LP 1, \dots , H_2 \RP \RP$ are fixed set of vectors and matrices. The residual vector $\Lp \epsilon(\bfs_1), \dots , \epsilon(\bfs_n) \Rp^\prime$ are sampled from a second-order stationary mean-zero Gaussian process with isotropic Mat\'ern covariance given by
\begin{equation}\label{eq:Matern}
\begin{split}
C(d;  \boldsymbol{\theta}) = \sigma_\epsilon^2 {\frac {2^{1-\kappa }}{\Gamma (\kappa )}}{\Bigg (}{\sqrt {2\kappa }}{\frac {d}{\rho }}{\Bigg )}^{\kappa }K_{\kappa }{\Bigg (}{\sqrt {2\kappa }}{\frac {d}{\rho }}{\Bigg )} + \tau^2 \mathbbm{1}(d = 0),
\end{split}
\end{equation} where $\Gamma$  is the gamma function, $K_{\kappa}$ is the modified Bessel function of the second kind with order $\kappa$ (\citealt{besselFn}). The covariance parameter vector of interest is $\boldsymbol{\theta} = \Lp \tau^2, \sigma_\epsilon^2, \rho, \kappa \Rp^\prime$, where $\tau^2$ is the nugget effect, $\sigma_\epsilon^2, \rho, \kappa$ are the partial sill, range and smoothness parameters respectively (\citealt{haskard07}, \citealt{gelfand10}).

To generate noise for the varying-quality observations, we use the following model for the additive and multiplicative components, denoted by $\bfeps_A \coloneqq \Lp \epsilon_{A_1}, \dots , \epsilon_{A_n} \Rp^\prime$ and $\bfeps_M \coloneqq \Lp \epsilon_{M_1}, \dots , \epsilon_{M_n} \Rp^\prime$ respectively:
\begin{equation}
\label{eq:error_dist}
\begin{split}
\epsilon_{M_i} \sim \begin{cases}
 \Delta(1) \;\; &\text{if}\;\; i \in G_n\\
 2 \times \text{Beta}(\alpha_{M}, \alpha_{M}) \;\; &\text{o.w.}
 \end{cases}\; ;\;\;
\epsilon_{A_i}  \sim \begin{cases}
 \Delta(0) \;\; &\text{if}\;\; i \in G_n\\
 N(0,\sigma_{A}^2) \;\; &\text{o.w.}\; ,
\end{cases}
\end{split}
\end{equation} where, $\Delta(x)$ denotes a degenerate distribution with point mass at $-\infty < x < \infty$; variance corresponding to the multiplicative component $\sigma_{M}^2 = \frac{1}{2\alpha_{M} + 1}$; $G_n \subset \LP 1, \dots , n \RP$ is a subset of indices and $\sigma_M, \sigma_A$ are positive constants. With this model, if $i \in G_n$, we have no noise associated with the observation, i.e., $Z(\bfs_i) = Y(\bfs_i)$. If $i \notin G_n$, then $Z(\bfs_i) = \epsilon_{M_i} Y(\bfs_i) + \epsilon_{A_i}$, where $\epsilon_{M_i}$ and $\epsilon_{A_i}$ have positive variance. Also, we have taken $\LP \epsilon_{M_i}\RP_{i = 1}^n$, $\LP \epsilon_{A_i}\RP_{i = 1}^n$ and $\LP \epsilon(\bfs_i) \RP_{i = 1}^n$ are independent of each other. We further assume that the proportion of ``good'' observations is a constant (w.r.t. $n$) denote by $q_e$, i.e., $\lvert G_n \rvert /n \approx q_e$, and $1 - q_e$ is the proportion of noisy observations in the data. This model is inspired by the crowdsourced data analysis scenario where only a proportion of observations are ``bad''. The choice of multiplicative error distribution in Equation~\ref{eq:error_dist} restricts its realizations to be in $[0,2]$ and also ensures that the multiplicative errors are symmetric around 1.

We set $\bfbeta = \Lp 55, 1.5, -1, -0.08 \Rp^\prime$, $\bftheta = \Lp 0, 6, 0.5, 3 \Rp^\prime$. To investigate the robustness of the VS with increasing noise in the data, we consider three contamination models specified by the following parameters:
(a) $\sigma_A = 5, \alpha_M = 2, q_e = 0.95$, (b) $\sigma_A = 50, \alpha_M = .5, q_e = 0.9$ and (c) $\sigma_A = 100, \alpha_M = 0.05, q_e = 0.8$. As we go from model (a) to (c), the noise in the data increases both in extent and magnitude. For example, with model (a), the variance of a noisy observation at location $\bfs$ is $0.2\Lp \bfx(\bfs)^\prime \bfbeta \Rp^2 + 28.6$, and the proportion of such observations is $5\%$; with model (c), the same variance will be $0.91\Lp \bfx(\bfs)^\prime \bfbeta \Rp^2 + 10005.73$, and the proportion of noisy observations rises to $20\%$.

Next we analyze the simulation results to compare the performances of VS-based and standard approach. The choices of the regularity parameters in the VS-based estimation like the baseline deviation $\alpha$ and the smoothing parameter $q$ are discussed in Section C.1 in the supplementary material.

In Figure~\ref{fig:SimMeanEst} we show boxplots of the VS-based estimator $\hat{\beta}_{\text{vs}}$ and the standard estimator $\hat{\beta}_{\text{ols}}$ for the four regression parameters based on $B = 200$ simulations with $n = 500$ samples.
\begin{figure}
  \centering
 \begin{subfigure}{0.5\textwidth}
     \centering
    \includegraphics[trim={1.5cm .1cm 1.5cm 0.1cm}, width=.75\textwidth, height=0.18\textheight]{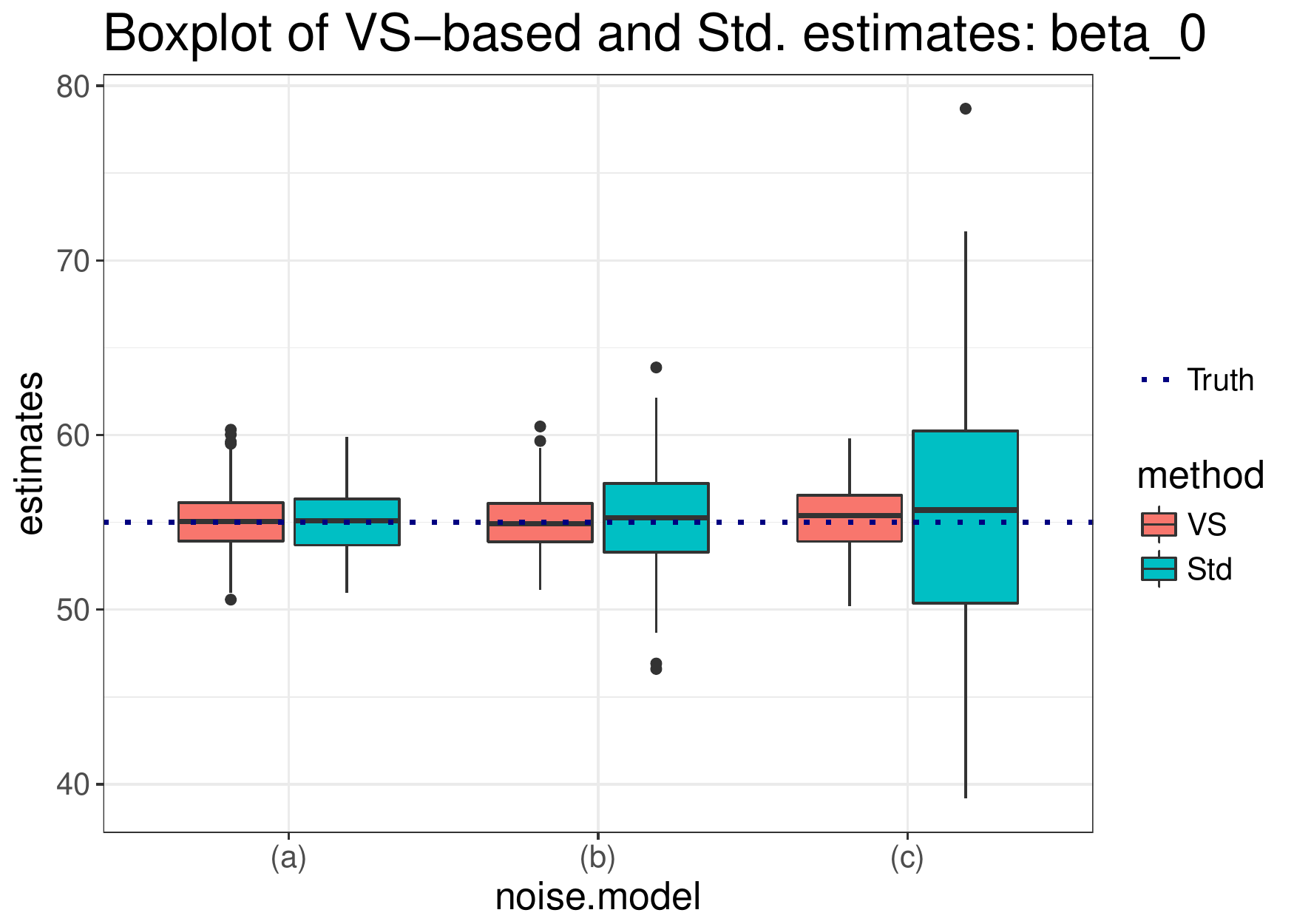}
 \end{subfigure}%
 \begin{subfigure}{0.5\textwidth}
     \centering
    \includegraphics[trim={1.5cm .1cm 1.5cm 0.1cm}, width=.75\textwidth, height=0.18\textheight]{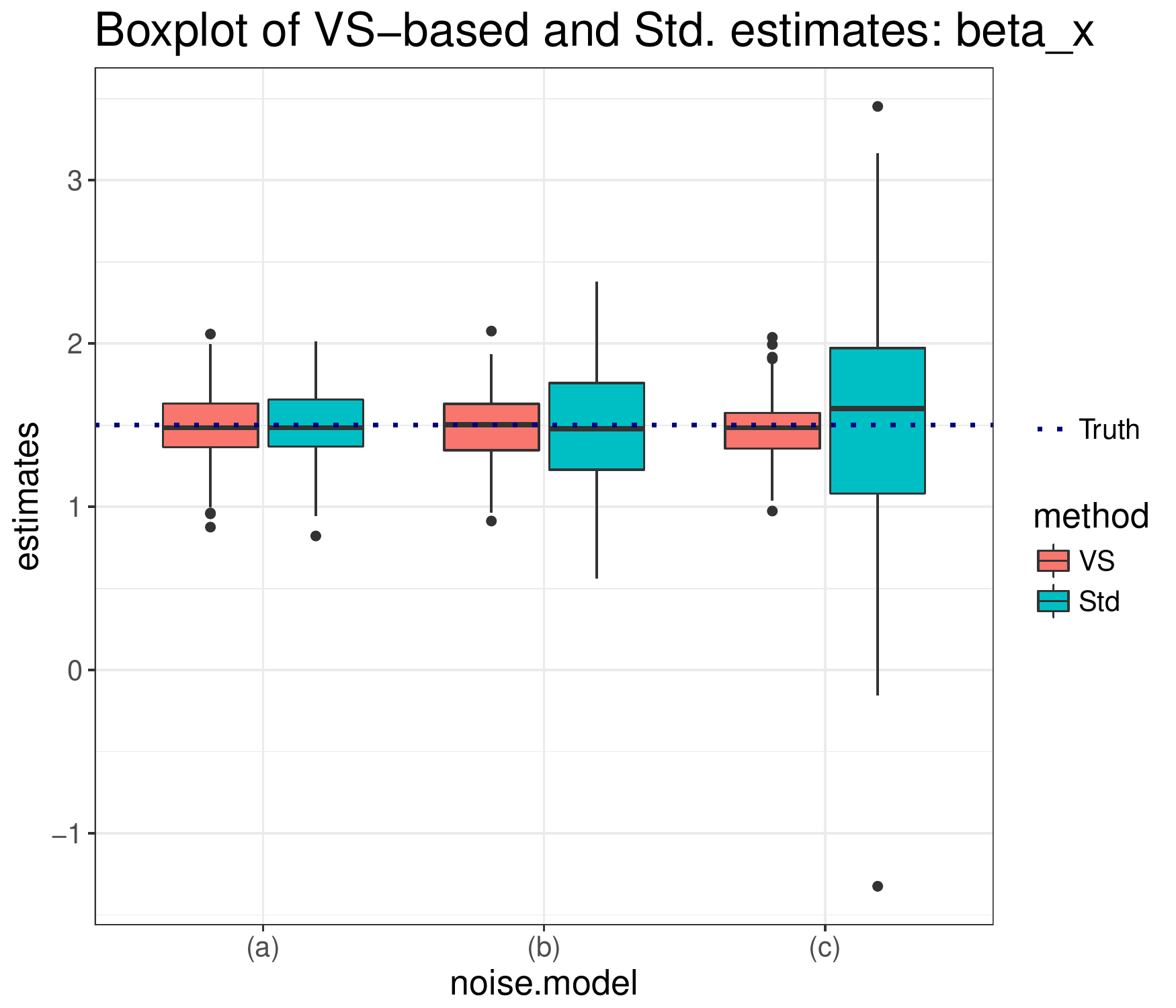}
 \end{subfigure}\vspace{2mm}
 \begin{subfigure}{0.5\textwidth}
     \centering
    \includegraphics[trim={1.5cm .1cm 1.5cm 0.1cm}, width=.75\textwidth, height=0.18\textheight]{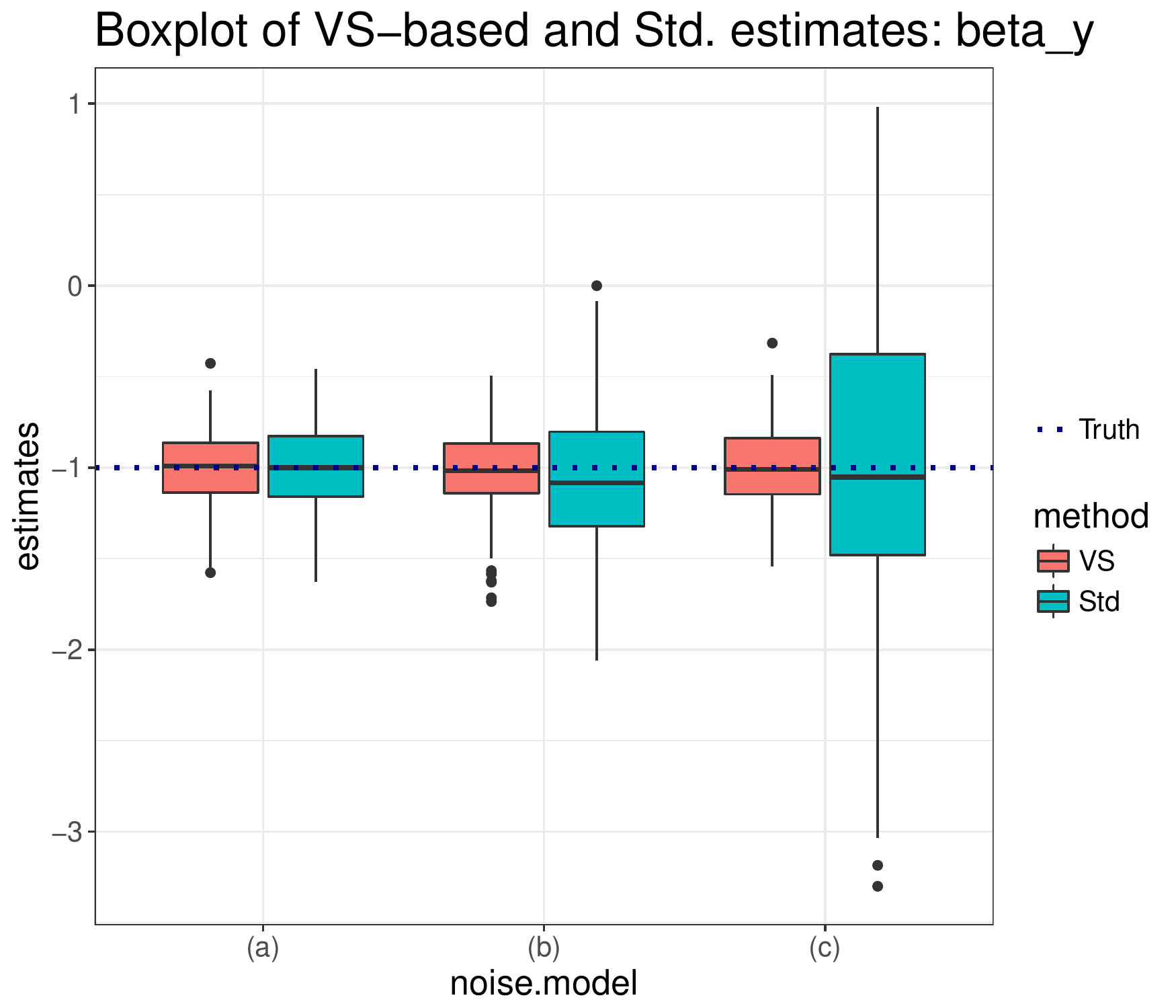}
 \end{subfigure}%
 \begin{subfigure}{0.5\textwidth}
     \centering
    \includegraphics[trim={1.5cm .1cm 1.5cm 0.1cm}, width=.75\textwidth, height=0.18\textheight]{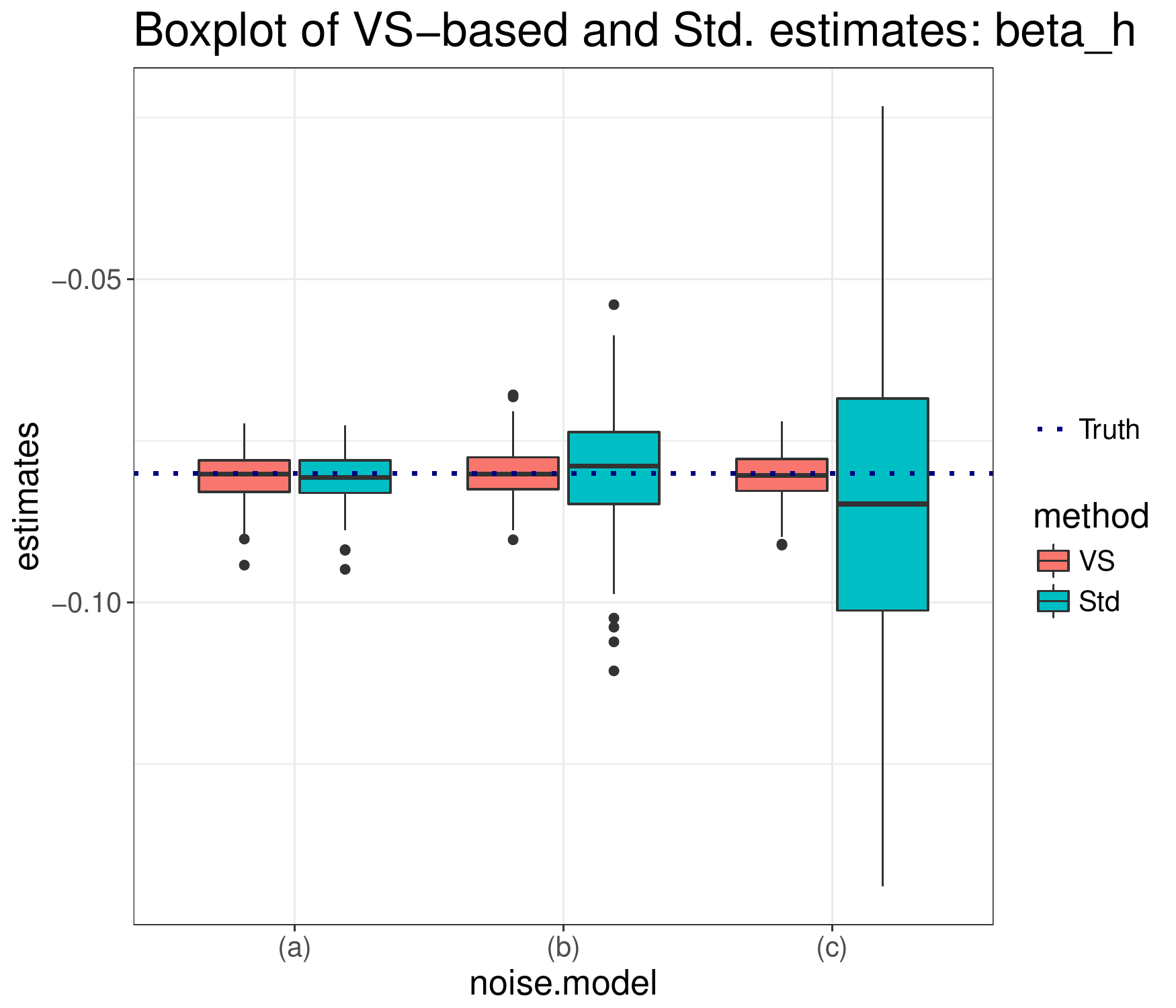}
 \end{subfigure}
 \caption{Performance of the VS-based and standard regression parameter estimators for analyzing varying-quality observations (sample size $n = 500$) without reference data.}\label{fig:SimMeanEst}
 \end{figure} 
 The VS-based technique shows more robustness towards the added noise in the observations. As we move from noise model (a) to (c), the efficiency of the o.l.s. estimator is heavily compromised, where as the spread of the VS-based estimates is hardly increased. Section C.2 in the supplementary material contains additional simulation results for regression parameter estimation: boxplots if the estimates for $n = 100, 3000$ (Figure S4 and S5). All of these simulations show similar results to justify the superiority of VS-based mean parameter estimation in the analysis of noisy spatial data as compared to the standard o.l.s method.
 
We also evaluate the VS-based and standard covariance parameter estimation and show the results in Table~\ref{tab:CovEstPerform_woRef}.
\begin{table}[ht]
\centering
\resizebox{.9\columnwidth}{.16\textwidth}{%
\begin{tabular}{cccccc}
\toprule
\textbf{Noise Model} & $\mathbf{n}$ & \textbf{bias.sill.VS} & \textbf{bias.sill.Std} &  \textbf{bias.range.VS} & \textbf{bias.range.Std}\\ 
  \hline
  \hline
 \multirow{3}{*}{(a)} & 100 & -0.313 (3.31) & 3837.513 (9867.28)  & -0.296 (0.13) & 6.671 (16.1) \\ 
  & 500 & 0.23 (1.16) & 623.629 (1644.56) & -0.114 (0.06) & 3.778 (9.91) \\ 
  & 3000 & 0.344 (0.62) & 36.098 (82.01)  & -0.026 (0.05) & 0.307 (3.2) \\ \hline
 \multirow{3}{*}{(b)} & 100 & 7.657 (8.13) & 17545.465 (58680) &  -0.357 (0.08) & 69.945 (454.78) \\ 
  & 500 & 1.747 (1.52) & 5135.181 (14207.51) &  -0.158 (0.06) & 3.711 (11.05) \\ 
  & 3000 & 0.48 (0.96) & 1108.544 (3515.95) &  -0.06 (0.05) & 8.377 (52.63) \\ \hline
 \multirow{3}{*}{(c)} & 100 & 32.774 (9.51) & 6606.713 (27599.4) &  -0.39 (0.03) & 130.833 (463.05) \\ 
  & 500 & 15.352 (6.23) & 21915.533 (63507.44) &  -0.241 (0.05) & 6.222 (47.09) \\ 
  & 3000 & 2.933 (1.14) & 5289.832 (12192.3) &  -0.111 (0.04) & 4.624 (19.92) \\ 
   \toprule
\end{tabular}%
}
\caption{\small Performance of the VS-based methodology and standard approach in estimating covariance parameters on varying-quality observations.}
\label{tab:CovEstPerform_woRef}
\end{table}
In each of the cases, the estimates of the sill parameter ($\sigma_\epsilon^2 + \tau^2$, the total variance the residual process) obtained by the VS-based methodology is more accurate by large margins as compared to standard variogram estimation. As the sample size increases both the bias and standard deviations of the VS-based estimators are closing towards $0$ under all the considered noise models. Table~\ref{tab:CovEstPerform_woRef} clearly establishes the efficiency of VS-based covariance estimation as compared to the standard approach when some of the observations are corrupted. 
For a fixed $n$, if we move from noise model $(a)$ to noise model $(c)$ the increase in bias and standard errors of the VS-based sill parameter estimator is prominent, though the magnitude of increment is much smaller as compared to the standard method of estimation.

\begin{figure}
  \centering
 \begin{subfigure}{0.45\textwidth}
     \centering
    \includegraphics[trim={2.5cm 1.1cm 3.5cm 1.1cm}, width=.65\textwidth, height=0.2\textheight]{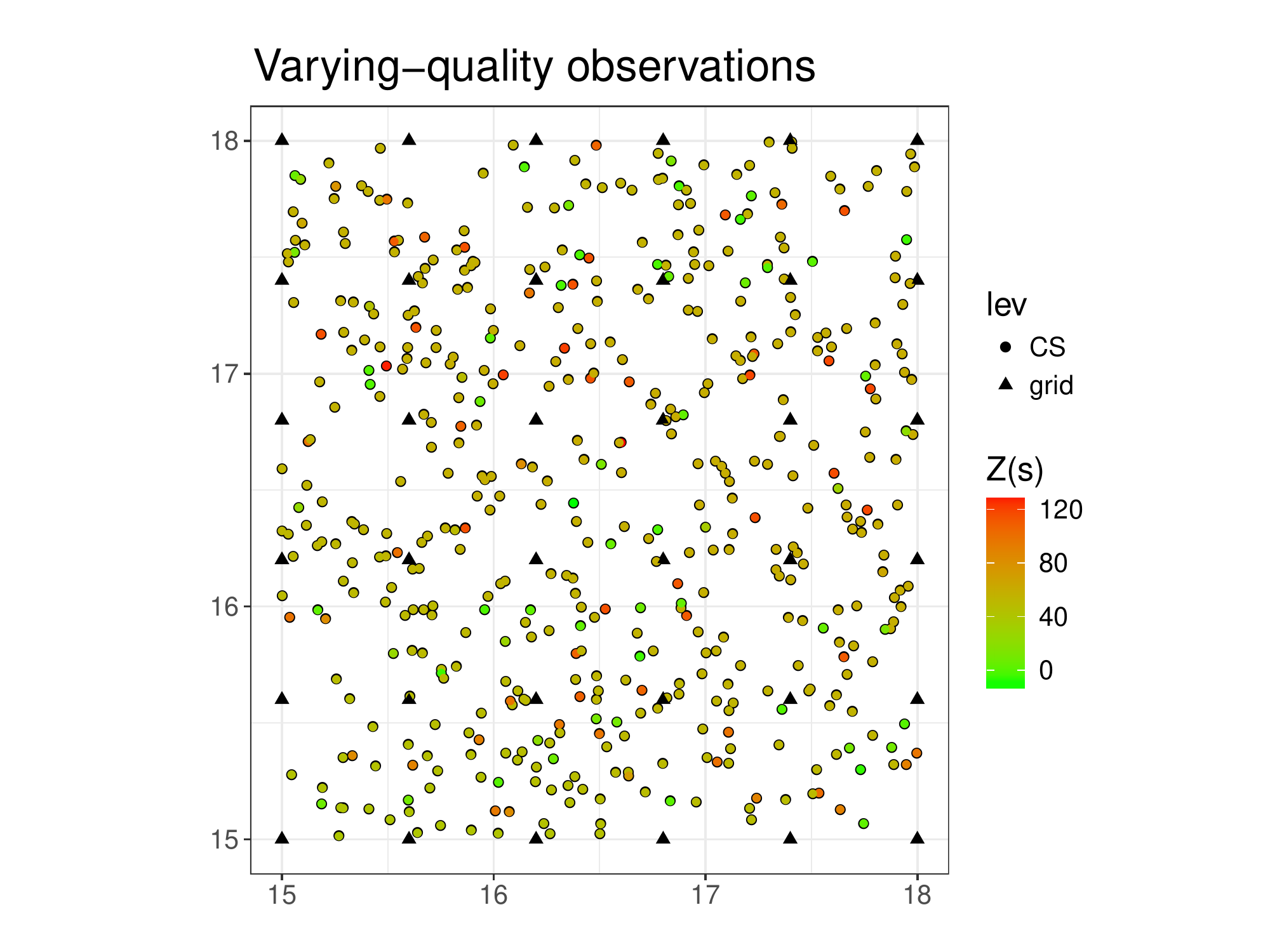}
    \subcaption{Varying-quality observations (CS = crowdsourced) and the grid to validate prediction.}\label{fig:cs_grid_sim_ex}
 \end{subfigure}%
 \hspace{6mm}
 \begin{subfigure}{0.45\textwidth}
     \centering
    \includegraphics[trim={3.5cm 1.1cm 2.5cm 1.1cm}, width=.7\textwidth, height=0.2\textheight]{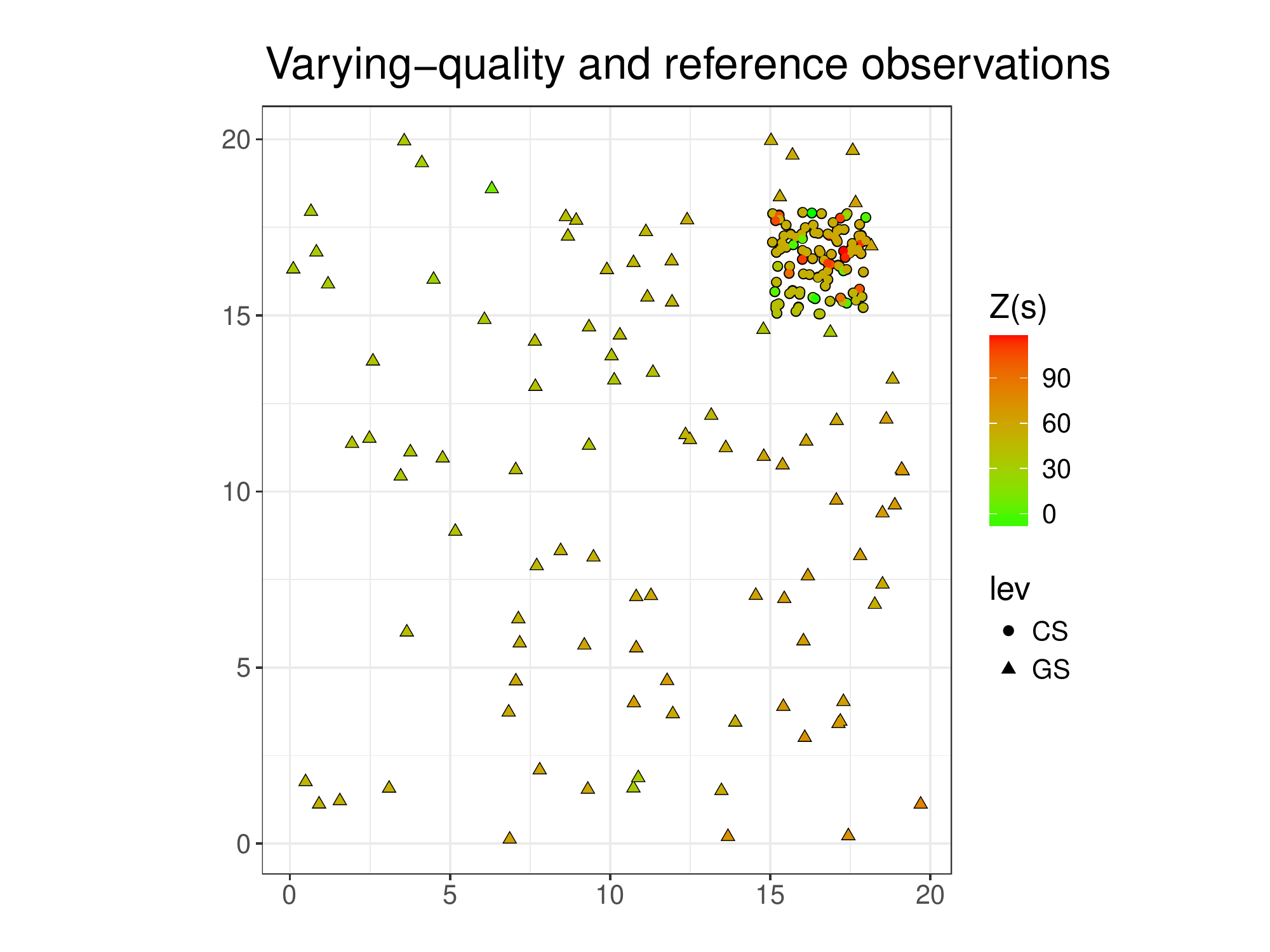}
    \subcaption{Varying-quality hyper-local observations (CS) with reference data (GS = ground-station).}\label{fig:cs_ref_sim_ex}
 \end{subfigure}\vspace{2mm}
 \caption{Example sampling points for the simulations.}\label{fig:SimSampEx}
 \end{figure}

Next we evaluate the VS-based spatial prediction using a $4 \lceil \lambda_n \rceil \times 4 \lceil \lambda_n \rceil$ grid over the sampling region $\mathcal{R}$ as shown in Figure~\ref{fig:cs_grid_sim_ex}.
We make predictions at these grid points using both the VS-based and standard approach and evaluate the predictions and kriging by the following metrics:
$$
\begin{aligned}
\text{RMSPE} &= \sqrt{\frac{1}{n}\sum_{\bfs^*}\Lp \hat{Y}_{\text{vs}}(\bfs^*) - Y(\bfs^*) \Rp^2}\; ;\;\;
\text{ResRMSPE} &= \sqrt{\frac{1}{n}\sum_{\bfs^*}\Lp \tilde{\epsilon}(\bfs^*) - \epsilon(\bfs^*) \Rp^2}\;,
\end{aligned}
$$ where the sum $\sum_{\bfs^*}$ is over the grid points. We define the performance metrics for the standard methods analogously. The Root-Mean-Squared-Prediction-Error (RMSPE) measures the average prediction error over the selected grid; and the Residual-Root-Mean-Squared-Prediction-Error (ResRMSPE) evaluates the accuracy and efficiency of the kriging on the selected grid for the spatially correlated residual process: $\{ \epsilon(\bfs) \}$. By Av.RMSPE we denote $\frac{1}{B}\sum_b \text{RMSPE}(b)$ where $\text{RMSPE}(b)$ is the prediction error in the $b$-th simulation iteration. We define Av.ResRMSPE similarly.

\begin{table}[ht]
\centering
\resizebox{.78\columnwidth}{.17\textwidth}{%
\begin{tabular}{|cc|cc|cc|}
\toprule
\multicolumn{2}{|c|}{} & \multicolumn{2}{c|}{\textbf{VS}} & \multicolumn{2}{c|}{\textbf{Std. App.}} \\ \cline{3 - 6}
 \textbf{Noise Model} & $\mathbf{n}$ & \textbf{Av.RMSPE} & \textbf{Av.ResRMSPE} & \textbf{Av.RMSPE} & \textbf{Av.ResRMSPE}\\ 
  \hline
  \hline
\multirow{3}{*}{(a)} & 100 & 5.29 (4.04) & 0.703 (0.22) & 8.61 (15.37) & 3.637 (1.82) \\ 
   & 500 & 4.046 (1.03) & 0.281 (0.03) & 4.826 (8.89) & 4.416 (1.33) \\ 
   & 3000 & 3.927 (1.07) & 0.141 (0.02) & 3.228 (1.37) & 5.306 (0.48) \\ \hline
\multirow{3}{*}{(b)} & 100 & 9.67 (6.11) & 1.796 (0.77) & 37.38 (92.72) & 14.717 (6.99) \\ 
  & 500 & 8.478 (5.04) & 0.358 (0.07) & 28.911 (75.28) & 14.267 (7.56) \\ 
   & 3000 & 5.196 (3) & 0.15 (0.02) & 20.546 (33.01) & 14.902 (8.07) \\ \hline
  \multirow{3}{*}{(c)} & 100 & 21.071 (11.29) & 5.833 (1.39) & 98.585 (206.89) & 38.74 (19.03) \\ 
   & 500 & 26.325 (14.35) & 1.376 (1.44) & 66.6 (152.04) & 36.354 (20.06) \\ 
   & 3000 & 13.722 (6.55) & 0.23 (0.04) & 94.429 (193.5) & 31.606 (23.25) \\ 
   \toprule
\end{tabular}%
}
\caption{\small Prediction performance of the VS-based methodology and standard approach on varying-quality observations without any reference data.}
\label{tab:PredSurfPerform_woRef}
\end{table} 
Table~\ref{tab:PredSurfPerform_woRef} summarizes the results which show that the VS-based predictions are much better than the standard analysis in almost all the cases. As we go from model (a) to model (c) the prediction accuracy has compromised for both the VS-based as well as the standard approach with much higher impact for the later one. However, in terms of residual kriging efficiency the VS-based methodology is highly robust as compared to the ordinary kriging using the residuals obtained from o.l.s.

\subsection{With Reference Data}
\label{subsec:SimStud_wRef}
In this subsection, we consider a situation that is more similar to our case study. In addition to the $n$ varying-quality observations in the hyper-local region $\mathcal{R} = [0, \lambda_n]^2$, we have $m$-many high-quality observations available over a larger region $\mathcal{D} = [0, \Lambda_m]^2$. One example of the sampling points is shown in Figure~\ref{fig:cs_ref_sim_ex}. Our goal is to predict the process within the hyper-local region $\mathcal{R}$ using the varying-quality observations. We again use a $4 \lceil \lambda_n \rceil \times 4 \lceil \lambda_n \rceil$ grid over the hyper-local region of interest $\mathcal{R}$ to evaluate the predictions. In addition to the predictions obtained by the VS-based and standard methodology on the varying-quality observations, we also consider the global predictions obtained by using only the reference data on the larger region as shown in Figure~\ref{fig:cs_ref_sim_ex}. For this simulations we have considered the sample sizes for varying-quality observations to be equal to $50, 100$ and $500$ because the hyper-local regions in our case studies do not contain very `large' (not more than $300$) number of crowdsourced observations. For the reference data the sample sizes we have taken $m = 100$. 

In Table~\ref{tab:PredSurfPerform}, first we compare the performances of the VS-based and standard predictions using hyper-local noisy data based on RMSPE for both at the response level (Av.RMSPE) and residual level (Av.ResRMSPE).
\begin{table}[ht]
\centering
\resizebox{\columnwidth}{.17\textwidth}{%
\begin{tabular}{|cc|cc|cc|cc|cc|}
\toprule
  & \multicolumn{1}{c|}{} &
  \multicolumn{2}{c|}{\textbf{VS}} &
  \multicolumn{2}{c|}{\textbf{Std. App.}} & \multicolumn{2}{c|}{\textbf{Ref. Only}} \\
  \cline{3-8}
  
 \textbf{Noise Model} & $\mathbf{n}$ &
  \textbf{Av.RMSPE} &
  \textbf{Av.ResRMSPE} &
  \textbf{Av.RMSPE} & \textbf{Av.ResRMSPE} & \textbf{Av.RMSPE} &  \textbf{Av.ResRMSPE} \\ 
  \hline
  \hline
\multirow{3}{*}{(a)} & 50 & 12.26 (12.71) & 7.084 (5.08) & 1740.696 (9518.03) & 1745.244 (9604.81) & \multirow{9}{*}{9.711 (8.54)} & \multirow{9}{*}{9.017 (7.46)} \\ 
 & 100 & 10.877 (11.61) & 6.104 (5.24) & 230.117 (918.91) & 224.56 (934.89) &  &  \\ 
 & 500 & 8.787 (8.05) & 6.287 (6.47) & 358.694 (1976.29) & 352.86 (1975.49) &  &  \\ \cline{1-6}
 \multirow{3}{*}{(b)} & 50 & 12.933 (13.1) & 8.206 (6.76) & 52829.372 (662485.91) & 52946.917 (664727.7) &  \\ 
& 100 & 9.907 (10.81) & 6.439 (5) & 115.222 (923.6) & 387.071 (915.98) &  &  \\ 
& 500& 9.005 (8.66) & 6.72 (5.06) & 26.31 (19.18) & 217.784 (15.88) &  &  \\ \cline{1-6}
\multirow{3}{*}{(c)} & 50 & 12.33 (18.51) & 8.7 (16.61) & 10198.908 (85831.41) & 9740.72 (85082.65) &  \\ 
& 100 & 10.131 (10.93) & 7.093 (5.04) & 155.796 (126.08) & 412.788 (31.68) &  &  \\ 
& 500 & 9.786 (8.45) & 6.402 (5.24) & 239.728 (29.49) & 27.335 (8.35) &  &  \\ 
   \hline
   \end{tabular}%
}
\caption{\small Performance of hyper-local predictions using the VS-based methodology, the standard approach and global predictions using reference data only. For these simulations we used reference data with sample size $m = 100$.}
\label{tab:PredSurfPerform}
\end{table}
Clearly, we can see that VS-based predictions are uniformly better than the standard ones in all the considered cases. Next, we compare the VS-based predictions using hyper-local noisy data and the predictions obtained by implementing the standard methodology on the high-quality reference data over a bigger region. We refer the later one as `Ref. Only'. From Table~\ref{tab:PredSurfPerform}, we see that, at response level (i.e. comparing Av.RMSPE), under all noise models, the performance of the VS-based predictor using varying-quality observations is similar or slightly worse to the `Ref. Only' predictor, when the number of hyper-local noisy data and the high-quality reference data are comparable (i.e. the case when both $n$ and $m$ = $100$.) In case we have larger sample size ($n = 500$) in the hyper-local regions, we see a little gain in prediction efficiency in terms of Av.RMSPE. However, if we consider the residual kriging performance i.e. the ResRMSPE, the VS-based technique has outperformed the `Ref Only' kriging for all the cases, even when we have only $n = 50$ many varying-quality observations. As the kriging is more efficient when we have observations closer to the locations of our interest, the varying-quality hyper-local observations along with the robust VS-based methodology improves the efficiency of the spatial prediction as compared to the corresponding `Ref. Only' version. Additional details regarding the simulation results, e.g. the parameters of the models and choices of the regularity parameters etc., are reported in Section C.1 of the supplementary material.

\section{Case Study: Spatial Analysis of WeatherSignal Data}
\label{sec:VS-OpneSignal}
In this section, we analyze the WeatherSignal data described in Section~\ref{subsec:OpenSignal} using the VS-based methodology (Section~\ref{sec:vsBased}). Our goal for this noisy crowdsourced data set is to perform structure exploration and then prediction of the daily average ambient temperature process in hyper-local regions of interest.

\subsection{Building Hyper-Local Prediction Surfaces}
\label{subsec:HyperLocalPred} 
\begin{figure}
  \centering
 \begin{subfigure}{0.32\textwidth}
     \centering
    \includegraphics[trim={3.5cm 2cm 2.5cm 0.5cm}, width=.7\textwidth, height=0.15\textheight]{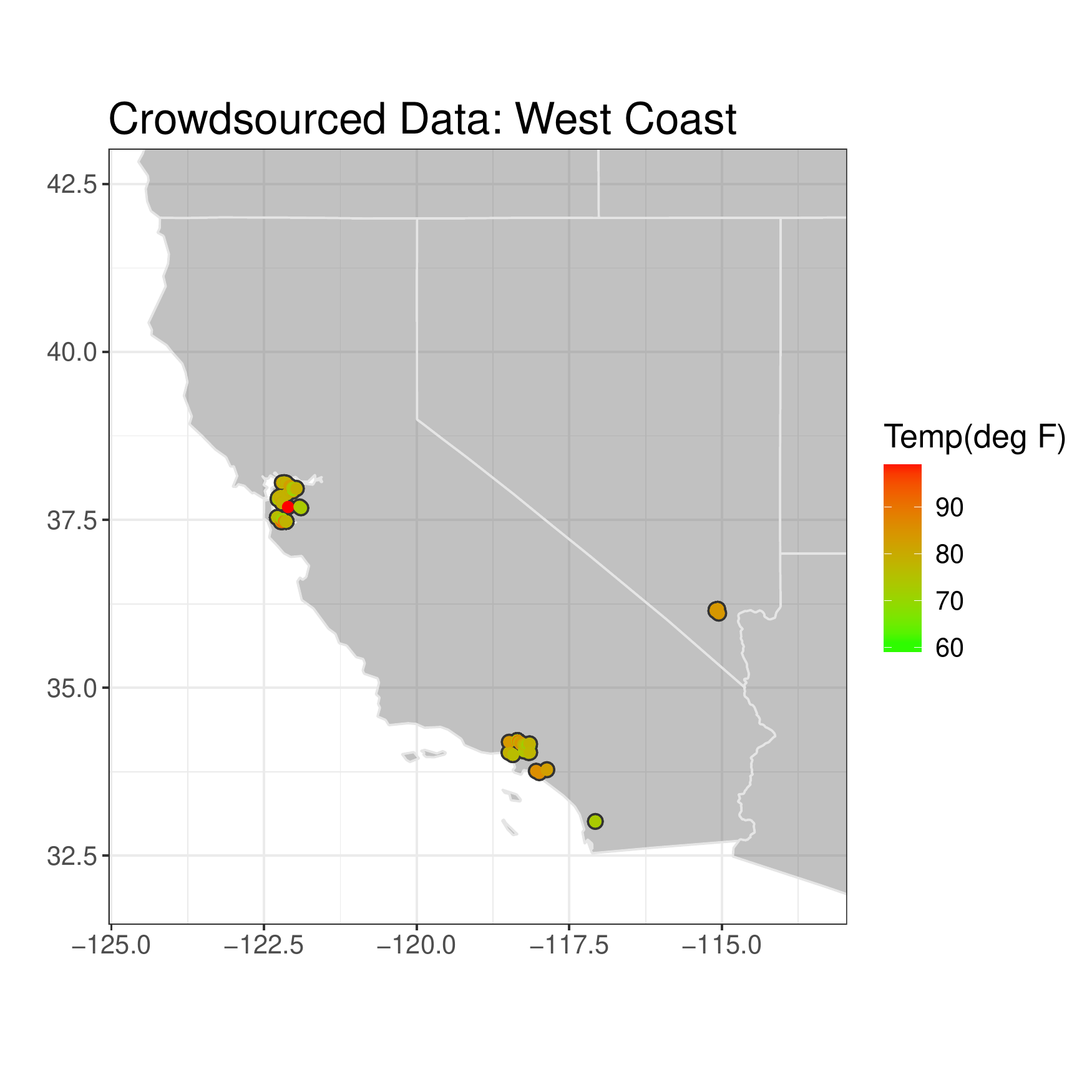}
    \subcaption{}\label{fig:CS_Data_West}
 \end{subfigure}\hspace{2mm}%
 \begin{subfigure}{0.32\textwidth}
     \centering
    \includegraphics[trim={3.5cm 2cm 2.5cm 0.5cm}, width=.7\textwidth, height=0.15\textheight]{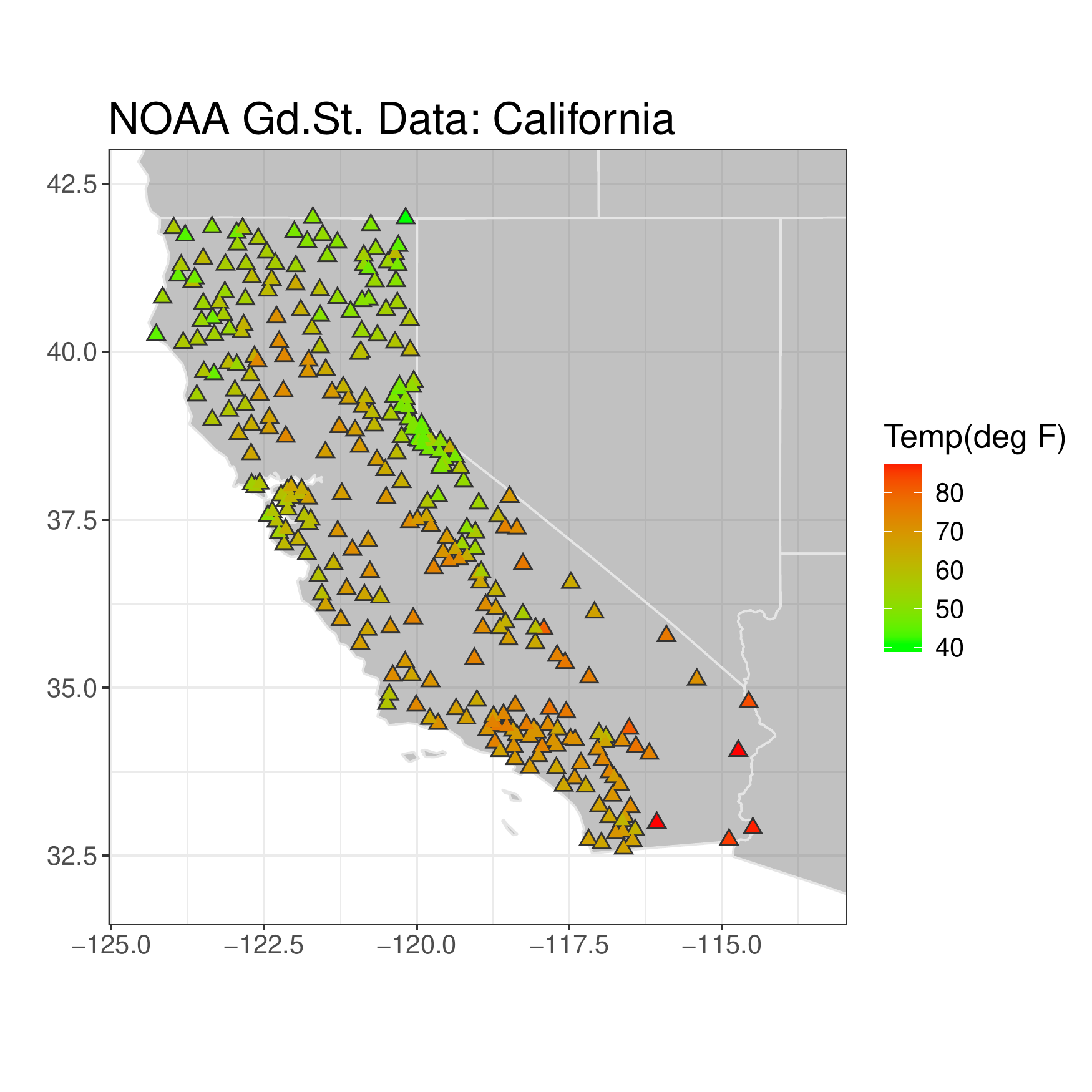}
    \subcaption{}\label{fig:GS_Data_California}
 \end{subfigure}\hspace{2mm}%
 \begin{subfigure}{0.32\textwidth}
     \centering
    \includegraphics[trim={3.5cm 2cm 2.5cm 0.5cm}, width=.7\textwidth, height=0.15\textheight]{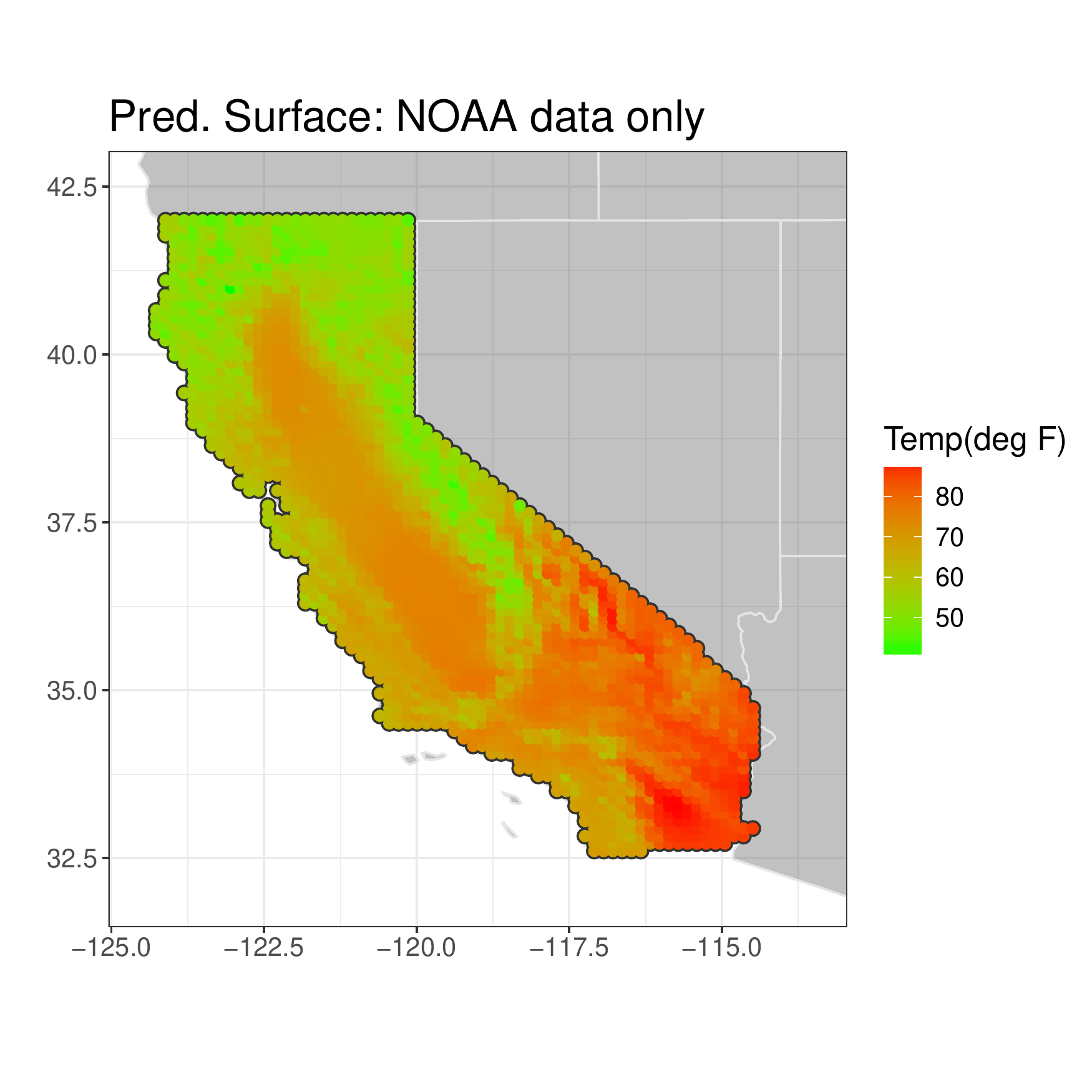}
    \subcaption{}\label{fig:PredSur_NoaaOnly_Cali}
 \end{subfigure}\vspace{-3mm}
  \begin{subfigure}{0.32\textwidth}
     \centering
    \includegraphics[trim={4.5cm 2.5cm 2.5cm 1.5cm}, width=.7\textwidth, height=0.16\textheight]{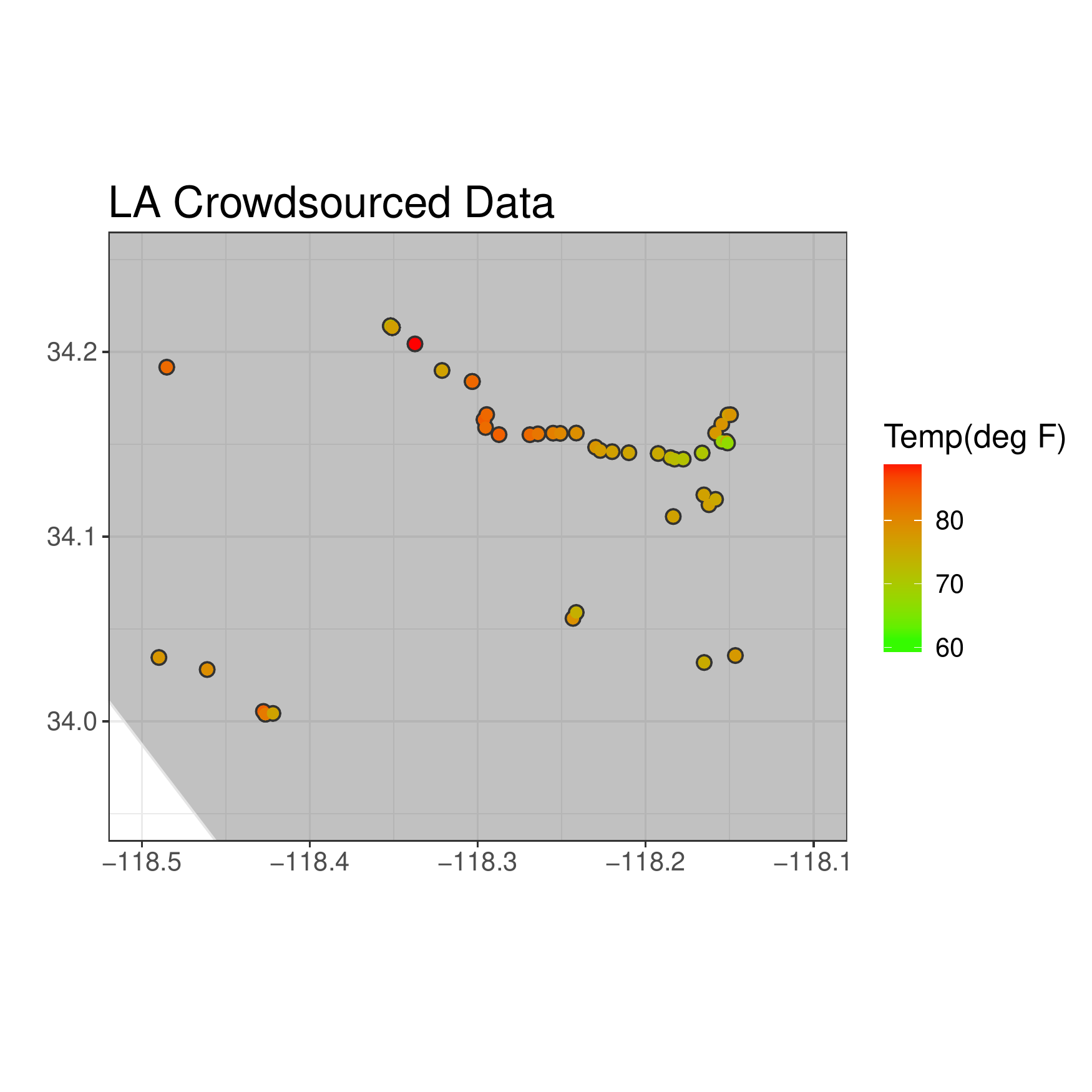}
    \subcaption{}\label{fig:CS_Data_LA}
 \end{subfigure}\hspace{10mm}%
 \begin{subfigure}{0.32\textwidth}
     \centering
    \includegraphics[trim={3.5cm 2.5cm 3.5cm 1.5cm}, width=.7\textwidth, height=0.16\textheight]{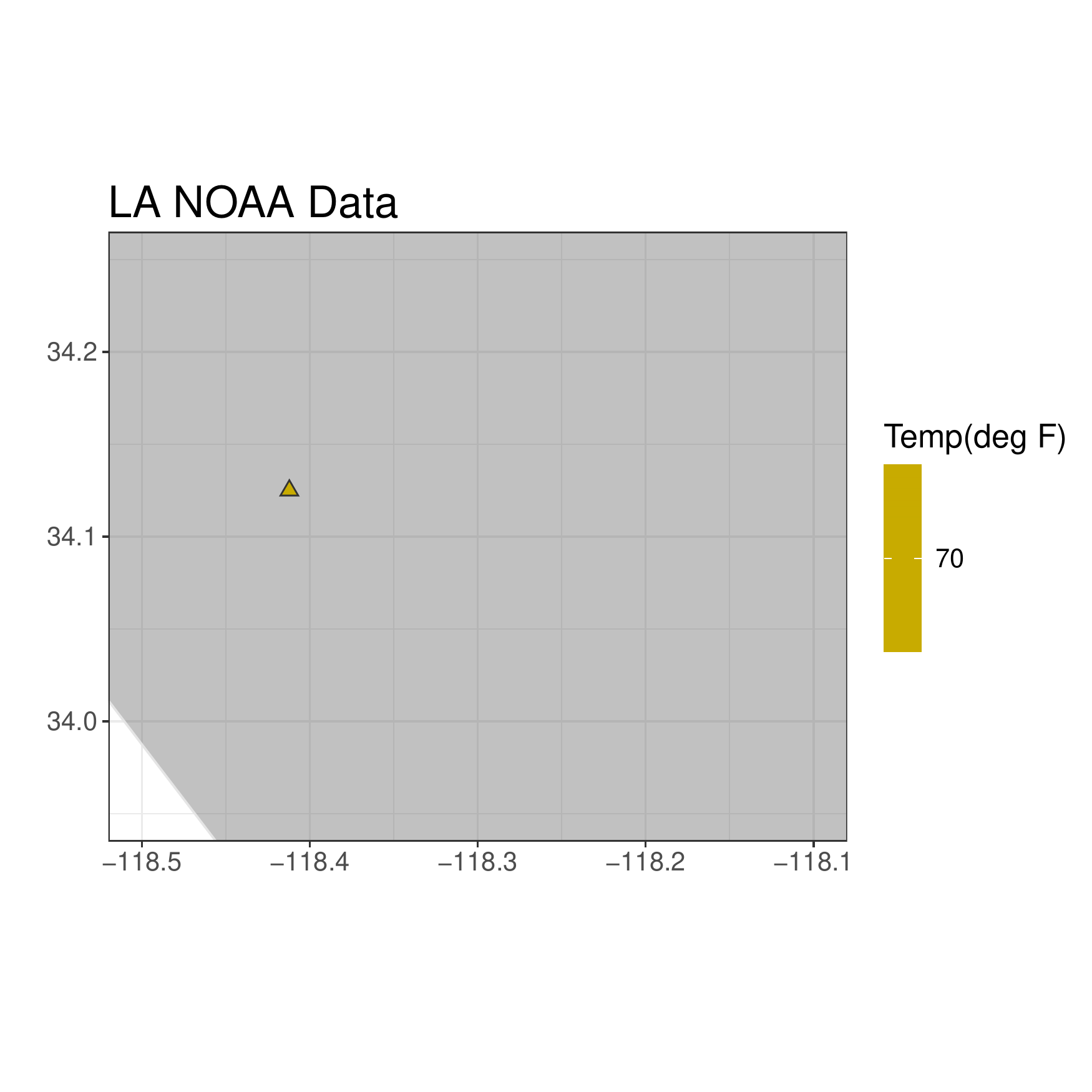}
    \subcaption{}\label{fig:GS_Data_LA}
 \end{subfigure}
 \caption{\small{(a) Crowdsourced observations in CA; (b) Available ground-station observations; (c) Prediction surface using the standard approach on the ground-station data; (d) Crowdsourced observations in a hyper-local region around Los Angeles; (e) ground-station observations in a hyper-local region around Los Angeles.}}\label{fig:LA_Dat_Plot}
 \end{figure}

Here we describe the VS-based analysis of the crowdsourced WeatherSignal data using the NOAA ground-station data as reference. We first select a hyper-local region, as denoted by $\mathcal{R}$ in Section~\ref{subsubsec:vsWref}, around Los Angeles, CA, as shown in Figure~\ref{fig:CS_Data_LA}.
 The analysis starts by defining a region large enough to have sufficient NOAA ground-station observations to build a reasonable global prediction surface around the region of interest. In Figure~\ref{fig:GS_Data_California}, we plot the $m = 310$ ground-station observations in California. Using the standard approach on the NOAA ground-station data, as described in Section~\ref{subsec:stanGeo}, we build a prediction surface for California and plot it in Figure~\ref{fig:PredSur_NoaaOnly_Cali}. The model we use to estimate the mean is given by
 \begin{equation}
 \label{eq:MeanModel_CaseStudy}
     \begin{split}
         \mu(\bfs) = \beta_0 + \beta_x \; s_x + \beta_y \; s_y + \beta_{xy} \; s_x s_y + \beta_h \;h(\bfs),
     \end{split}
\end{equation} where $\bfs \coloneqq \Lp s_x, s_y \Rp^\prime$ and $h(\bfs)$ denotes the elevation of the point $\bfs$. The mean model explains $79\%$ (adjusted $\text{R}^2$) of the variability in the ground-station ambient temperatures in California. 

We then fit a Mat\'ern covariance to the observed residuals from the mean model estimation. Details of the variogram estimation are given in Table~\ref{tab:EstMat_Noaa_cali} and Figure~\ref{fig:VarioEst_Noaa_Cali}. We then use standard kriging methodology with the estimated mean and covariance model to create the prediction surface $\LP (\bfs, \hat{Y}(\bfs)): \bfs \in \mathcal{D} \RP$, as shown in Figure~\ref{fig:PredSur_NoaaOnly_Cali}.

\begin{minipage}[c]{0.49\textwidth}
\centering
\resizebox{.6\columnwidth}{.18\textwidth}{%
\begin{tabular}{cc}
\hline
\hline
\textbf{Parameters} & \textbf{Estiamtes} \\ 
\hline
 partial sill ($\sigma^2$)    & 13.78              \\ 
range ($\rho$)        & 0.36               \\ 
nugget ($\tau^2$)     & 7.95               \\ 
smoothness ($\kappa$) & 2.45                  \\ 
   \hline
   \hline
\end{tabular}%
}
\captionof{table}{\small{Estimated Mat\'ern parameters.}}
\label{tab:EstMat_Noaa_cali}
\end{minipage}
\begin{minipage}[c]{0.4\textwidth}
\includegraphics[trim={1.5cm 0.9cm 1.5cm 0.5cm}, width=.95\textwidth, height = .19\textheight]{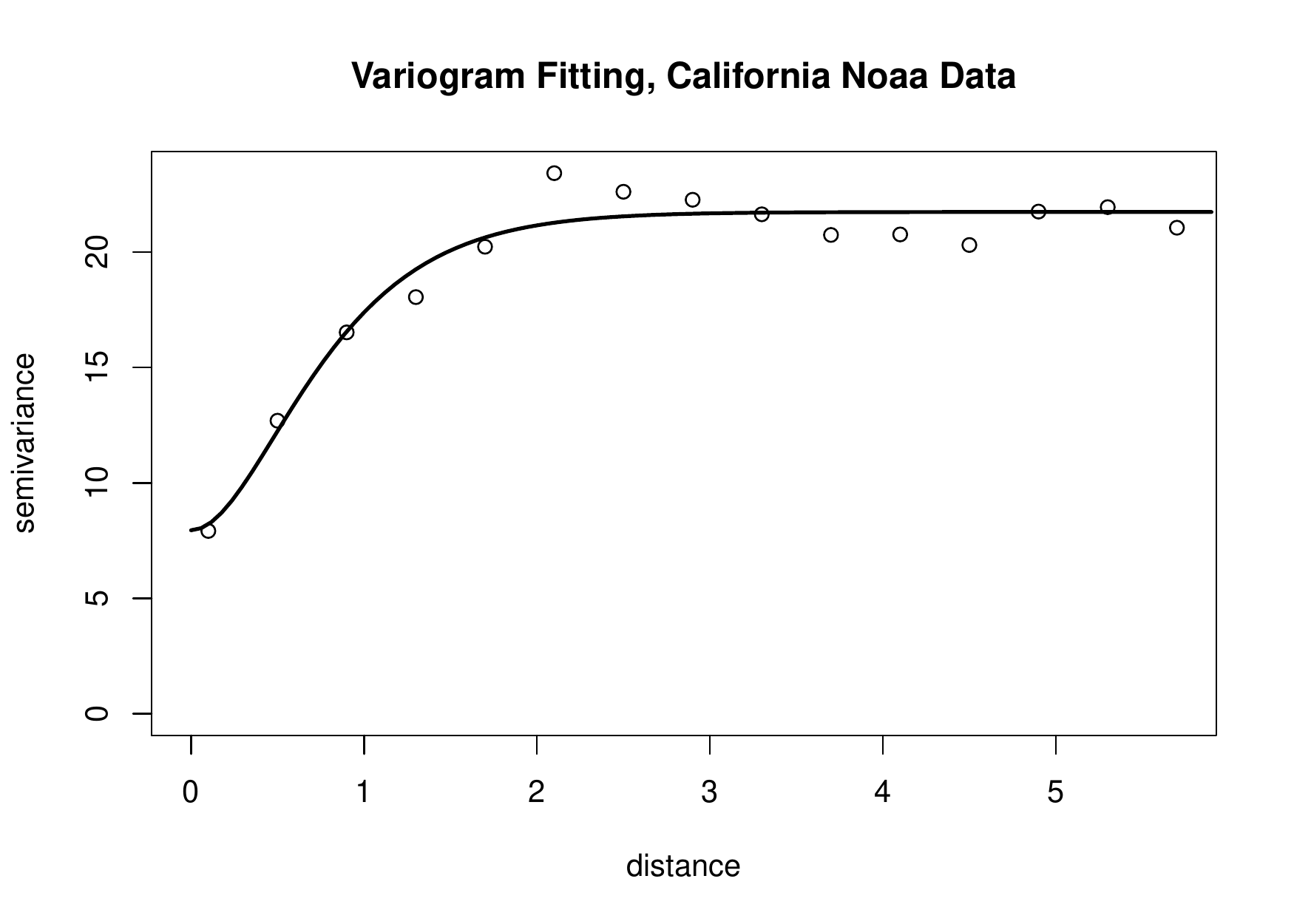}
\captionof{figure}{\small{Variogram estimation}}
\label{fig:VarioEst_Noaa_Cali}
\end{minipage}

As we can see in Figure~\ref{fig:CS_Data_West}, the spatial coverage of the crowdsourced data does not support a global prediction surface over California or even the coast of California. However, if we consider the $25 \times 25$ mile region ($\mathcal{R}$) in LA, as shown in Figure~\ref{fig:CS_Data_LA}, the density of crowdsourced data is much higher as compared to only one ground-station observation (Figure~\ref{fig:GS_Data_LA}). While there is only one ground-station available at Los Angeles International Airport, the number of crowdsourced observations, $\LP Z(\bfs_1), \dots , Z(\bfs_n) \RP$, in $\mathcal{R}$ is $n = 80$. 

The next part of the analysis examines whether we can leverage the additional crowdsourced information through the VS-based methodology. We want to explore whether we can create a more reasonable and efficient prediction surface $\LP (\bfs, \hat{Y}_{\text{vs}}(\bfs)): \bfs \in \mathcal{R} \RP$ over the region $\mathcal{R}$ in Los Angeles as compared to the surface obtained from the analysis of the ground-station data only, $\LP (\bfs, \hat{Y}(\bfs)): \bfs \in \mathcal{R} \RP$.

The VS-based analysis starts by computing the veracity score of the crowdsourced observations using the definition in Equation~\ref{eq:VerSc_wRef}. We set the baseline deviation $\alpha=3$. In an ideal scenario, when the corresponding $\delta$-neighborhood has very little variation and $\text{IQR}\Lp \boldsymbol{\xi}_i \Rp \approx 0$, an observation with $3^\circ$F deviation from the corresponding benchmark value has a VS approximately equal to $\exp(-1) \approx 0.368$, while an observation with a $1^\circ$F deviation has a VS $\approx 0.716$. To define the neighborhood for computation of the VS, we take $\delta = 0.08$ in the units of latitude and longitude. To choose a suitable mixing parameter $\nu$, we use the function
$$
\nu(\bfs_i) = 1 - \exp\Lp\frac{-1}{(1 - \text{R}^2)\; \sqrt{n(i)}}\Rp,
$$ where $\text{R}^2$ is the adjusted R-squared for the estimation of the mean surface using NOAA ground-station data only and $n(i)$ is the number of crowdsourced data in the $\delta$-neighborhood.
\begin{figure}
  \centering
   \begin{subfigure}{0.45\textwidth}
     \centering
    \includegraphics[trim={3.5cm .5cm 2.5cm 0.5cm}, width=.7\textwidth, height=0.19\textheight]{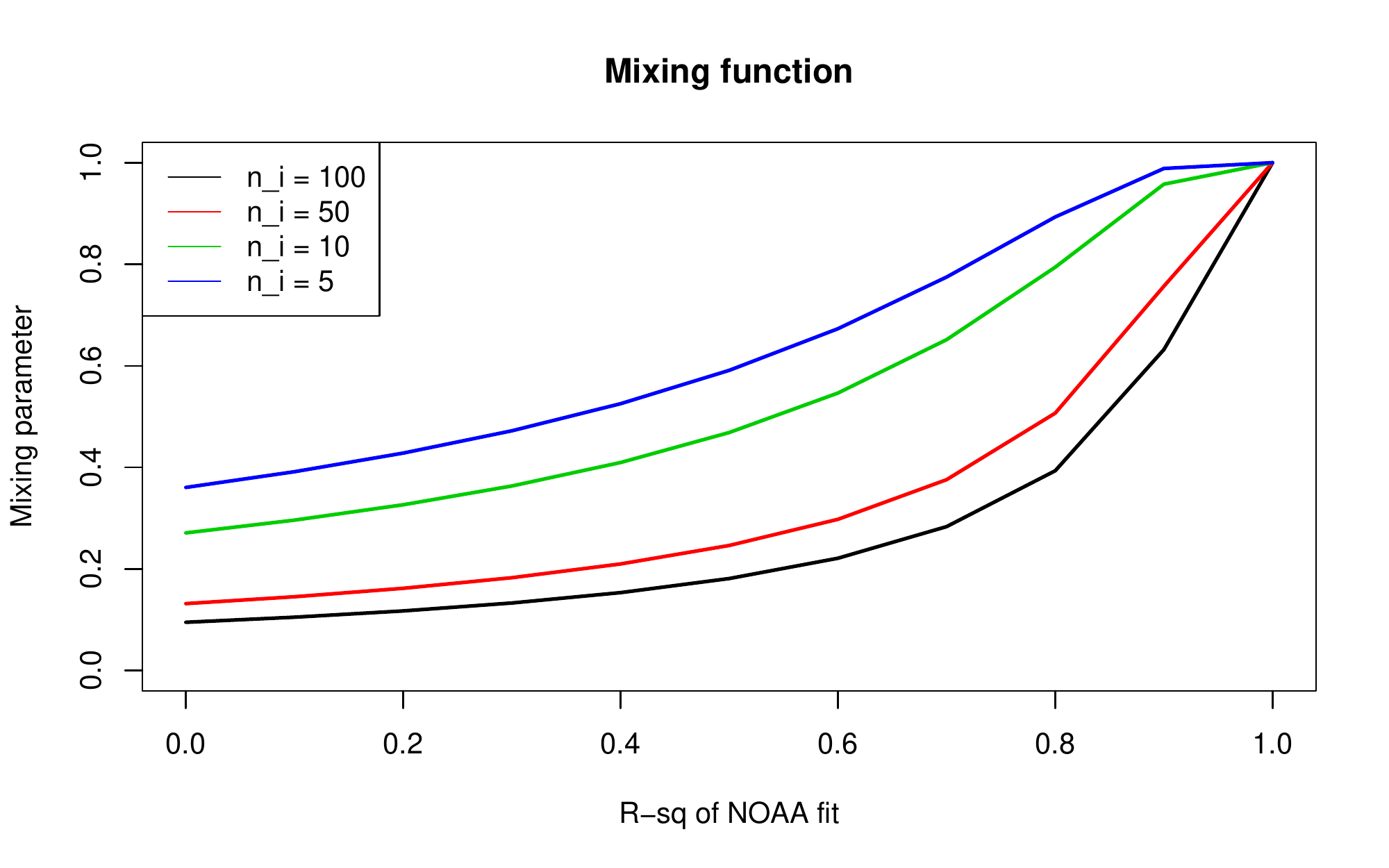}
    \subcaption{}\label{fig:mixing_param}
 \end{subfigure}\hspace{3mm}%
 \begin{subfigure}{0.45\textwidth}
     \centering
    \includegraphics[trim={3.5cm .5cm 2.5cm 0.5cm}, width=.65\textwidth, height=0.18\textheight]{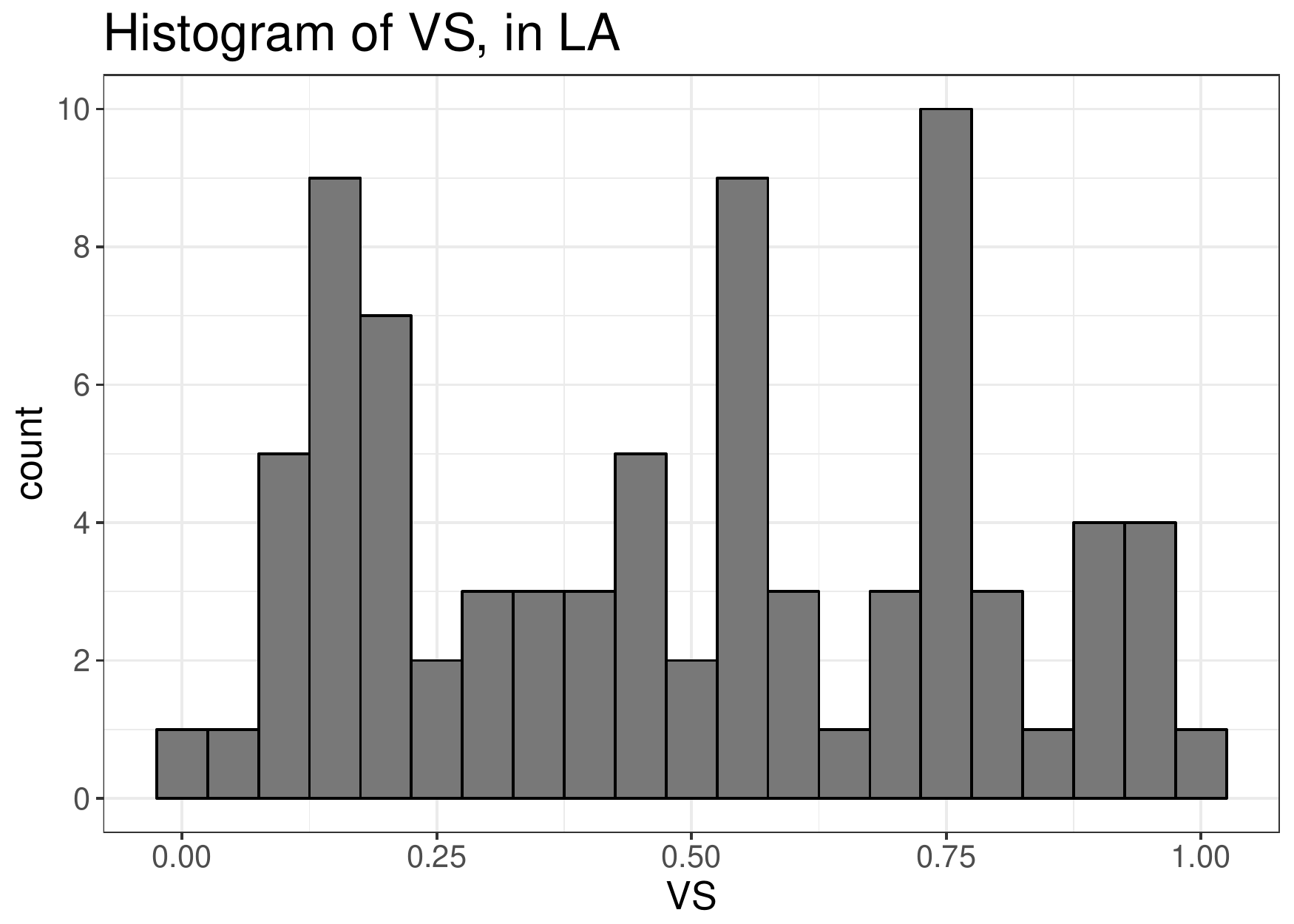}
    \subcaption{}\label{fig:vs_la}
 \end{subfigure}
 \caption{\small{Mixing function (a) and the histogram of the veracity scores (b) for the crowdsourced observations in Los Angeles.}}
\end{figure}
As Figure~\ref{fig:mixing_param} shows, this function is increasing in $\text{R}^2$ and decreasing in $n(i)$. $\nu(\bfs_i) = 1$ if $\text{R}^2 = 1$ and $\nu(\bfs_i) = 0$ if $n(i) = \infty$. With this formulation, the mixing parameter takes both the goodness of fit for the ground-station data and the number of crowdsourced observations used for local approximation of the target value into account. Using the specified parameters, we compute the VS for the crowdsourced observations in $\mathcal{R}$ and plot their empirical distribution in Figure~\ref{fig:vs_la}.

We next estimate the mean and covariance of the process. For robust estimation of the mean function, we use the weighted MM-type estimator, as discussed in Section~\ref{subsec:RobMeanEst} with the VS of the observations as the corresponding weights. Once the regression parameters are estimated, for a given smoothing parameter $q$ in Equation~\ref{eq:NewRes_wRef}, we use the VS-based smoothing technique to reduce the effects of noise in the residual process as discussed in Section~\ref{subsec:CovEst}. Using the smoothed residuals, we  estimate the covariance parameters and use the estimates to create a prediction surface using VS-based kriging as discussed in Section~\ref{subsec:VSKrig}. 

To make an optimal choice for $q$, we use the reference data. For a pre-specified set of values of $q \in [0.05, 3]$ the covariance estimation and kriging are executed at the ground-station locations that are inside the hyper-local region $\mathcal{R}$, and the $q$ that minimizes the mean squared error of prediction at the stations is chosen to be optimal. In the analysis for the hyper-local region around Los Angeles, there is only one station available, so we use the set of points with VS greater than or equal to $0.8$ as test data and minimize leave-one-out cross-validated mean squared prediction error, i.e. $n_*^{-1}\sum_j \Lp Z(\bfs_j) - \hat{Y}^{(-j)}_{\text{vs}}(\bfs_j) \Rp^2$ where $\hat{Y}^{(-j)}_{\text{vs}}(\bfs_j)$ is the predicted value at $\bfs_j$ obtained using $\LP Z(\bfs_1), \dots, Z(\bfs_{j-1}), Z(\bfs_{j+1}), \dots Z(\bfs_n)  \RP$ as the training data and the sum is over the test data set whose cardinality is denoted by $n_*$.
\begin{figure}
  \centering
   \begin{subfigure}{0.32\textwidth}
     \centering
    \includegraphics[trim={3.5cm .5cm 2.5cm 0.5cm}, width=.65\textwidth, height=0.18\textheight]{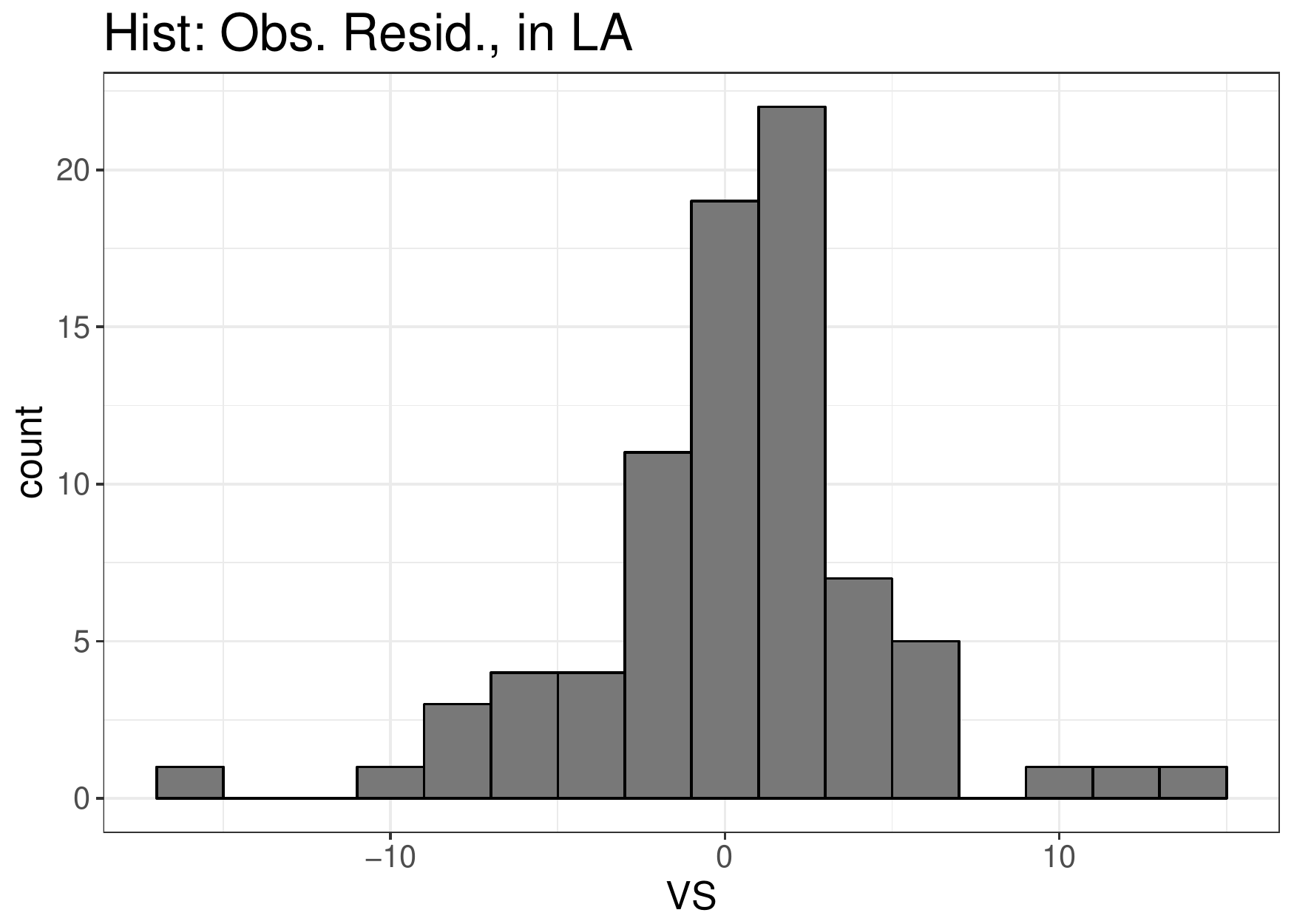}
    \subcaption{}\label{fig:Hist_Obs_Res_LA}
 \end{subfigure}\hspace{3mm}%
 \begin{subfigure}{0.32\textwidth}
     \centering
    \includegraphics[trim={3.5cm .5cm 2.5cm 0.5cm}, width=.65\textwidth, height=0.18\textheight]{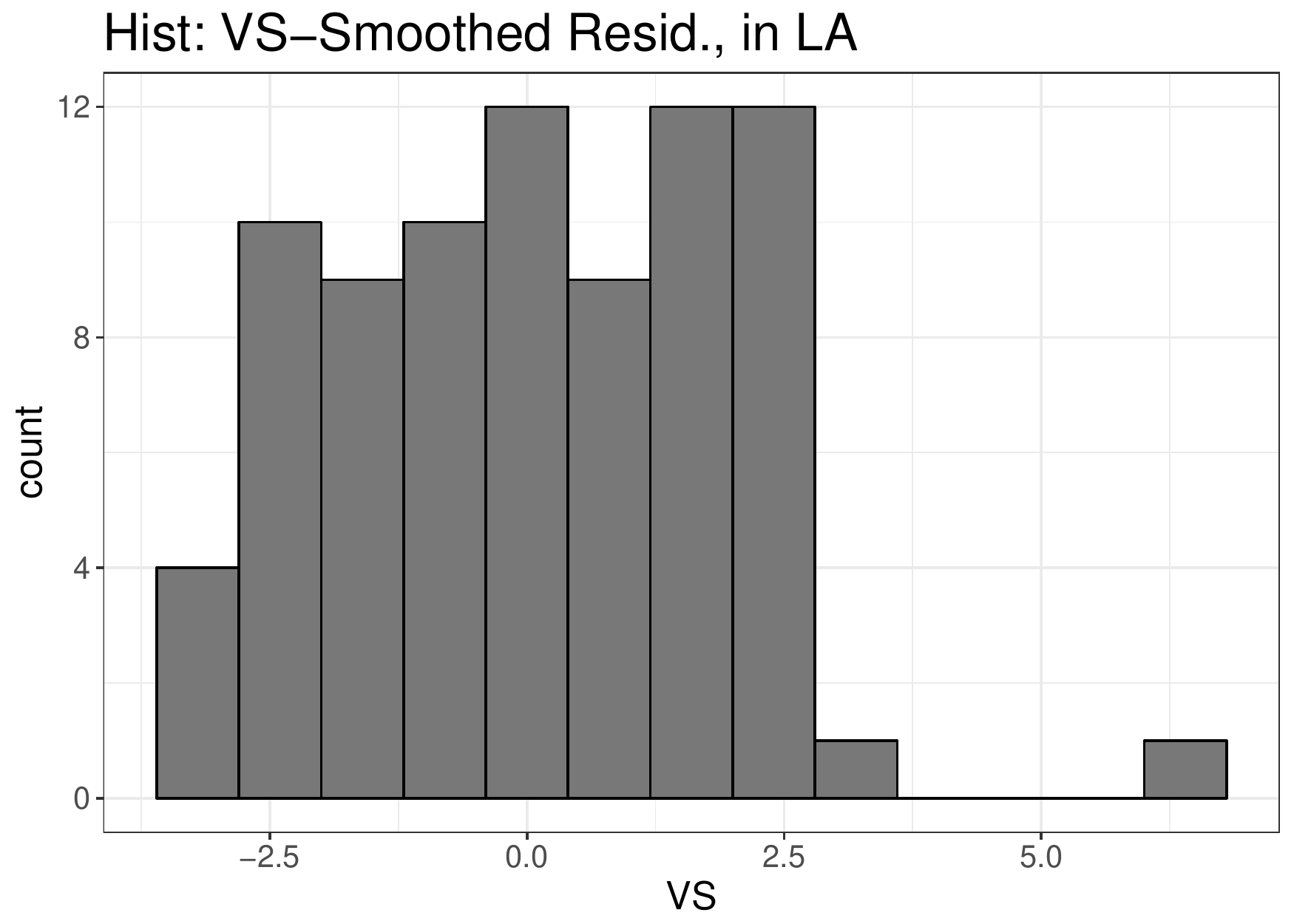}
    \subcaption{}\label{fig:Hist_VS_SM_Res_LA}
 \end{subfigure}\hspace{3mm}%
 \begin{subfigure}{0.32\textwidth}
     \centering
    \includegraphics[trim={3.5cm .5cm 3.5cm 0.5cm}, width=.65\textwidth, height=0.18\textheight]{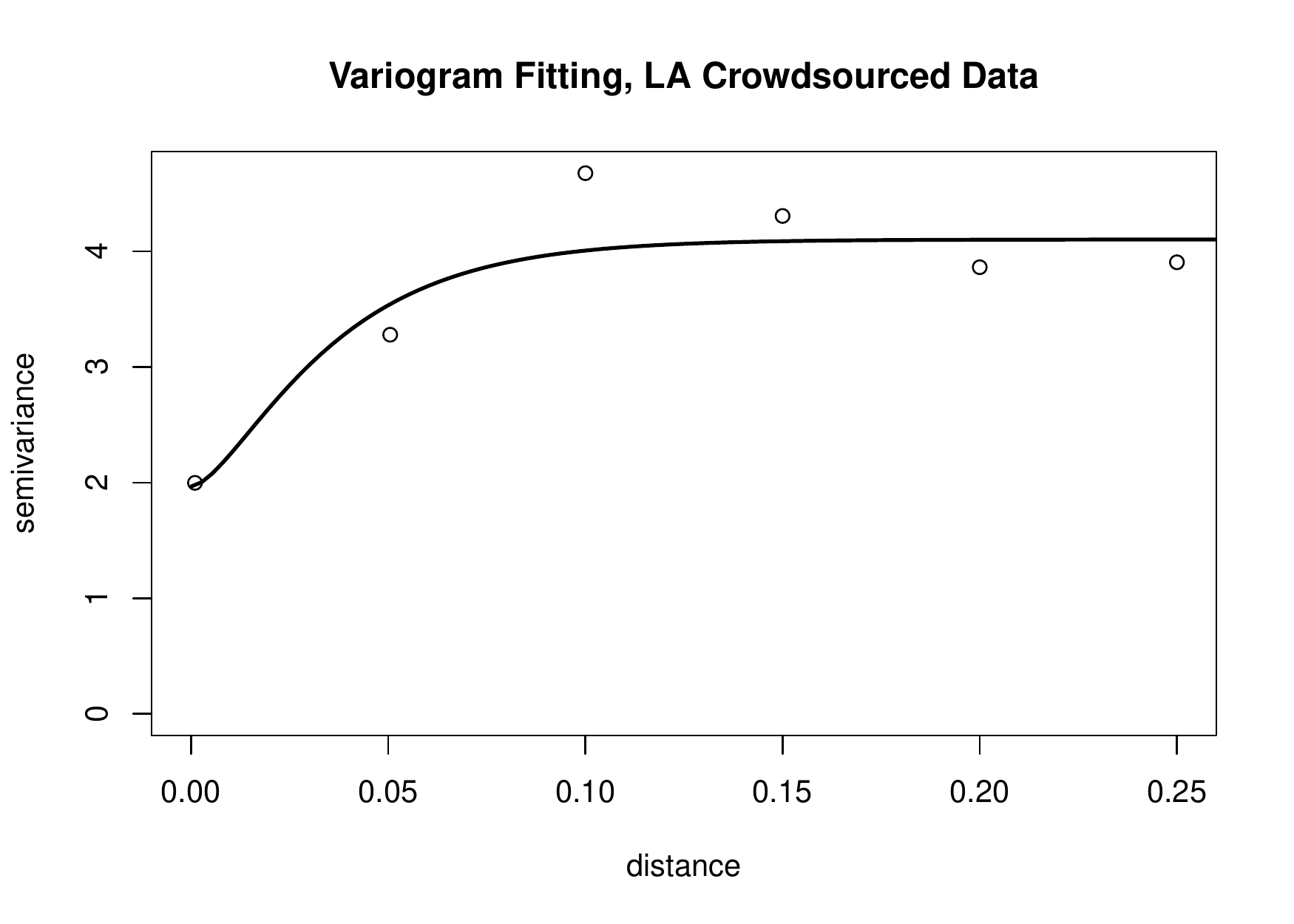}
    \subcaption{}\label{fig:Vario_Est_VS_LA}
 \end{subfigure}
 \caption{\small{Histograms of the observed residuals (a) and VS-based smoothed residuals (b) and the VS-based variogram fitting (c) for optimal $q = 0.8$.}}
\end{figure} 
In Figures~\ref{fig:Hist_Obs_Res_LA} and \ref{fig:Hist_VS_SM_Res_LA}, we plot the histograms of the observed residuals from the VS-based robust regression and the residuals after the VS-based smoothing. The VS-based smoothing clearly reduces the spread of the residual values by smoothing out the large errors. In Figure~\ref{fig:Vario_Est_VS_LA}, we show the robust variogram fitting of the VS-based smoothed residuals for the optimal choice of the smoothing parameter $q = 0.8$.

Given these analyses, we construct a prediction surface over the region $\mathcal{R}$ using Equation~\ref{eq:KrigEq_Z}.
\begin{figure}
  \centering
 \begin{subfigure}{0.45\textwidth}
     \centering
    \includegraphics[trim={4.5cm 1cm 1.5cm 1.5cm}, width=.6\textwidth, height=0.13\textheight]{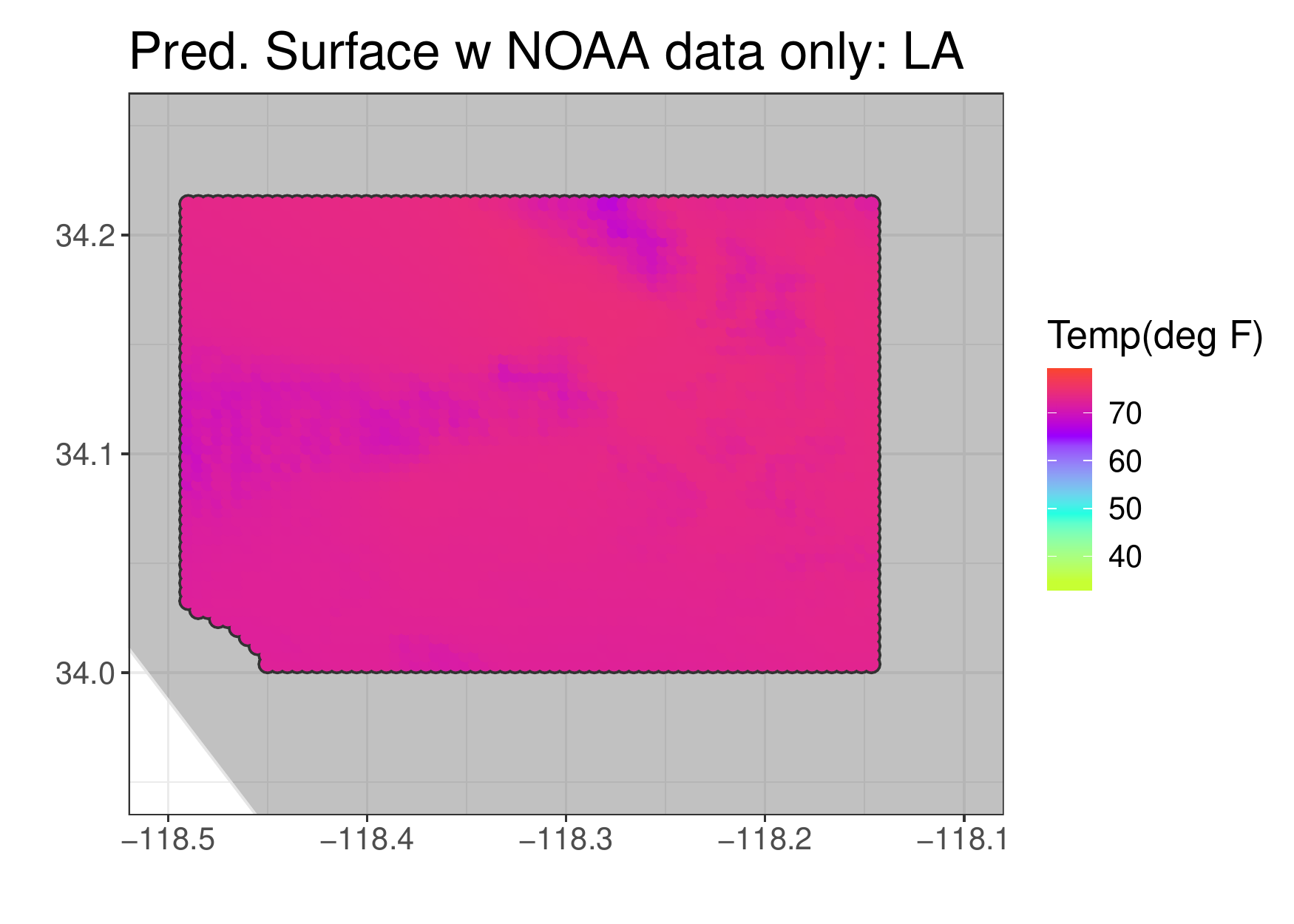}
    \subcaption{}\label{fig:PredSurf_Noaa_LA}
 \end{subfigure}\hspace{6mm}%
 \begin{subfigure}{0.45\textwidth}
     \centering
    \includegraphics[trim={3.5cm 1cm 2.5cm 1.5cm}, width=.6\textwidth, height=0.13\textheight]{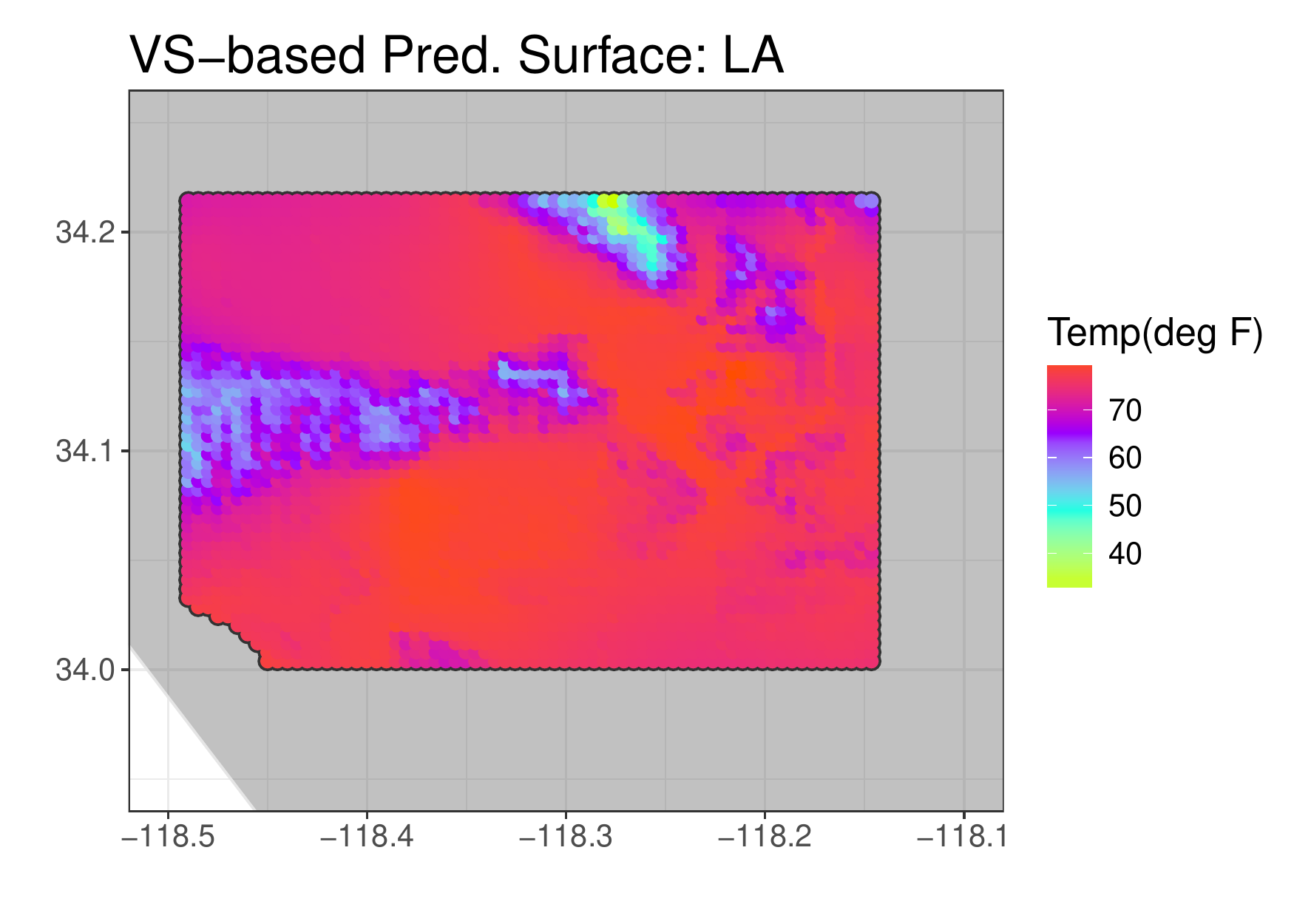}
    \subcaption{}\label{fig:PredSurf_VS_LA}
 \end{subfigure}
 \begin{subfigure}{0.33\textwidth}
     \centering
    \includegraphics[trim={4.5cm .5cm 1.5cm 0.5cm}, width=.65\textwidth, height=0.13\textheight]{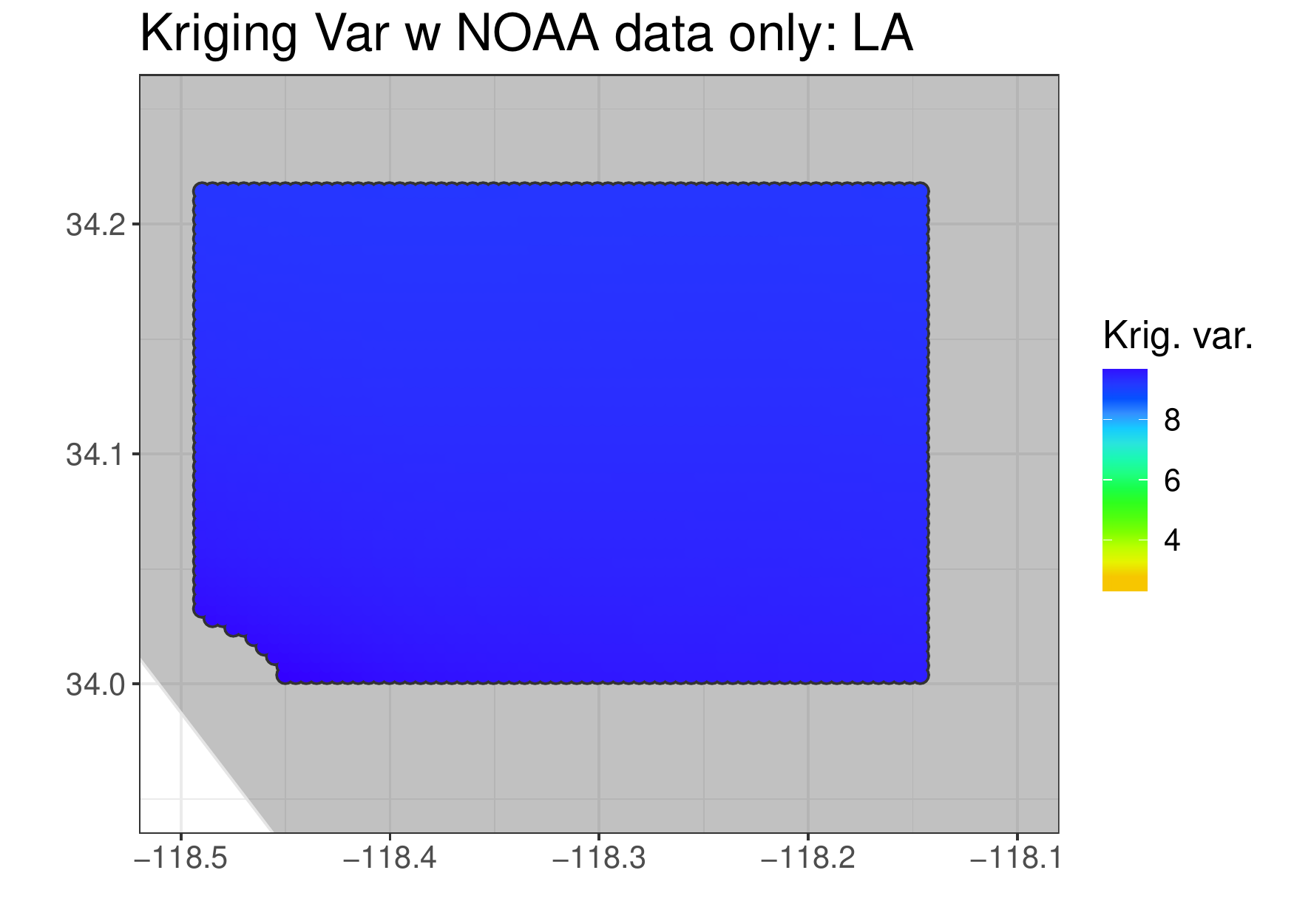}
    \subcaption{}\label{fig:KrigVar_Noaa_LA}
 \end{subfigure}%
  \begin{subfigure}{0.33\textwidth}
     \centering
    \includegraphics[trim={3cm 0.5cm 3cm 0.5cm}, width=.65\textwidth, height=0.13\textheight]{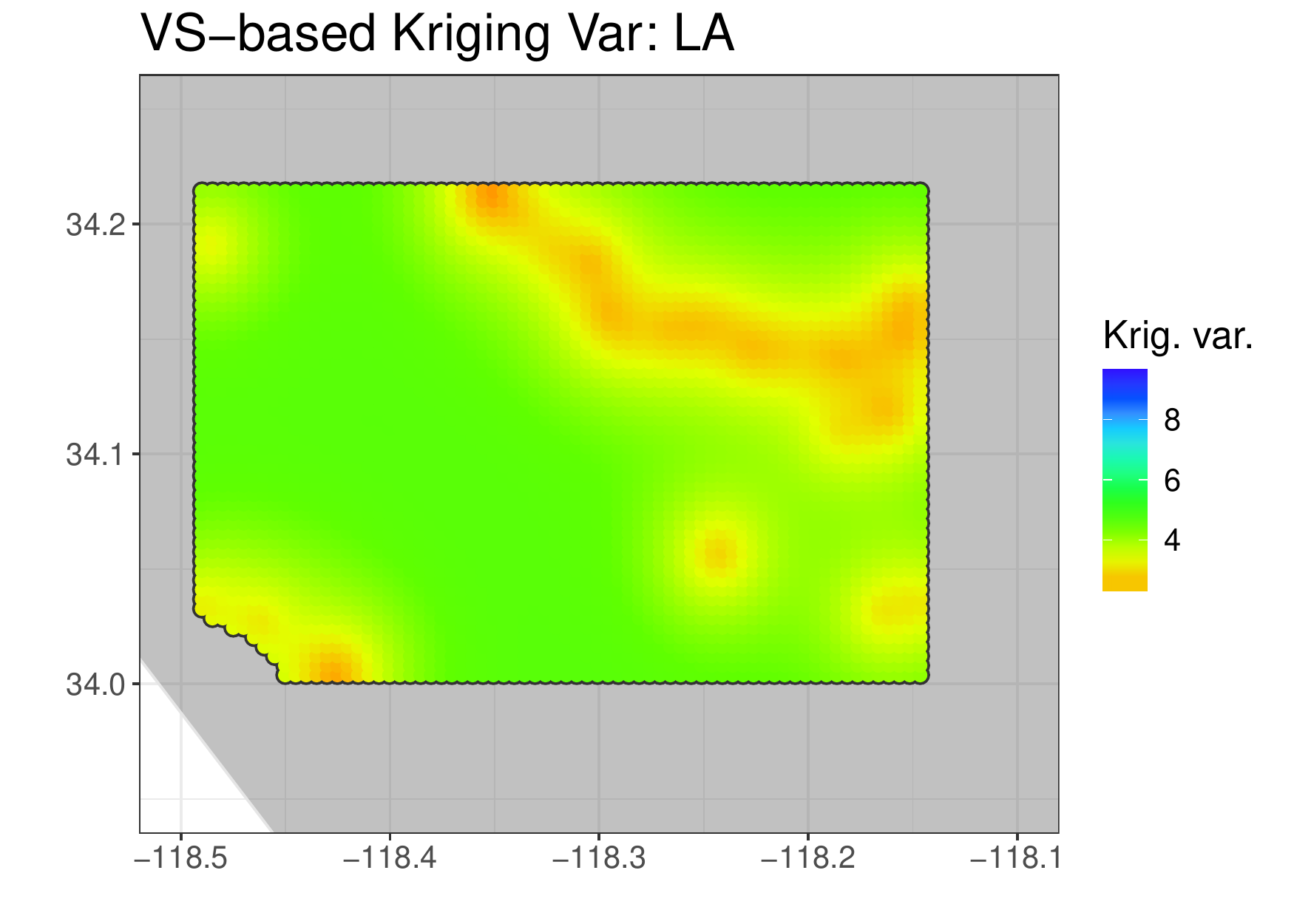}
    \subcaption{}\label{fig:KrigVar_VS_LA}
 \end{subfigure}%
 \begin{subfigure}{0.33\textwidth}
     \centering
    \includegraphics[trim={1.5cm 0.5cm 4.5cm 0.5cm}, width=.65\textwidth, height=0.13\textheight]{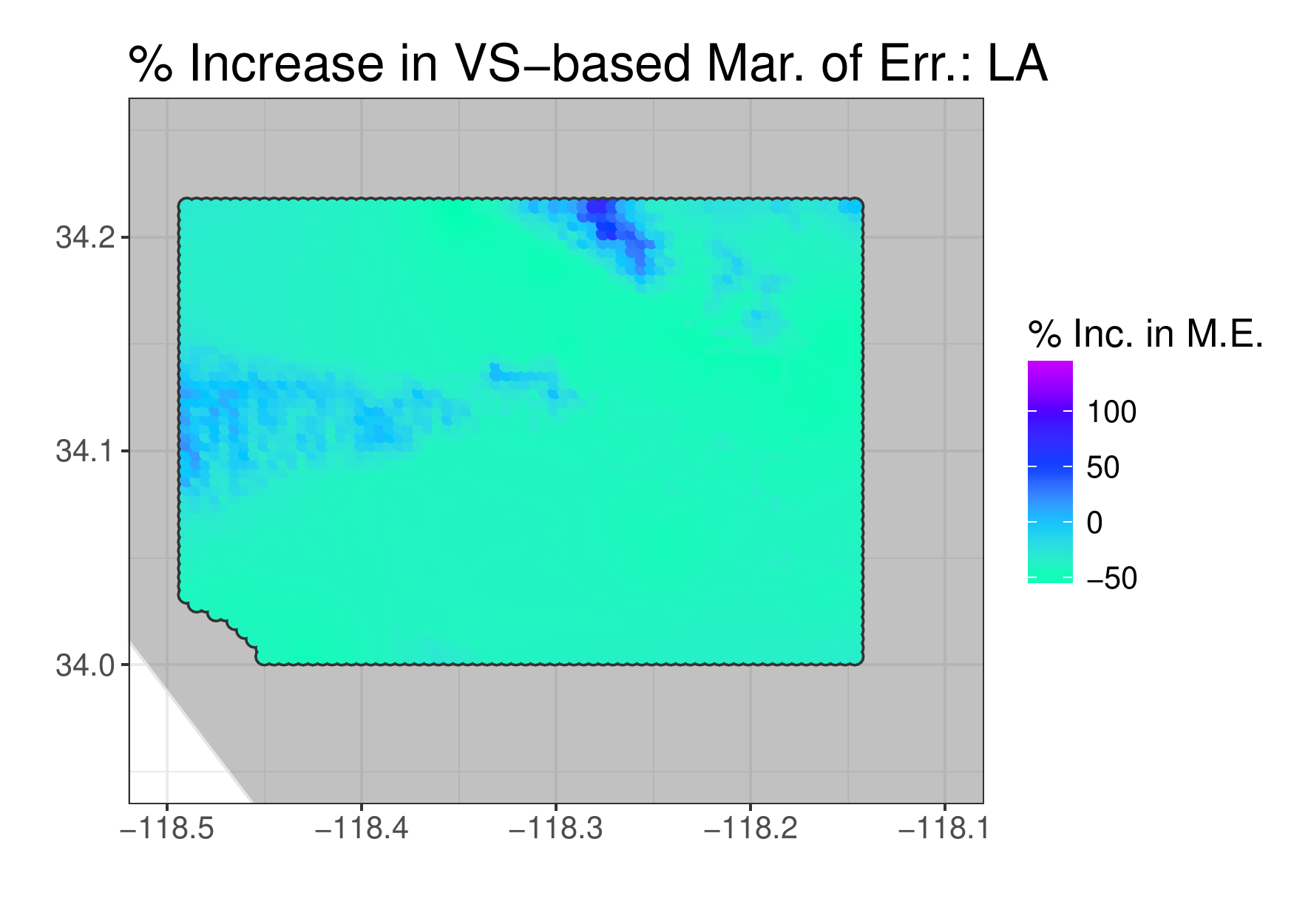}
    \subcaption{}\label{fig:ME_Inc_VS_NOAA_LA}
 \end{subfigure}
 \caption{\small{(a) Hyper-local version of the same surface as in Figure~\ref{fig:PredSur_NoaaOnly_Cali}; (b) Prediction surface obtained by the VS-based technique on the crowdsourced data in Los Angeles; (c) Residual kriging variance for the predictions using NOAA data only (d) Residual kriging variance for the predictions using the VS-based predictions with crowdsourced data; (e) the \% increase in the margin of error for the VS-based predictions as compared to the predictions with NOAA data}}\label{fig:LA_Pred_Surf_Plot}
 \end{figure}
 In Figure~\ref{fig:LA_Pred_Surf_Plot}, we plot the hyper-local prediction surfaces obtained by the standard analysis with the NOAA ground-station data only as well as the one obtained by implementing the VS-based technique on the crowdsourced observations with the ground-station data as the reference. Clearly the prediction surface obtained from standard analysis of the ground-station data (Figure~\ref{fig:PredSurf_Noaa_LA}) is too smooth to capture the local variability accurately. The prediction surface obtained by the VS-based analysis on crowdsourced data shows more variation across the space. To highlight the advantage of having crowdsourced observations, we compare the residual kriging variance surfaces in Figures~\ref{fig:KrigVar_Noaa_LA} and \ref{fig:KrigVar_VS_LA}. It is prominent from Figure~\ref{fig:KrigVar_VS_LA} that, the VS-based kriging variance is much smaller as compared to the global kriging using only the ground-station data, especially at locations that are close to the crowdsourced observations.
 
 In addition, we illustrate the gain in efficiency by plotting the percentage increase in margin of error (at $95\%$ confidence) for the VS-based predictions from the hyper-local crowdsourced information as compared to the global prediction using ground-station data only, i.e., $100 \times \Lp \text{M.E.}(\hat{Y}_\text{vs}(\bfs)) - \text{M.E.}(\hat{Y}(\bfs)) \Rp/\Lp \text{M.E.}(\hat{Y}(\bfs)) \Rp$, where M.E. denotes the `margin of error' (half of the length of the prediction interval) to predict the target response $Y(\bfs)$. To compute the margin of error, we use ad hoc confidence intervals for the residual kriging predictor with $\rpm 1.96$ as the corresponding quantiles and then add the margin of error of the mean ($1.96 \times \text{s.e.}(\bfx(\bfs)^\prime \hat{\bfbeta}_{\text{vs}})$) and the margin of error of the residual kriging predictor ($1.96 \times \sqrt{\text{Krig.Var.}(\tilde{\epsilon}(\bfs))}$). The margin of error for the standard predictor is computed similarly. A more theoretically justifiable interval can be obtained through spatial re-sampling technique as discussed in \cite{lahiri03}, but that requires further research and is beyond the scope of this study. In Figure~\ref{fig:ME_Inc_VS_NOAA_LA}, for most of the locations where the predictions have been carried out, there are decrease in the margin of errors for the VS-based predictions as compared to the global predictions using ground-station data only. At the locations that are close the crowdsourced observations, the VS-based prediction technique has achieved up to a $50\%$ gain in efficiency.
 
 The disadvantage of VS-based hyper-local analysis is that the model is estimated very regionally and hence extrapolation of the estimated mean model outside the sample space is likely to give misleading and inefficient predictions. For example, in Figure~\ref{fig:PredSurf_VS_LA} there are locations with elevations of more than 500 meters, while the maximum elevation in the crowdsourced sample is 350 meters. This leads to poor predictions (e.g., ambient temperature less than $50^\circ$F) at some locations as can be seen in Figure~\ref{fig:ME_Inc_VS_NOAA_LA}. Note that, though in those regions the efficiency of VS-based predictions fall short, the residual kriging variance (Figure~\ref{fig:KrigVar_Noaa_LA} and ~\ref{fig:KrigVar_VS_LA}) for the VS-based kriging predictor is still less than the global kriging with NOAA data only. So, the loss in efficiency in VS-based predictions is solely due to the the extrapolation of the hyper-locally estimated mean function at points outside the covariate sample space.
 
\begin{figure}
  \centering
   \begin{subfigure}{0.45\textwidth}
     \centering
    \includegraphics[trim={3.5cm 6cm 2.5cm 4cm}, width=.65\textwidth, height=0.2\textheight]{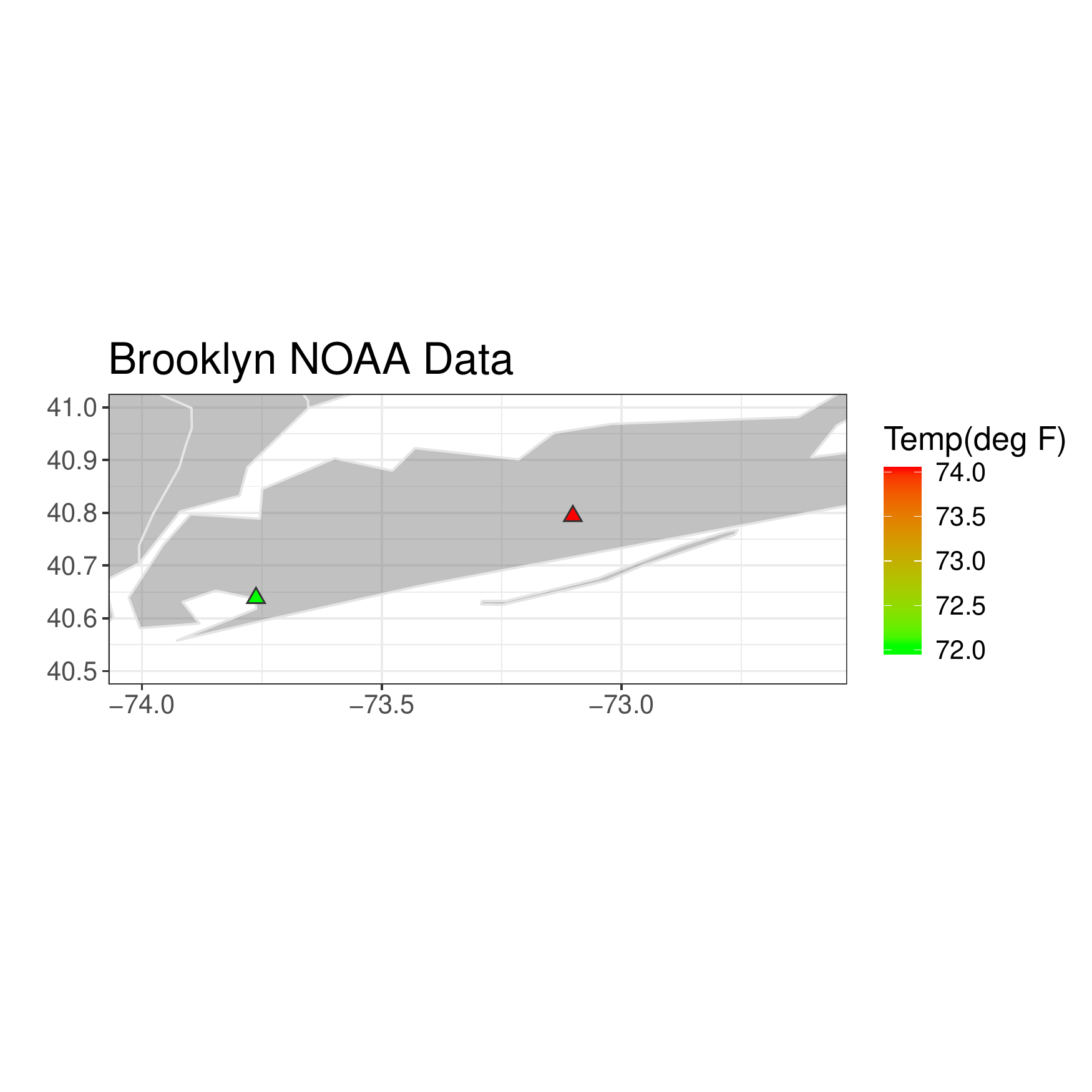}
    \subcaption{}\label{fig:GS_Data_Nyc}
 \end{subfigure}\hspace{6mm}%
 \begin{subfigure}{0.45\textwidth}
     \centering
    \includegraphics[trim={2.5cm 6cm 3.5cm 4cm}, width=.65\textwidth, height=0.2\textheight]{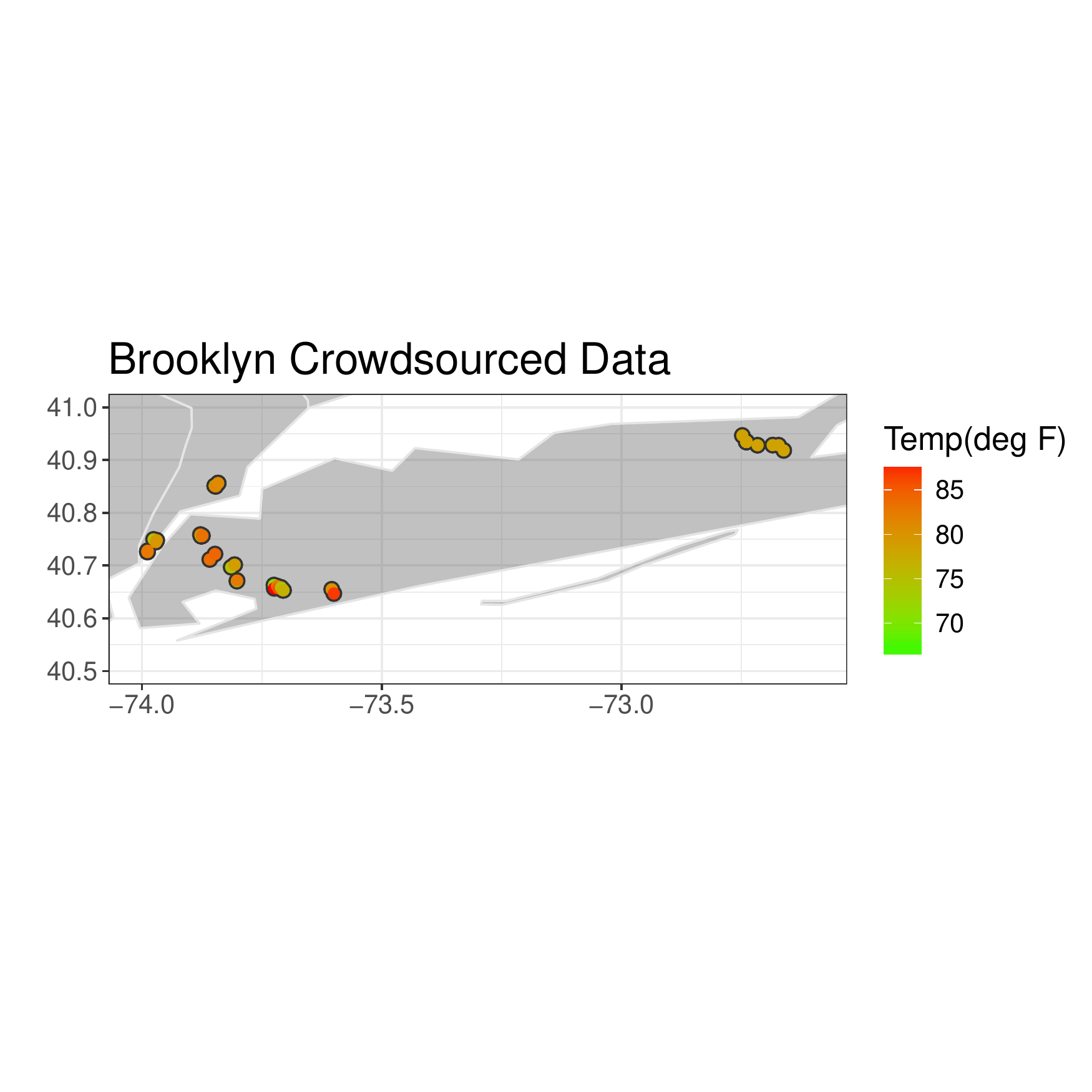}
    \subcaption{}\label{fig:CS_Data_Nyc}
 \end{subfigure}
 \begin{subfigure}{0.45\textwidth}
     \centering
    \includegraphics[trim={3.5cm 3.5cm 2.5cm 2cm}, width=.65\textwidth, height=0.2\textheight]{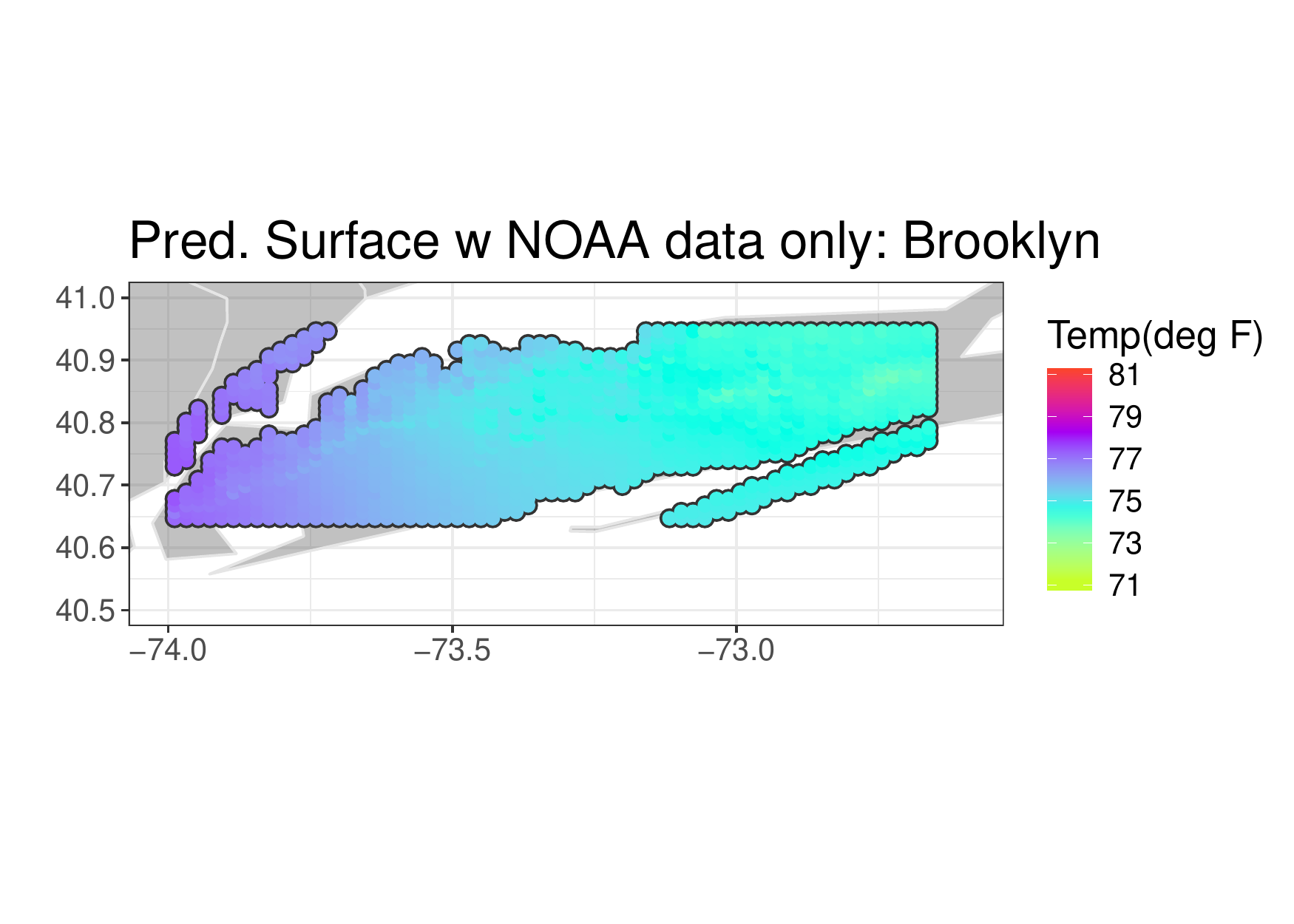}
    \subcaption{}\label{fig:Pred_Surf_Noaa_Nyc}
 \end{subfigure}\hspace{10mm}%
 \begin{subfigure}{0.45\textwidth}
     \centering
    \includegraphics[trim={2.5cm 3.5cm 3.5cm 2cm}, width=.65\textwidth, height=0.2\textheight]{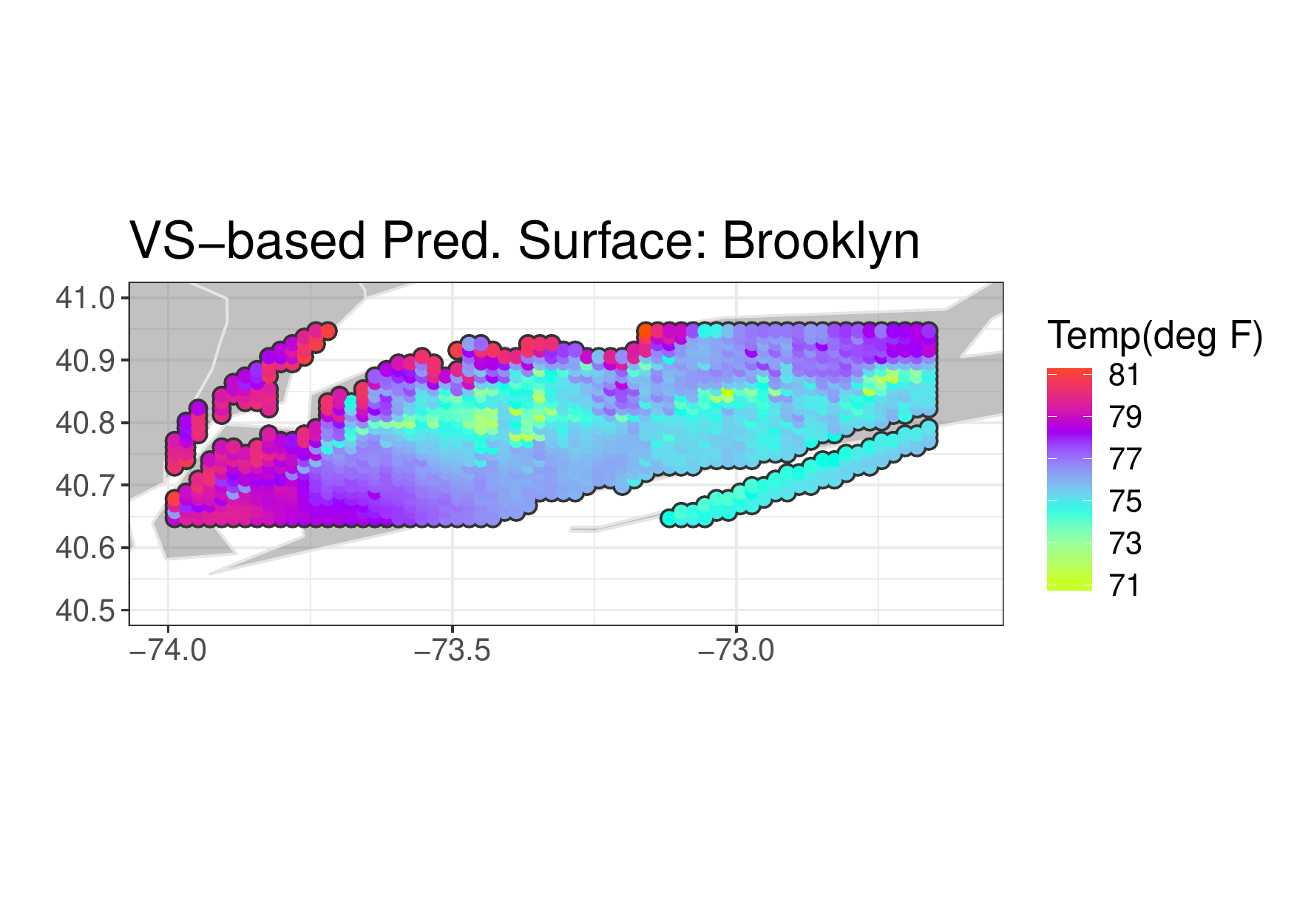}
    \subcaption{}\label{fig:Pred_Surf_VS_Nyc}
 \end{subfigure}
 \begin{subfigure}{0.33\textwidth}
     \centering
    \includegraphics[trim={3.5cm 3.5cm 1.5cm 1.5cm}, width=.7\textwidth, height=0.18\textheight]{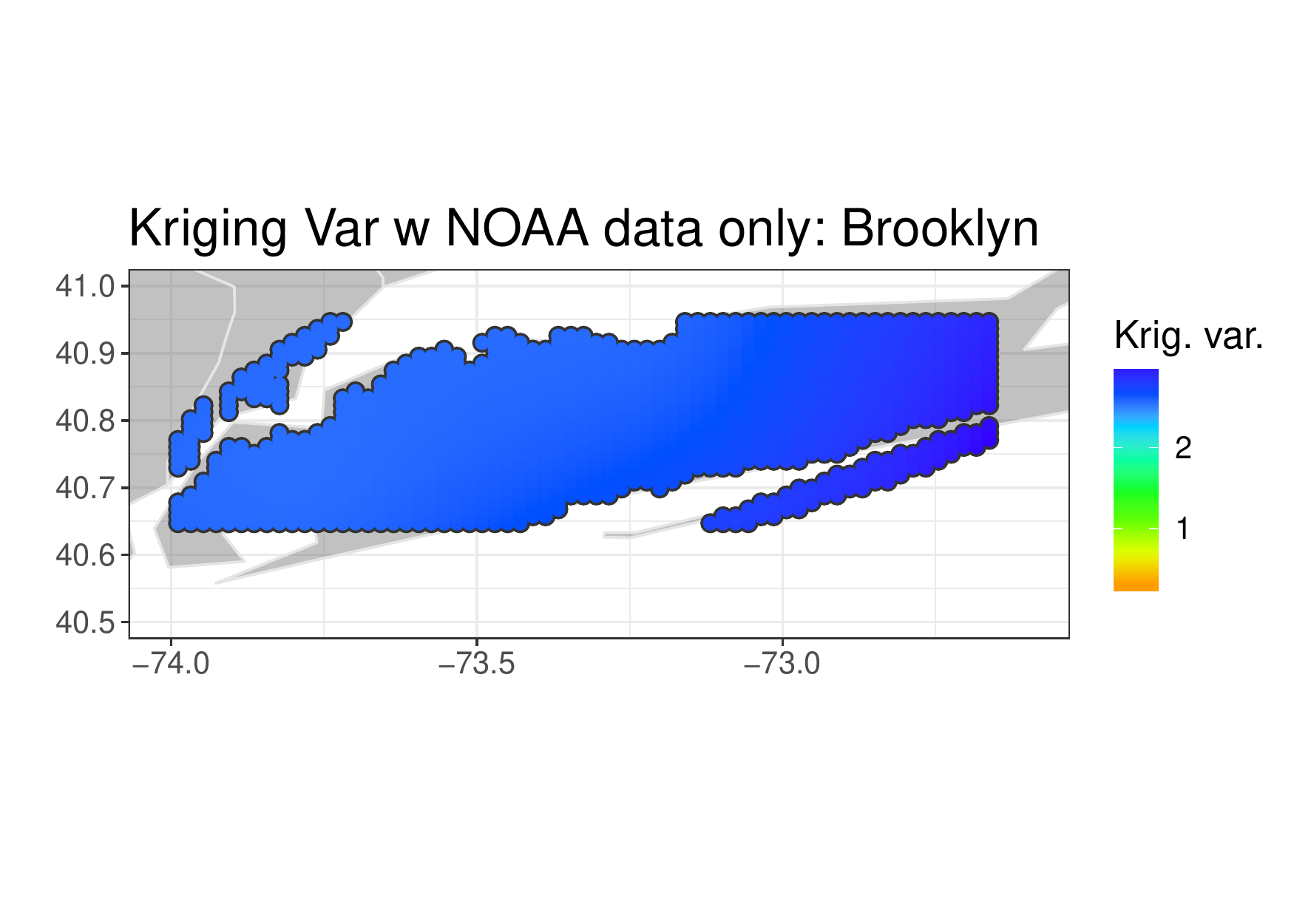}
    \subcaption{}\label{fig:Krig_Var_Noaa_Nyc}
 \end{subfigure}%
  \begin{subfigure}{0.33\textwidth}
     \centering
    \includegraphics[trim={2.5cm 3.5cm 2.5cm 1.5cm}, width=.7\textwidth, height=0.18\textheight]{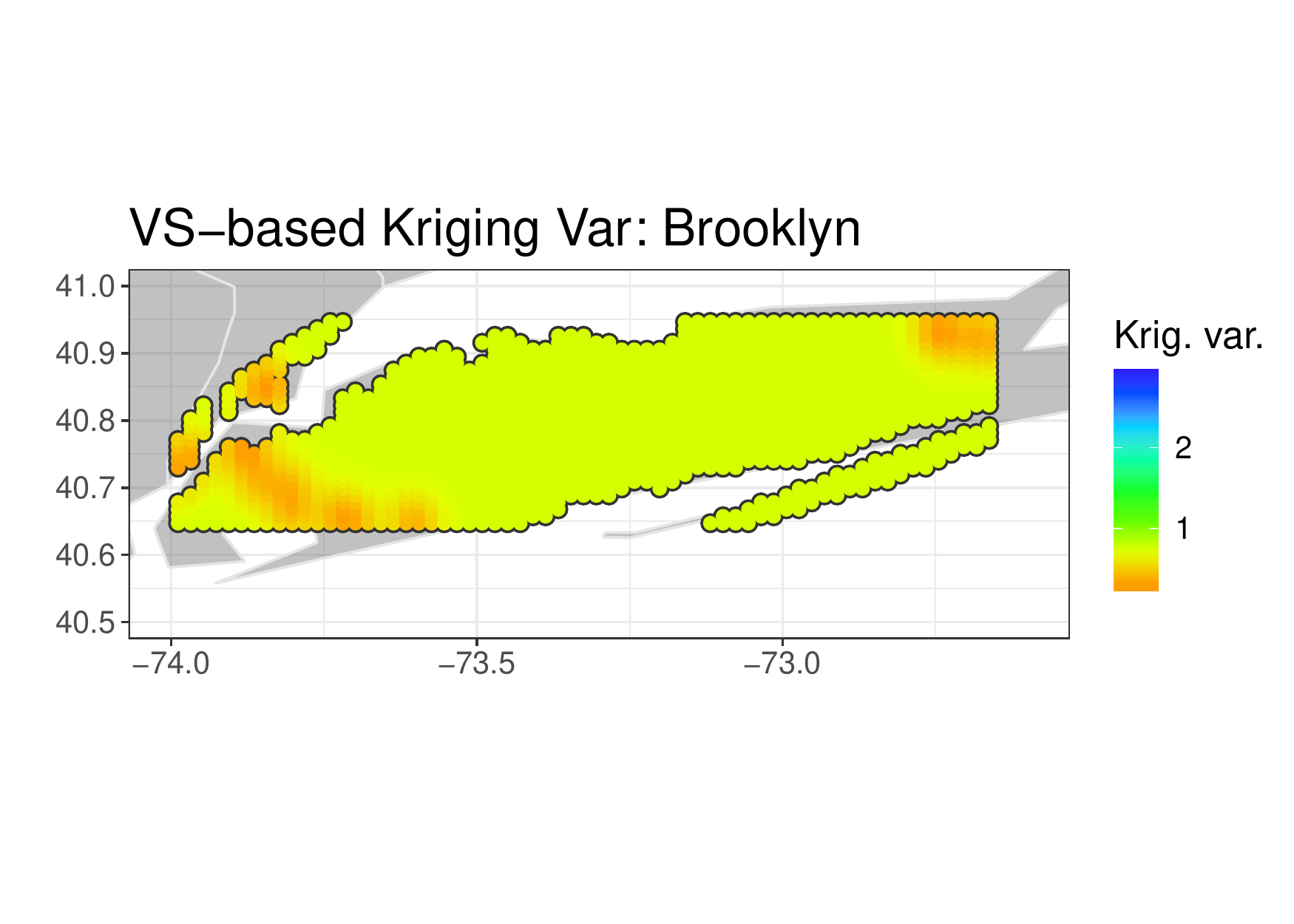}
    \subcaption{}\label{fig:Krig_Var_VS_Nyc}
 \end{subfigure}%
 \begin{subfigure}{0.33\textwidth}
     \centering
    \includegraphics[trim={1.5cm 3.5cm 3.5cm 1.5cm}, width=.7\textwidth, height=0.18\textheight]{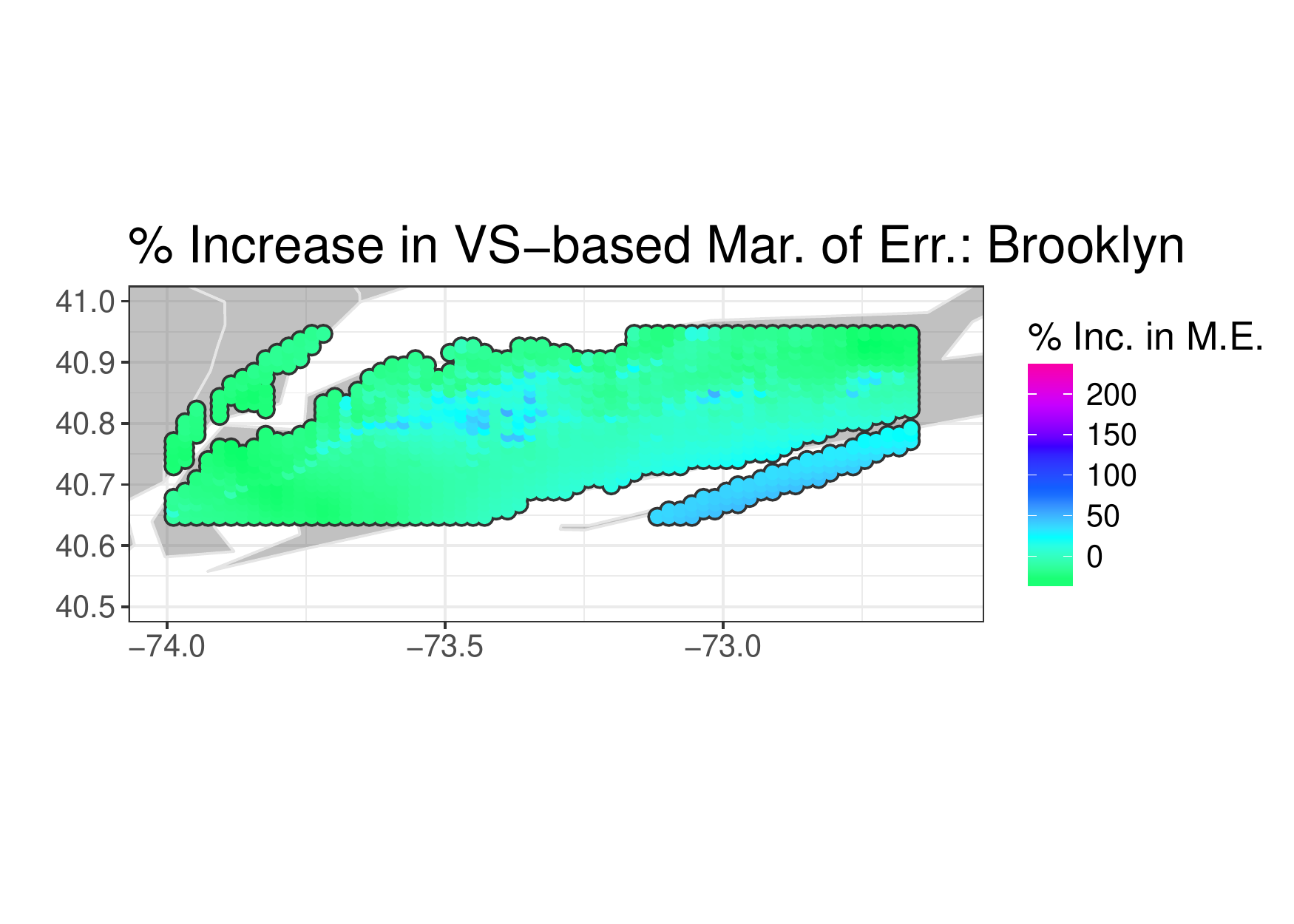}
    \subcaption{}\label{fig:ME_Inc_VS_Noaa_Nyc}
 \end{subfigure}
 \caption{\small{(a) Ground-station observations in the selected hyper-local region; (b) Crowdsourced observations in the same region (c) Prediction surface obtained by standard analysis of NOAA ground-station data; (d) Prediction surface obtained by the VS-based technique on the crowdsourced data; (e) Residual kriging variance for predictions using NOAA data only; (f) Residual kriging variances for the predictions using the crowdsourced data; (g) Percent increase in the margin of error for the VS-based predictions compared to the predictions with NOAA data}}\label{fig:NYC_Analysis_All_Plot}
 \end{figure}

We conduct a similar analysis for another hyper-local region close to Brooklyn, NY and plot the results in Figure~\ref{fig:NYC_Analysis_All_Plot}. 
 The prediction surface in Figure~\ref{fig:Pred_Surf_Noaa_Nyc} is obtained by using standard methodology on $120$ ground-station observations over the east-coast; and the suface in Figure~\ref{fig:Pred_Surf_VS_Nyc} is generated through VS-based hyper-local analysis of the crowdsourced observations in Figure~\ref{fig:CS_Data_Nyc}. Comparing these two prediction surfaces, we again see that the regional variation is prominent for the prediction surface obtained from VS-vased hyper-local analysis where as, the global analysis generates a surface that is too smooth to accurately capture local variations. In Figure~\ref{fig:Krig_Var_VS_Nyc}, the advantage of having crowdsourced data for hyper-local prediction of the process is visible, as the kriging variance of the VS-based methodology is much smaller compared to Figure~\ref{fig:Krig_Var_Noaa_Nyc}, especially in locations close to the crowdsourced observations. In Figure~\ref{fig:ME_Inc_VS_Noaa_Nyc}, we see up to $33\%$ gain in margin of error by implementing the VS-based methodology on the crowdsourced data in locations close to the crowdsourced observations. Similar to the previous analysis of the Los Angeles data, the advantage of the VS-based hyper-local predictions is lost if the predictions are attempted at locations too far from the crowdsourced observations or at locations with elevations outside the range of crowdsourced sample.
 
 In addition to the VS-based hyper-local analysis, we have also conducted the analysis for both of the hyper-local regions in Los Angeles and Brooklyn with the standard approach without considering the veracity of the crowdsourced observations and then compared the predictions with the global prediction surface obtained using reference data only.
 \begin{figure}
  \centering
   \begin{subfigure}{0.45\textwidth}
     \centering
    \includegraphics[trim={3.5cm 3.8cm 2.5cm 1.2cm}, width=.65\textwidth, height=0.1\textheight]{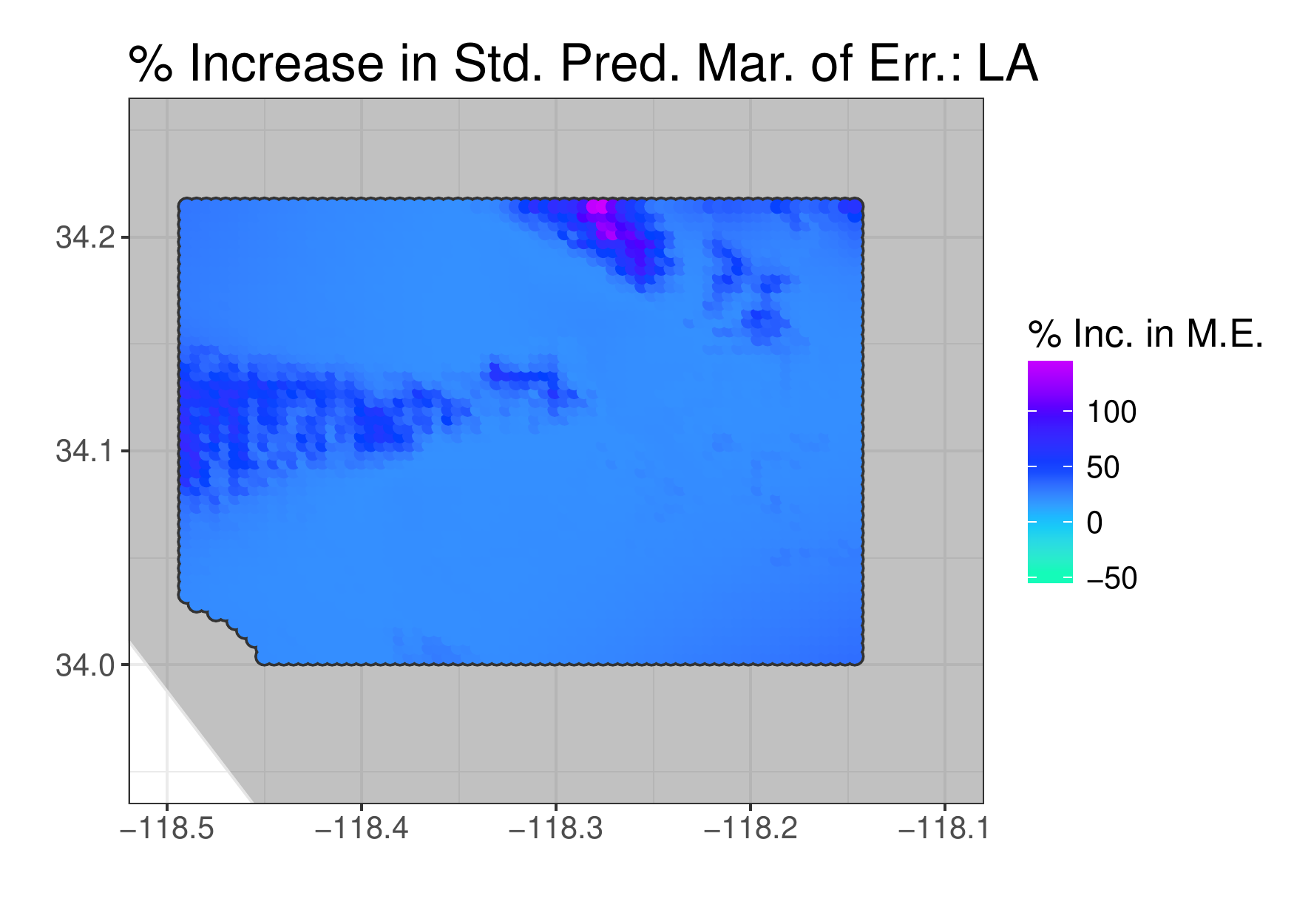}
 \end{subfigure}\hspace{5mm}%
 \begin{subfigure}{0.45\textwidth}
     \centering
    \includegraphics[trim={3.5cm 3.8cm 2.5cm 1.2cm}, width=.68\textwidth, height=0.18\textheight]{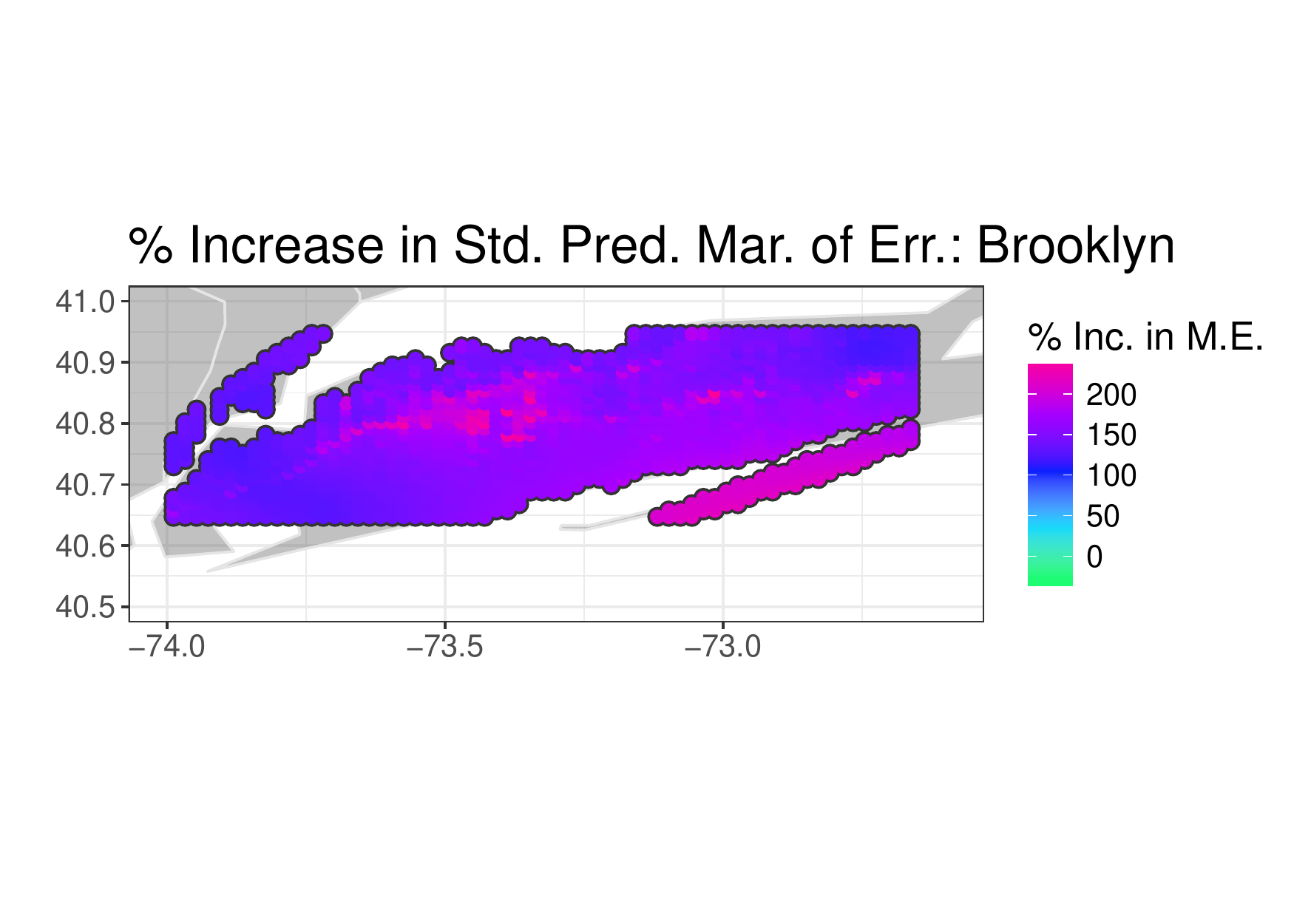}
 \end{subfigure}
 \caption{\small{The increase in margin of error for the standard approach in hyper-local regions in Los Angeles (left) and Brooklyn (right).}}\label{fig:ME_Inc_Std_Noaa}
\end{figure} Comparing the plots in Figure~\ref{fig:ME_Inc_Std_Noaa} with Figure~\ref{fig:ME_Inc_VS_NOAA_LA} and Figure~\ref{fig:ME_Inc_VS_Noaa_Nyc} we can see that, in both Los Angeles and Brooklyn, the margins of error for the predictions using the standard approach are larger in all the locations as compared to the global predictions using ground-station data. In Brooklyn, even at the locations around the crowdsourced observations, with reference to the global prediction using ground-station data, the margin of error of standard predictions using the crowdsourced observations have increased by at the least $120\%$, where as, as we have mentioned already, the VS-based methodology has achieved a decrease in the margin of error up to $33\%$ (Figure~\ref{fig:ME_Inc_VS_Noaa_Nyc}). Clearly, no gain from the `hyper-local' analysis is achieved, as compared to the `global' prediction from the ground-station data, unless the robust VS-based methodology is employed on the varying-quality crowdsourced data.

\subsection{Validation at the ground-stations}
\label{subsec:PredAtGr.St.} The goal of the analysis in this section is to validate the predictions obtained by hyper-local analysis of crowdsourced data using VS-based methodology. To do so, we have selected a set of 14 ground-stations that satisfy the following criteria: (1) there are at least 30 crowdsourced data points available nearby with at least 20 observations with a VS greater or equal to $0.4$; (2) the elevation of those stations is not too far from the range of the local crowdsourced samples. We have conducted 14 hyper-local analyses, as described in Section~\ref{subsec:HyperLocalPred}, for hyper-local structure exploration of the ambient temperature and then predicted at those selected ground-station locations to validate the VS-based predictions. We have omitted these 14 stations before-hand so that these are not used in defining the `benchmark' value at the crowdsourced data locations to compute VS; this way the validation data has no effect on the training phase of the predictions. We have also conducted the same hyper-local analyses using the standard technique without taking the quality of the observations into account.  
\begin{table}[ht]
\resizebox{1\columnwidth}{.18\textwidth}{%
\begin{tabular}{|l|c|c|c|c|c|}
\hline
\multicolumn{1}{|c|}{\textbf{STATION\_NAME}}    & \textbf{Target Temp.} & \textbf{PredTemp.VS} & \textbf{VS.ME} & \textbf{PredTemp.Std} & \textbf{Std.ME} \\ \hline\hline
CHICAGO OHARE INTERNATIONAL AIRPORT IL US & 76 & 76.01 & 6.22 & 76.75 & 8.74 \\ 
WASHINGTON DULLES INTERNATIONAL AIRPORT VA US & 79 & 82.80 & 5.13 & 75.33 & 18.35 \\ 
WASHINGTON REAGAN NATIONAL AIRPORT VA US & 80 & 81.95 & 7.77 & 75.52 & 31.64 \\ 
MIAMI INTERNATIONAL AIRPORT FL US & 79 & 77.81 & 0.50 & 78.57 & 1.23 \\ 
LITTLE TUJUNGA CALIFORNIA CA US & 68 & 64.78 & 6.01 & 63.74 & 7.41 \\ 
LOS ANGELES INTERNATIONAL AIRPORT CA US & 68 & 68.91 & 3.31 & 67.87 & 4.26 \\ 
BEVERLY HILLS CALIFORNIA CA US & 70 & 67.94 & 6.27 & 68.12 & 7.54 \\ 
TOLEDO EXPRESS AIRPORT OH US & 75 & 79.45 & 5.72 & 79.39 & 8.18 \\ 
DETROIT METROPOLITAN AIRPORT MI US & 76 & 78.79 & 6.66 & 80.74 & 9.73 \\ 
MINNEAPOLIS ST PAUL INTERNATIONAL AIRPORT MN US & 70 & 77.03 & 6.67 & 76.97 & 10.96 \\ 
CARLOS AVERY MINNESOTA MN US & 69 & 73.33 & 11.46 & 74.65 & 19.05 \\ 
JFK INTERNATIONAL AIRPORT NY US & 72 & 78.24 & 3.06 & 80.82 & 3.82 \\ 
ISLIP LI MACARTHUR AIRPORT NY US & 74 & 75.10 & 6.29 & 75.61 & 8.79 \\ 
AUSTIN BERGSTROM INTERNATIONAL AIRPORT TX US & 81 & 78.87 & 2.24 & 86.50 & 10.19\\
\hline
\end{tabular}%
}
\caption{\small{Predictions using both the VS-based and standard approach at the ground-stations with crowdsourced observations in proximity.}}\label{tab:PredAtGrSt}
\end{table} The results are compiled in Table~\ref{tab:PredAtGrSt}. The advantage of using the VS-based techniques as compared to the standard methodology is clear from the results. The RMSPE of the VS-based predictor for these 14 ground-stations is 3.71, while for the standard approach it is 4.54. More importantly, the average margin of error (at $95\%$ confidence) for standard predictor is 13.61 and for the VS-based methodology it is 6.28. Relative to the standard methodology, on average, the VS-based technique has achieved approximately $54\%$ gain in efficiency of the predictions.

\section{Summary and Conclusions}
\label{sec:conc} In this paper, we have introduced the veracity score to assess the quality of observations in geostatistical setting. The VS is defined by comparing the varying quality observations with a benchmark. We used the ground-station data as our reference to define the benchmark values in the case studies. The similar scoring approach to assess the veracity of the observations can be used in other contexts as well. We have also discussed the case when no other reference information is available and propose a version of VS using locally and robustly estimated measure of center as the benchmark. A robust approach for modeling varying-quality spatial data using the VS has been proposed and evaluated. We have illustrated the VS-based methodology on a crowdsourced data set coming from the mobile app WeatherSignal using NOAA ground-station data as the reference. Both the simulation studies in Section~\ref{sec:simStudy} and the case studies in Section~\ref{subsec:HyperLocalPred} show the advantages of the VS-based methodology over the standard geostatistical approach when dealing with noisy spatial data. In addition, by implementing the VS-based methodology on the varying-quality local crowdsourced data, we can achieve a more accurate and efficient hyper-local predictions as compared to the global prediction obtained from the analysis of ground-station data only.

In the analysis of crowdsourced data using the VS-based methodology, the model is estimated using observations in a hyper-local region. Predicting at more distant locations or with covariates outside the range of the sample may provide misleading predictions, as we have seen for some of the locations in Figure~\ref{fig:PredSurf_VS_LA} and Figure~\ref{fig:Pred_Surf_VS_Nyc}. The mean and covariance models used to explore the structures of the average temperature process are quite simple, yet reasonable and effective for hyper-local analysis of ambient temperature. More complex models like nonlinear regression models \citep{frei14} and anisotropic covariance \citep{haskard07} can be incorporated in the VS-based technique to increase flexibility of the analysis. The VS-based kriging automatically reduces the impact of the corrupted observations and thus, it does not require removing the outliers manually (e.g. see \citealt{frei14}) -- which is often not feasible when dealing with large crowdsourced spatial data. In addition, as the veracity of the observations has been measured non-parametrically using `local' summaries, the proposed VS-based kriging does not require any distributional assumption (e.g. Gaussian, see \citealt{lussana10}) on the underlying spatial process or the noise associated with it. The analysis presented in this paper shows that the systematic incorporation of VS in the geostatistical analysis helps us capture the local variability of the ambient temperature field by considering crowdsourced data in hyper-local regions. The VS-based kriging decreases the margin of prediction errors up to $50\%$ as compared to the global predictions from ground-station data only. On the other hand, if the same analysis is carried out on the noisy crowdsourced data with standard kriging, there is no gain in efficiency. In fact, there are locations, even close to the crowdsourced observations, where the margin of prediction errors by standard methods are more than $80\%$ higher than the corresponding global predictions.

There are several interesting future directions for this work. First, we have not provided theoretical justification for the superiority of the VS-based methodology as compared to the standard approach in the analysis of noisy spatial data. Inspired by the simulations executed in this work, we believe that under a suitable spatial asymptotic framework (e.g. \textit{mixed-increasing domain}, \citealt{hall94}; \citealt{lahiri02}) and a fairly general non-stationary noise model (e.g. the \textit{additive-multiplicative} model defined in Equation~\ref{eq:AddMultNoiseMod}), we can theoretically justify the robustness and efficiency of the VS-based methodology (for details, see \citealt{chak-prop-vs}). Second, the methodology discussed in this article can be systemically extended to develop a more sophisticated VS-based kriging technique that incorporates both the ground-station data and the crowdsourced data for spatial prediction. Third, a spatio-temporal VS and corresponding methods for real-time crowdsourced data can be developed by considering neighborhoods in both space and time.

\section*{Acknowledgements}
This material is based upon work supported in whole or in part with funding from the Laboratory for Analytic Sciences (LAS). This research is also partially funded by National Science Foundation (NSF) grant DMS-1613192. Any opinions, findings, conclusions, or recommendations expressed in this material are those of the author(s) and do not necessarily reflect the views of the LAS and/or any agency or entity of the United States Government. Special thanks are extended to the OpenSignal team that was in-charge of the academic partnership program in 2015 for making the data available to the authors. The authors also thank the Editor, the Associated Editor and three anonymous referees for a number of thoughtful comments that significantly improved the paper.

\begin{supplement}
\sname{Supplement A}\label{suppA}
\stitle{Supplementary Material for Spatial Analysis of Noisy Crowdsourced Mobile Data}
\sdescription{This file contains additional details on data preprocessing, the simulations, and the case study. It contains additional plots, tables and discussions to support our claims and findings in the main article.}
\end{supplement}

\clearpage
\bibliographystyle{imsart-nameyear}
\bibliography{CrowdSourced_AoAS}

\end{document}